\RequirePackage{silence}
\documentclass[fleqn,usenatbib,useAMS]{mnras} 

\usepackage{newtxtext}

\usepackage{graphicx}
\usepackage{amsmath}

\usepackage{amsfonts}
\usepackage{float}
\usepackage{bm}
\usepackage[T1]{fontenc}
\usepackage{ae,aecompl}
\usepackage{array}
\usepackage{soul}
\usepackage{mathtools}
\usepackage{multirow}
\usepackage[T1]{fontenc}
\usepackage{lmodern}
\usepackage[utf8]{inputenc}
\usepackage{booktabs}
\usepackage{graphicx}
\usepackage{subcaption}
\usepackage[normalem]{ulem}
\usepackage{anyfontsize}

\DeclareRobustCommand{\VAN}[3]{#2}
\let\VANthebibliography\thebibliography
\def\thebibliography{\DeclareRobustCommand{\VAN}[3]{##3}\VANthebibliography}

\DeclareRobustCommand{\appropto}{\mathrel{\vcenter{
		\offinterlineskip\halign{\hfil$##$\cr 
			\propto\cr\noalign{\kern2pt}\sim\cr\noalign{\kern-2pt}}}}}

\defcitealias{Hernandez_2012}{H12}
\defcitealias{Hernandez_2019_WB}{H19}
\defcitealias{Pittordis_2019}{P19}
\defcitealias{Hernandez_2022}{H22}
\defcitealias{Pittordis_2023}{P23}
\defcitealias{Hernandez_2023}{H23}
\defcitealias{Chae_2023}{C23}
\defcitealias{Chae_2024a}{C24a}
\defcitealias{Chae_2024b}{C24b}
\defcitealias{Banik_2024_WBT}{B24}
\defcitealias{Cookson_2024}{Co24}

\title[A quality framework for the wide binary test]{A Quality Framework for Testing Gravity with Wide Binaries: No Evidence for MOND}
\author[S. A. Cookson et al.]{
Stephen A. Cookson$^{1}$\thanks{E-mail: \href{mailto:stephen.cookson@sca-uk.com}{stephen.cookson@sca-uk.com} (SAC)},
Indranil Banik$^{2}$\thanks{E-mail: \href{mailto:indranil.banik@port.ac.uk}{indranil.banik@port.ac.uk} (IB)},
Kareem El-Badry$^{3}$\thanks{E-mail: \href{mailto:kelbadry@caltech.edu}{kelbadry@caltech.edu} (KEB)},
\newauthor
Will Sutherland$^{4}$\thanks{E-mail: \href{mailto:w.j.sutherland@qmul.ac.uk}{w.j.sutherland@qmul.ac.uk} (WS)},
Zephyr Penoyre$^{5}$\thanks{E-mail: \href{mailto:zephyrpenoyre@gmail.com}{zephyrpenoyre@gmail.com} (ZP)},
Charalambos Pittordis$^{4}$\thanks{E-mail: \href{mailto:cp.pittordis@gmail.com}{cp.pittordis@gmail.com} (CP)},
Cathie J. Clarke$^{6}$\thanks{E-mail: \href{mailto:cclarke@ast.cam.ac.uk}{cclarke@ast.cam.ac.uk} (CJC)}\\
$^{1}$Independent Researcher, Crawley, UK\\
$^{2}$Institute of Cosmology \& Gravitation, University of Portsmouth, Dennis Sciama Building, Burnaby Road, Portsmouth PO1 3FX, UK\\
$^{3}$Division of Physics, Mathematics and Astronomy, California Institute of Technology, Pasadena, CA 91125, USA\\
$^{4}$Astronomy Unit, School of Physical and Chemical Sciences, Queen Mary University of London, Mile End Road, London E1 4NS, UK\\
$^{5}$Leiden Observatory, Leiden University, PO Box 9513, 2300 RA Leiden, The Netherlands\\
$^{6}$Institute of Astronomy, University of Cambridge, Madingley Road, Cambridge CB3 0HA, UK}

\date{Accepted XXX. Received YYY; in original form ZZZ}

\pubyear{\the\year{}}

\begin{document}
\label{firstpage}
\pagerange{\pageref{firstpage}--\pageref{lastpage}}
\maketitle

\begin{abstract} 
Wide binaries (WBs) offer a unique opportunity to test gravity in the low-acceleration regime, where modifications such as Milgromian dynamics (MOND) predict measurable deviations from Newtonian gravity. We construct a rigorous framework for conducting the wide binary test (WBT), emphasizing high quality sample selection, filtering of poor astrometric solutions, contamination mitigation, and uncertainty propagation. We show that undetected close binaries, chance alignments, and improper treatment of projection effects can mimic MOND-like signals. We introduce a checklist of best practices to identify and avoid these pitfalls. Applying this framework to \emph{Gaia} DR3 data, we compile a high-purity sample of WBs within 130 pc with projected separations of $1-30$~kAU, spanning the transition between the Newtonian and MOND regimes. We find that the scaled relative velocity distribution of wide binaries does not exhibit the 20\% enhancement expected from MOND and is consistent with Newtonian gravity across all separations. A meta-analysis of previous WBTs shows that apparent MOND signals diminish as methodological rigour improves. We conclude that when stringent quality controls are applied, there is no observational evidence for MOND-induced velocity boosts in wide binaries. Our results place strong empirical constraints on modified gravity theories operating between $a_0/10$ and $200 \, a_0$, where $a_0$ is the MOND acceleration scale. Across this range of internal accelerations, Newtonian gravity is up to $1500\times$ more likely than MOND for our cleanest sample.

\end{abstract}


\section{Introduction}
\label{Introduction}

It is well known that more distant planets in the Solar System orbit the Sun at a lower velocity, a fact which can be explained by the Newtonian inverse-square law of gravity. However, applying this law to the observed matter distribution in galaxies leads to disagreement with observations, which instead show that the rotation velocity remains constant beyond the bulk of the visible mass \citep[][and references therein]{Faber_1979}. 

This missing gravity problem has led to the suggestion that galaxies are embedded in massive haloes of dark matter \citep*{Ostriker_Peebles_1973, Ostriker_Peebles_Yahil_1974, White_Rees_1978}. The discovery of missing gravity in other contexts like galaxy clusters \citep*{Zwicky_1933, Clowe_2004, Pointecouteau_2005} and the cosmic microwave background \citep[CMB;][]{Planck_2020, Tristram_2024} lent further support to the idea of dark matter that has little or no interaction with ordinary matter besides gravity \citep[for a review, see][]{Peebles_2017_DM_review}. Nowadays, this hypothesis is an integral part of the standard cosmological paradigm known as $\Lambda$-Cold Dark Matter \citep*[$\Lambda$CDM;][]{Efstathiou_1990, Ostriker_Steinhardt_1995}.

An alternative solution to the missing gravity problem is known as Milgromian dynamics \citep[MOND;][]{Milgrom_1983}. Its conventional modified gravity form \citep{Bekenstein_Milgrom_1984, QUMOND} postulates that the Newtonian gravity $g_{_N}$ due to the baryons alone must be enhanced by some acceleration-dependent factor $\nu$ with argument $g_{_N}$ \citep[for extensive reviews, see][]{Famaey_McGaugh_2012, Banik_Zhao_2022}. To remain consistent with Solar System ephemerides while reproducing the observed flat galaxy rotation curves \citep{Rubin_1970, Rogstad_1972, Roberts_1973, Roberts_1975, Bosma_1978, Rubin_1978_NGC4378, Rubin_1978, Bosma_1981_21cm, Bosma_1981}, the interpolating function $\nu$ must have the following asymptotic limits:
\begin{eqnarray}
    \nu \left( g_{_N} \right) ~\to~ \begin{cases}
    1 \, , & \textrm{if} ~g_{_N} \gg a_{_0} \, , \\
    \sqrt{\frac{a_{_0}}{g_{_N}}} \, , & \textrm{if} ~g_{_N} \ll a_{_0} \, .
    \end{cases}
    \label{nu_cases}
\end{eqnarray}
The new physics is encapsulated by $a_{_0}$, a new fundamental constant of nature with dimensions of acceleration. The latest data require a fairly gradual transition function \citep[see the discussion in section~5.3 of][]{Banik_2024_WBT}. We will assume $\nu$ takes the MLS form \citep*{McGaugh_Lelli_2016} because it works well with the Spitzer Photometry and Accurate Rotation Curves \citep[SPARC;][]{SPARC} catalogue of galaxy photometry in the near-infrared and their corresponding rotation curves taken from the literature. Interpreting their equation~4 as the MOND interpolating function, MOND predicts that for an isolated spherically symmetric system, the actual gravity
\begin{eqnarray}
    g ~=~ \nu g_{_N} ~=~ \frac{g_{_N}}{1 - \exp \left(-\sqrt{g_{_N}/a_{_0}} \right)}  \, ,
    \label{eq:observed_gravity}
\end{eqnarray}
with galaxy rotation curves implying that $a_0 = 1.2 \times 10^{-10}$~m/s$^2$ \citep*{Begeman_1991, Gentile_2011, McGaugh_Lelli_2016}. MOND has enjoyed unparalleled predictive success with the rotation curves of disc galaxies \citep{McGaugh_2020}. Its prediction of a tight radial acceleration relation (RAR) between $g$ and $g_{_N}$ also works in ellipticals \citep{Lelli_2017, Shelest_2020} and with stacked galaxy-galaxy weak lensing, which can probe $g$ hundreds of kpc from a galaxy in a statistical sense \citep{Brouwer_2017, Brouwer_2021, Mistele_2024}.

An interesting aspect of MOND is that it lacks a fundamental length scale. For any isolated point mass $M$, deviations from Newtonian gravity set in at distances $r \ga r_{_{\mathrm{M}}}$, where the MOND radius
\begin{eqnarray}
    r_{_{\mathrm{M}}} ~\equiv~ \sqrt{\frac{GM}{a_{_0}}} \, ,
    \label{r_M}
\end{eqnarray}
the radius at which $g_{_N} = a_{_0}$. By construction, the MOND radius of a typical galactic mass is several kpc. However, the MOND radius of the Sun is only $7$~kAU or 0.034~pc. MOND effects should be detectable at even smaller distances, especially in `saddle regions' where the gravity from the Sun cancels that of a giant planet \citep[][and references therein]{Penner_2020_SP}. MOND would also affect the orbits of Kuiper Belt Objects and comets in the Oort cloud, thus affecting the energy distribution of long-period comets that come close enough to the Sun to be detectable \citep{Penner_2020_comets, Vokrouhlicky_2024}. Even the orbit of Saturn would be slightly anomalous in MOND \citep{Hees_2014, Hees_2016}. The lack of any such anomaly in Cassini radio tracking data combined with galaxy rotation curve constraints leads to an $8.7\sigma$ failure of classical modified gravity MOND as the primary cause of the missing gravity problem \citep{Desmond_2024}.

Another consequence of MOND is that wide binary stars (WBs) in the Solar neighbourhood with separations of several kAU should orbit each other about 20\% faster than in Newtonian gravity once the Galactic gravity is taken into account \citep{Banik_2018_Centauri}. Although WBs are quite loosely bound, they can easily survive for several Gyr \citep{Jiang_2010, Feng_2018}. WBs are quite common, with even the nearest star to the Sun actually being part of a WB \citep{Beech_2009, Beech_2011, Kervella_2016, Kervella_2017}. The typical separation between field stars in the Solar neighbourhood is about 1~pc or 210~kAU, which greatly exceeds the MOND radius of a typical star. This leads to a range of WB separations which are ideal for testing MOND because $g_{_N} \ll a_{_0}$. Due to the long orbital periods, we have to make do with only the present separation and relative velocity of a WB, which cannot provide strong constraints from any individual system \citep[unless we reach $\mu$as astrometric precision; see][]{Banik_2019_Proxima}. However, a careful statistical analysis of many WBs could provide a very powerful test of MOND \citep*[as first proposed by][]{Hernandez_2012}. The kAU scales of WBs are much smaller than the hypothetical Galactic halo of CDM, so this would negligibly affect the internal dynamics of any WB \citep{Acedo_2020}. Consequently, any detected deviation from the Newtonian expectation cannot plausibly be ascribed to CDM. Conversely, a null detection of the predicted 20\% enhancement to WB orbital velocities would cast doubt on MOND given that the Galactic rotation curve is enhanced over the Newtonian baryonic expectation by about this much \citep{Zhou_2023, Zhu_2023}.

All these factors set the stage for the wide binary test (WBT), a comparison between the observed relative velocity distribution of WBs and the predictions of Newtonian and Milgromian dynamics. Thanks to exquisite astrometric data from the \emph{Gaia} mission \citep{Perryman_2001}, the WBT has rapidly emerged as a potentially decisive test of the most promising solution to the missing gravity problem that makes do with only actually detected mass and is consistent with the standard model of particle physics. There have been conflicting reports in the literature about the outcome of the WBT, with some studies favouring MOND \citep{Hernandez_2019_WB, Hernandez_2022, Hernandez_2023, Chae_2024a, Chae_2024b, Hernandez_2024_statistical, Hernandez_2025} and others favouring Newton \citep{Pittordis_2023, Cookson_2024, Banik_2024_WBT}. This motivates us to set up a standardized checklist for the WBT to improve its robustness and reliability, covering issues of both sample selection and data analysis. We do not discuss what would constitute a suitable forward model for the WB population, focusing instead on the features of the observed velocity distribution most relevant to the WBT.

After discussing theoretical expectations for the WBT in greater detail (Section~\ref{Theoretical_expectations}), we focus on best practices for selecting a suitable sample of WBs (Section~\ref{Sample_selection}) and analysing their observed properties in catalogues like \emph{Gaia} (Section~\ref{Data_analysis}). We then apply these techniques to obtain our main result (Section~\ref{Results}). We go on to consider if prior attempts at the WBT implemented the best practices we identify (Section~\ref{Previous_studies}). We discuss our findings in Section~\ref{Discussion} and conclude in Section~\ref{Conclusions}. In Appendix~\ref{Other_studies}, we briefly review other studies that did not implement the WBT but set the groundwork for it, for instance by preparing catalogues of WBs, identifying important data analysis issues, or clarifying model predictions.

We use three star catalogues in this paper: the C10 and C25 catalogues from \citet{Cookson_2024} extended out to 150 pc, and H22 \citep[the catalogue from][]{Hernandez_2022}. C10 and C25 only differ in the maximum allowed parallax uncertainty, which is 1\% (2.5\%) for C10 (C25). We use C25 almost everywhere, but briefly use C10 in Section~\ref{Nulls} and H22 in Section~\ref{sec:Hernandez_2022}. The Chae papers are emulated in Section~\ref{sec:appendix_spher_proj} using a modification to C25.

\section{Theoretical expectations}
\label{Theoretical_expectations}

The WBT is sensitive to the factor by which MOND boosts $g_{_N}$ when averaging over all angles, accounting for the fact that the potential of a point mass is not spherically symmetric once we include the external field effect \citep[EFE;][]{Milgrom_1986, Banik_2020_M33, Oria_2021}. It is recognised by most authors that the EFE is required to calculate the escape velocity in MOND \citep{Famaey_2007, Wu_2008, Banik_2018_escape}. For WBs in the Solar neighbourhood, the EFE arises mainly from the Galactic external field, which we assume has strength $g_e = 1.785 \, a_{_0}$ based on the rotation curve amplitude at the Solar circle \citep[section 3.1 of][and references therein]{Banik_2024_WBT}. Since the $\nu$ function takes as argument $g_{_N}$ rather than the actual gravity $g$ (Equation~\ref{nu_cases}), we need to find the corresponding Newtonian-equivalent external field $g_{_{N,e}}$. We find that
\begin{eqnarray}
    g_{_{N,e}} ~=~ 1.184 \, a_{_0} \, ,
\end{eqnarray}
which gives the correct external field $\nu g_{_{N,e}} ~=~ 1.785 \, a_{_0}$ using Equation~\ref{eq:observed_gravity} \citep[see equation~14 of][]{Banik_2024_WBT}. 

The internal Newtonian gravity $g_{_{N,i}}$ of a WB with separation $r$ is given by the standard inverse-square law:
\begin{eqnarray}
    g_{_{N,i}} ~=~ \frac{a_{_0}}{x^2} \, ,
\end{eqnarray}
where $x \equiv r/r_{_{\mathrm{M}}}$ is the separation in units of the MOND radius $r_{_{\mathrm{M}}}$ (Equation~\ref{r_M}). We can approximate that the total Newtonian gravity $g_{_{N,t}}$ is given by the quadrature sum of the internal and external contributions \citep[see equation~24 of][]{Zonoozi_2021}:
\begin{eqnarray}
    g_{_{N,t}} ~=~ \sqrt{g_{_{N,i}}^2 + g_{_{N,e}}^2} \, .
    \label{gN_t}
\end{eqnarray}

Since the circular orbital velocity $v_c \propto \sqrt{g}$, MOND enhances $v_c$ compared to the Newtonian expectation $v_{c,N}$ by the square root of the factor by which MOND enhances the angle-averaged inward radial gravity. Following equations~23 and 24 of \citet{Zonoozi_2021}, we obtain
\begin{eqnarray}
    \label{v_ratio_MOND}
    \frac{v_c}{v_{c,N}} ~&=&~ \sqrt{\nu \left( 1 + \frac{K}{3} \left[ \tanh \left( \frac{0.825 \, g_{_{N,e}}}{g_{_{N,i}}} \right) \right]^{3.7} \right)} \, , \\
    K ~&\equiv&~ \frac{\partial \ln \nu}{\partial \ln g_{_N}} \, .
\end{eqnarray}
The MOND boost factor $\nu$ and its logarithmic derivative $K$ must be evaluated with argument $g_{_{N,t}}$ (Equation~\ref{gN_t}). 

Equation~\ref{v_ratio_MOND} shows that given the floor on $g_{_{N,t}}$ set by the Galactic gravity on the Solar neighbourhood, WBs with an asymptotically large separation should orbit each other about 20\% faster than the Newtonian expectation \citep[as first shown in][]{Banik_2018_Centauri}. This expectation must be tested statistically given the very long orbital periods of WBs, which mean that observations can only measure the velocity at the present orbital phase (e.g., see their figure~14). Given the limits of currently available data, we also have to make do with only the separation and relative velocity of WBs within the plane of the sky. These considerations motivate the use of the scaled velocity parameter $\widetilde{v}$, defined as follows:
\begin{eqnarray}
    \widetilde{v} ~\equiv~ v_{\mathrm{sky}} \div \overbrace{\sqrt{\frac{GM}{r_{\mathrm{sky}}}}}^{\text{Newtonian}~v_c},
    \label{v_tilde}
\end{eqnarray}
where $M$ is the total mass of a WB with sky-projected separation $r_{\mathrm{sky}}$ and relative velocity within the sky plane of $v_{\mathrm{sky}}$. The $\widetilde{v}$ parameter takes account of the expected dependence on mass and separation. It was first introduced in 3D form \citep{Pittordis_2018} and then in the above 2D form \citep{Pittordis_2019}. We do not discuss the 3D form further because observations would have to advance significantly for it to become useful, in particular requiring very precise parallax distances and radial velocities (RVs) for both stars \citep[see section~5.5 of][]{Banik_2024_WBT}. Available attempts at the WBT use the above 2D form. Since Equation~\ref{v_ratio_MOND} requires the actual 3D separation $r$, we approximate that
\begin{eqnarray}
    r ~\approx~ r_{\mathrm{sky}} \sqrt{\frac{3}{2}} \, . 
\end{eqnarray}

Regardless of the gravity law and even if there were no observational uncertainties, we expect to observe WBs with a range of $\widetilde{v}$ because of projection and orbital phase effects in the inevitably somewhat eccentric orbits. Since $\widetilde{v}$ combines velocities along both directions within the sky plane, we expect the distribution of $\widetilde{v}$ to rise approximately linearly with $\widetilde{v}$ at first, before tailing off \citep{Banik_2018_Centauri, Pittordis_2018, Pittordis_2019}. A bound Newtonian orbit must have $\widetilde{v} < \sqrt{2}$ if observational errors and perspective rotation are negligible \citep{Shaya_2011}. Orbits in MOND can reach about 1.7, but binaries above $\sqrt{2}$ are rare and are likely to be greatly outnumbered by triples and other contaminating effects, so simply counting systems with $\widetilde{v} > \sqrt{2}$ is not a practical MOND test. The key consequence in MOND is that the $\widetilde{v}$ distribution for WBs in the MOND regime would be about 20\% broader (i.e. $1.7/\sqrt{2}$) than in the Newtonian regime, but no such broadening is expected in Newtonian gravity \citep[e.g., see figure~3 of][]{Banik_2018_Centauri}. Their result nicely demonstrates that this effect is clearly distinct to anything that can be achieved in the same gravity law by altering the distribution of orbital eccentricity \citep[see also][]{Pittordis_2018}.

MOND therefore predicts that if we split WBs into bins in $r_{\mathrm{sky}}/r_{_{\mathrm{M}}}$, then the WBs in the more widely separated bins should have a $\widetilde{v}$ distribution that is 20\% broader than in the Newtonian bins. This should be evident in some convenient measure of the typical $\widetilde{v}$, with the median being a particularly robust choice given the inevitable outliers when handling real data. We would expect the median $\widetilde{v}$ to follow a curve given by Equation~\ref{v_ratio_MOND}: it should follow a step-like behaviour against $r_{\mathrm{sky}}/r_{_{\mathrm{M}}}$, with a rise of about 20\% when going from $r_{\mathrm{sky}}/r_{_{\mathrm{M}}} \ll 1$ to $r_{\mathrm{sky}}/r_{_{\mathrm{M}}} \ga 1$ \citep[see figure~1 of][]{Banik_2018_Centauri}. We would need to conduct additional detailed modelling to know the Newtonian normalization, which would depend on additional assumptions like the eccentricity distribution. Such detailed modelling is beyond the scope of this contribution, though the issue has been explored in the literature. The median $\widetilde{v}$ in Newtonian orbit integrations is about 0.6 in \citet[][]{Banik_2024_WBT} (see their figure~15) and about 0.5 here. We take the more general prediction of MOND that there should be a rise with shape given approximately by Equation~\ref{v_ratio_MOND}, but with arbitrary Newtonian normalization. In other words, the median $\widetilde{v}$ could just as well rise from $0.5 \to 0.6$ as from $0.7 \to 0.84$, with both results lending strong support for MOND. Alternatively, a flat trend would favour Newtonian gravity, regardless of the particular value.

\section{Sample selection and quality cuts}
\label{Sample_selection}

The WBT requires a clean sample of WBs. In this section, we discuss several problems that must be overcome through an optimal choice of quality cuts, for which we suggest best-practice techniques. Whenever we discuss quality cuts at the individual star level, we require both stars in each WB to pass the quality cut for the WB to be considered in the analysis.

\subsection{Flybys and historical encounters}
\label{Flybys}

Following Kepler's Third Law, WBs have an orbital period of order 1~Myr, which is still much shorter than the Galactocentric orbit of the Sun \citep{McMillan_2017, Klioner_2021}. As a result, WBs dynamically relax in only a few Myr, though their orbits may remain eccentric due to past interactions. It is not very likely that a WB encountered a third star or molecular cloud in this timeframe \citep[see section~8.1 of][]{Banik_2018_Centauri}. If an interaction longer ago disrupted the WB, then its members would have become very widely separated in only a few Myr, making it very unlikely that the WB enters into any analysis. For bound but perturbed WBs, the resulting eccentricity distribution can be empirically modeled without requiring detailed knowledge of its origin. However, unbound pairs—arising from cluster dissolution, diffusive perturbations by field stars, or Galactic tides \citep{Jiang_2010}—could contaminate samples near the Jacobi radius. This underscores the importance of restricting tests to projected separations well within the Jacobi radius, as further discussed in Section~\ref{Limited_separation}. If the interaction affected the WB without disrupting it completely, this would be no more relevant to the WBT than details of galaxy formation are to testing MOND using the presently observed rotation curve and mass distribution of a galaxy.

However, we do need to be concerned with systems that appear to be gravitationally bound WBs but are not. Chance alignments of field stars are unlikely to masquerade as a WB given the need for both stars to also have a similar proper motion \citep[see figure~7 of][]{Pittordis_2019}. The level of contamination increases with $r_{\mathrm{sky}}$ because more widely separated WBs are less common \citep*{Lepine_2007, Andrews_2017}. Moreover, stars are thought to form in star clusters rather than completely independently \citep*{Kroupa_1993, Kroupa_2001}, with the star cluster slowly dissolving once the gas has been expelled \citep{Wu_2018}. Because a star cluster might only dissolve rather slowly, its initially most gravitationally bound stars may form WBs that are only marginally unbound, thus taking many orbital times to disperse \citep*[a similar phenomenon arises in globular clusters; see][]{Claydon_2017}. In short, the correlated nature of star formation can make chance alignments more common. It is also possible that two stars are genuinely close to each other but are not gravitationally bound, a phenomenon known as a flyby.

\subsubsection{Using a \texorpdfstring{$\widetilde{v}$}{vtilde} cutoff}
\label{Flyby_cutoff_line}

\begin{figure*}
    \includegraphics[width=0.49\linewidth]{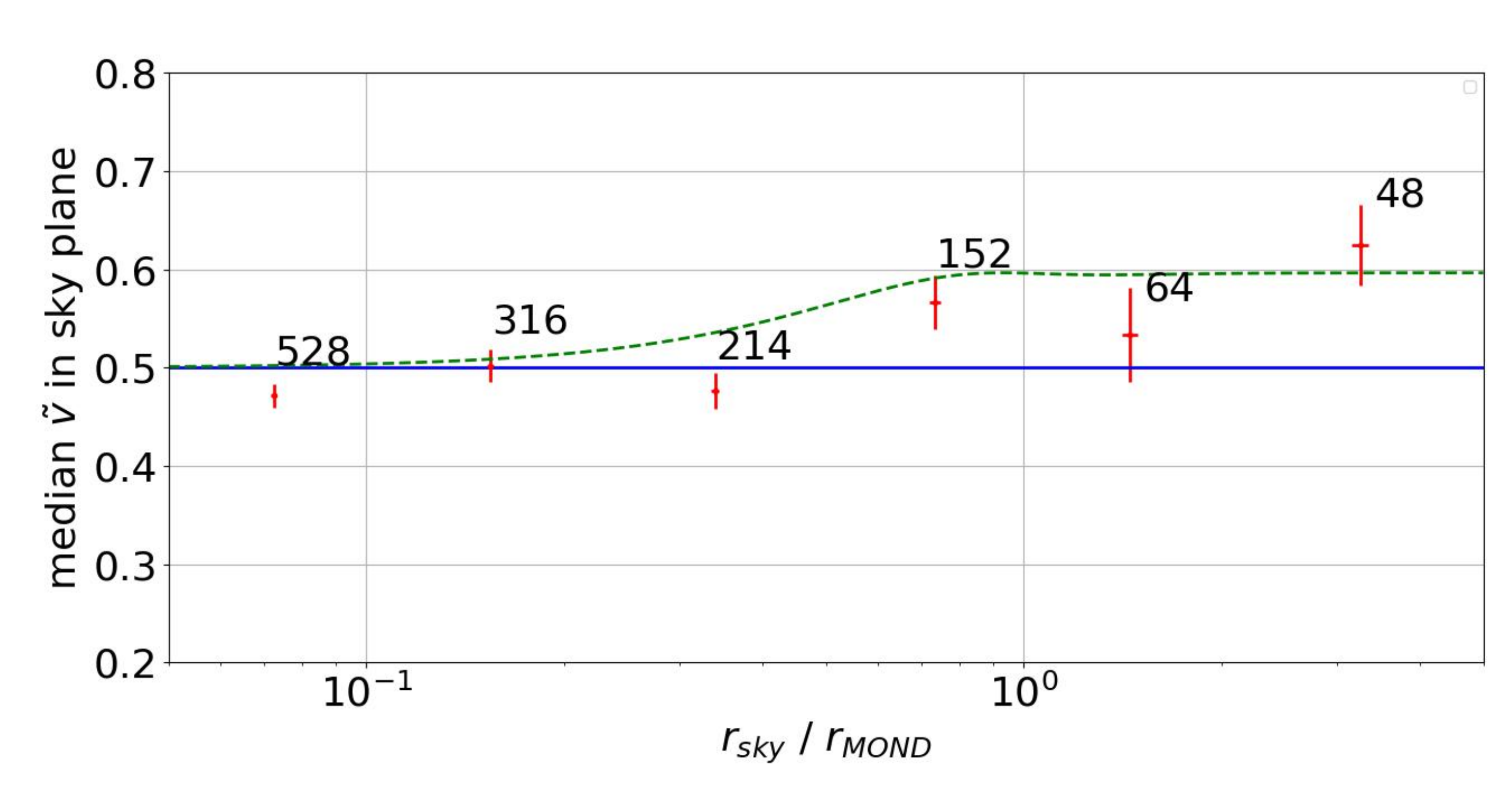}
    \hfill
    \includegraphics[width=0.49\linewidth]{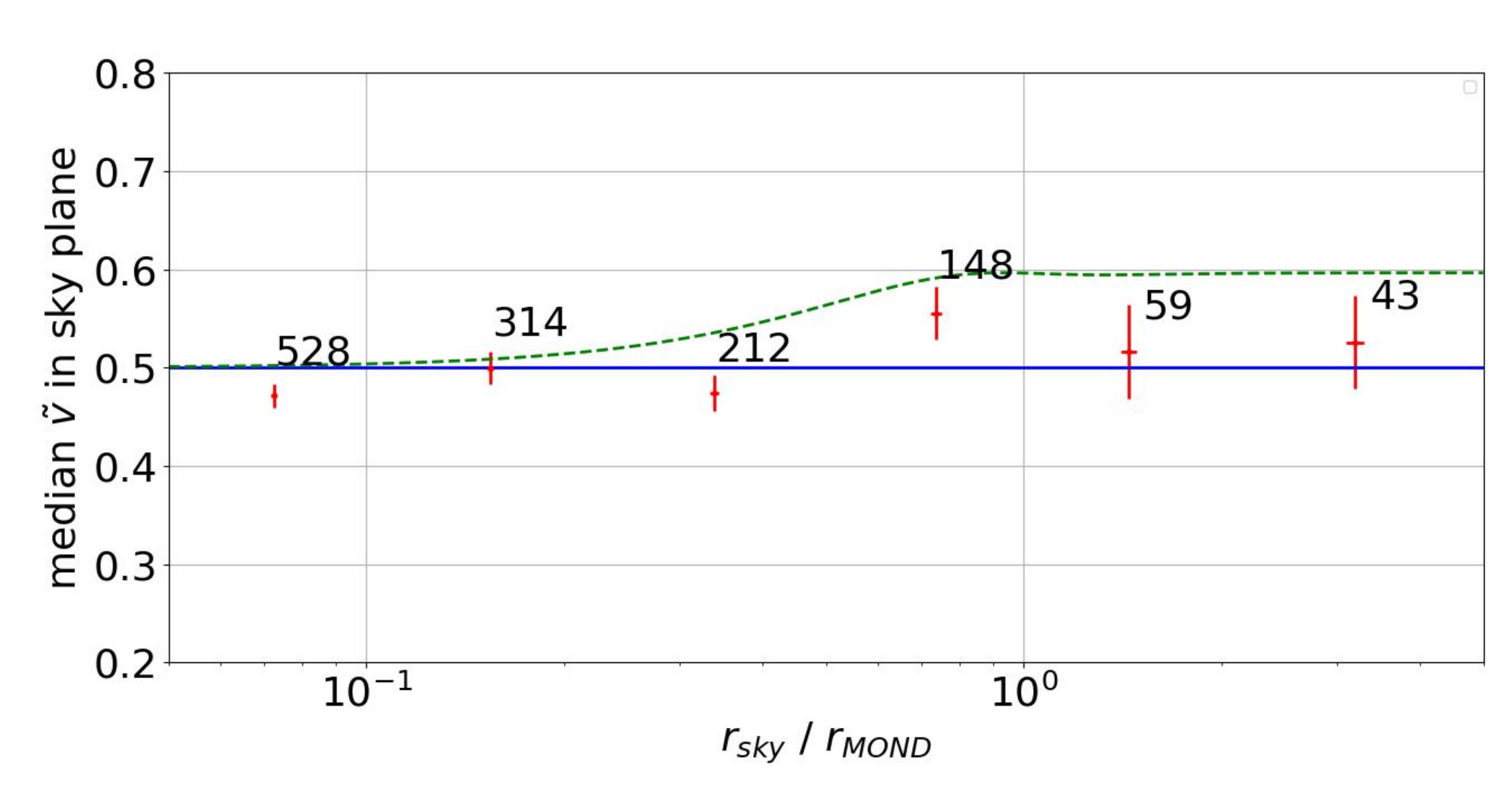} 
    \caption{The $\left( r_{\mathrm{sky}}/r_{\mathrm{M}}, \widetilde{v} \right)$ distribution of WBs within 150~pc, RUWE $< 1.25$ (Section~\ref{RUWE}), and $\Delta v_{\mathrm{sky}} < 10$~$\mathrm{km~s}^{-1}$ \citep[as in][]{Cookson_2024}. The dashed line is the approximate MOND expectation (Equation~\ref{v_ratio_MOND}), while the continuous line is the Newtonian prediction normalised to 0.5. The right panel is similar to the left, but imposes an additional filter of $\widetilde{v} < 2.5$. This removes systems with large separations and relative velocities, which are likely chance alignments or flybys.}
    \label{fig:dist_ruwe_RV_filter}
\end{figure*}

These issues can be mitigated by defining a downward sloping cutoff line in the space of $r_{\mathrm{sky}}$ and $v_{\mathrm{sky}}$ \citep*{Badry_2021}. The idea is to focus on systems towards the lower left of this flyby cutoff line, as any system towards the upper right is most probably perturbed rather than a gravitationally bound WB suitable for the WBT.

The introduction of $\widetilde{v}$ (Equation~\ref{v_tilde}) by \citet{Pittordis_2018} superseded the need for the flyby cutoff line used by some authors \citep[e.g., see figure~1 of][]{Cookson_2024}. This is because \citet{Pittordis_2018} showed that the Newtonian limit of $\sqrt{2}$ only rises to about 2.5 if we consider a more general class of non-Newtonian theories, including MOND \citep{Milgrom_1983}, TeVeS \citep{Bekenstein_2004}, and Emergent Gravity \citep{Verlinde_2017}. Therefore, we would not lose any non-Newtonian signal by imposing a cut of $\widetilde{v} < 2.5$. This cut would have the major advantage of reducing contributions from WBs whose dynamics may be influenced by perturbations or external gravitational effects besides the Galactic EFE. The efficiency of this approach is evident in figure~11 of \citet{Banik_2024_WBT}, which shows that an apparent MOND signal is present in the data when limiting to $\widetilde{v} < 5$, but this disappears when requiring $\widetilde{v} < 2.5$, even though this would still be sufficient to include any genuine MOND signal. We demonstrate this in Figure \ref{fig:dist_ruwe_RV_filter}.

When we come to assess previous implementations of the WBT (Section~\ref{Previous_studies}), we require the study to impose an upper limit on $\widetilde{v}$ \citep[even though this parameter was only introduced in][]{Pittordis_2018}. In Section \ref{sec:Hernandez_2022}, we reanalyse the data from \citet{Hernandez_2022} including our quality cuts, showing that the non-Newtonian signal completely disappears.

\subsubsection{Maximum RV difference}
\label{Maximum_RV_difference}

So far, we have discussed only the relative velocity in the sky plane. An unbound system can also be expected to have a large relative RV. This suggests that we should set limits to both $v_{\mathrm{sky}}$ and $\Delta \mathrm{RV}$, the maximum allowed RV difference. Workers have used caps on $\Delta \mathrm{RV}$ of $4-10$~$\mathrm{km~s}^{-1}$ \citep[e.g.,][]{Hernandez_2022, Hernandez_2024_statistical, Cookson_2024}. This is substantially smaller than the velocity dispersion of stars in the Solar neighbourhood \citep{Aumer_2016_AVR, Anguiano_2020}, making it more likely that stars with such a small $\Delta \mathrm{RV}$ are in fact bound. We expect a larger $\Delta \mathrm{RV}$ for a more tightly bound and more massive WB, so the $\Delta \mathrm{RV}$ cut might need to depend on $r_{\mathrm{sky}}$ or be somewhat relaxed if we aim to include WBs with a heavier primary or smaller separation. Additionally, \emph{Gaia} RVs can have uncertainties of a few $\mathrm{km~s}^{-1}$, which is greater than the orbital velocity of WBs, so in many cases the measured $\Delta \mathrm{RV}$ is dominated by uncertainties. We therefore require $\Delta \mathrm{RV} < 10$~$\mathrm{km~s}^{-1}$. This implicitly imposes the quality cut that both stars must have a measured RV.

\subsubsection{Limiting the WB separation}
\label{Limited_separation}

The predicted MOND enhancement to $g_{_N}$ remains fixed near 1.4 even if we consider WBs in the Solar neighbourhood with an arbitrarily large separation \citep[see figure~1 of][]{Banik_2018_Centauri}. These limited gains combined with the declining WB separation distribution \citep{Lepine_2007, Andrews_2017} suggest that we should limit the maximum allowed $r_{\mathrm{sky}}$. Studies have shown that WBs with separation $\la 31$~kAU (10\% of the tidal radius) would typically not have experienced a destructive encounter with a third star in the last 10~Gyr \citep*{Bahcall_1985, Jiang_2010}. The likelihood of a chance alignment also rises if considering WBs with a larger projected separation \citep[see section~3.3 of][]{Banik_2024_WBT}. Finally, although we correct $\bm{v}_{\mathrm{sky}}$ for perspective rotation effects, these grow linearly with WB angular separation, so their effect on $\widetilde{v}$ grows as $r_{\mathrm{sky}}^{1.5}$ (Section~\ref{Spherical_projection_correction}). For all these reasons, the WBT separation should be limited to $r_{\mathrm{sky}} \la 30$~kAU.

\subsection{Close binaries (CBs)}
\label{Close_binaries}

Even with perfect measurements of separations and relative velocities, the WBT is subject to various astrophysical systematics, which were discussed in detail in section~8 of \citet{Banik_2018_Centauri}. Those authors concluded that ``the most serious issue would probably be a low-mass undetected companion to at least one of the stars'' in some WBs, setting out a clear expectation ahead of \emph{Gaia} data suitable for the WBT. Undetected CBs are a problem because stars have a highly non-linear mass-luminosity relationship \citep{Pecaut_2013}, creating a mismatch between the photocentre and barycentre of the CB. Roughly speaking, we can neglect any light from its undetected component, but not the hidden mass. Underestimating the total mass similarly underestimates the Newtonian orbital velocity. Furthermore, the motion of the inner binary induces a recoil velocity on the brighter star in the CB, which we detect as part of a WB. The relative velocity of the WB is then biased, generally towards higher values because the orbital velocity of the CB can be much faster than that of the WB. We therefore expect undetected CBs to create an extended tail in the $\widetilde{v}$ distribution out to $\widetilde{v} \gg 2.5$.

Observations do in fact show the presence of such a tail \citep{Pittordis_2019}, with other workers soon pointing out that it might be due to undetected CBs rather than non-Newtonian dynamics \citep{Belokurov_2020, Clarke_2020}. Consequently, it is important to mitigate contamination from CBs as far as possible. The WBT can either rely on careful modelling of the remaining CB contamination, or rely more heavily on strict quality cuts that reduce the contamination. This section provides guidelines for the latter.

\subsubsection{Degrouping}
\label{Degrouping}

The WBT relies on a sample of WBs where the only external influence is the Galactic EFE. This makes it important to remove WBs where a star is common to two different WBs in the catalogue \citep[e.g., see section~2.2 of][]{Pittordis_2023}. This process of ``degrouping'' was discussed  in section~2.3 of \citet{Cookson_2024}. Degrouping ensures that triples, higher-order multiples, star clusters, and moving groups are removed, so we can be confident that there is no third star or CB within $100$~kAU of the WB, which we consider sufficient because we only use WBs with $r_{\mathrm{sky}} < 30$~kAU.

\subsubsection{Handling null values and high-noise stars}
\label{Nulls}

Relaxing the initial selection criteria before the degrouping process increases the initial number of candidate binaries at the beginning of the process. This leads to a \emph{reduction} in the number of clean pairs post-degrouping. If strict cuts are used early on, it cannot be recognised that some pairs are actually members of clusters/moving groups, because other members of the same groups are excluded. A common pitfall with the WBT is the removal at an early stage of faint stars with parameters that are too inaccurate for the WBT, or indeed with some required parameters not known at all. An important example is stars with unknown RV or with null values for $RP$ and $BP$, the photometric magnitudes in the \emph{Gaia} red and blue pass filters, respectively. While it is true that such a star is unsuitable for the WBT, it might still contaminate an otherwise suitable WB. For instance, one can easily envisage that one of the stars in a WB has a faint companion that is detected in \emph{Gaia}, but whose RV is not known. It is important to include such stars at the degrouping stage (Section~\ref{Degrouping}). An alternative approach would be to compare each WB with a faint star catalogue that contains stars otherwise unsuited to the WBT, thereby more robustly checking for possible CB contamination \citep[see section~2.4 of][]{Pittordis_2023}. If these precautions are not taken, the faint star might be removed from the dataset at an early stage, leading to the incorrect conclusion that any nearby bright star is `clean' and thus well suited to the WBT \citep[this issue is discussed in detail in][]{Cookson_2024}.

\begin{figure*}
    \includegraphics[width=0.49\linewidth]{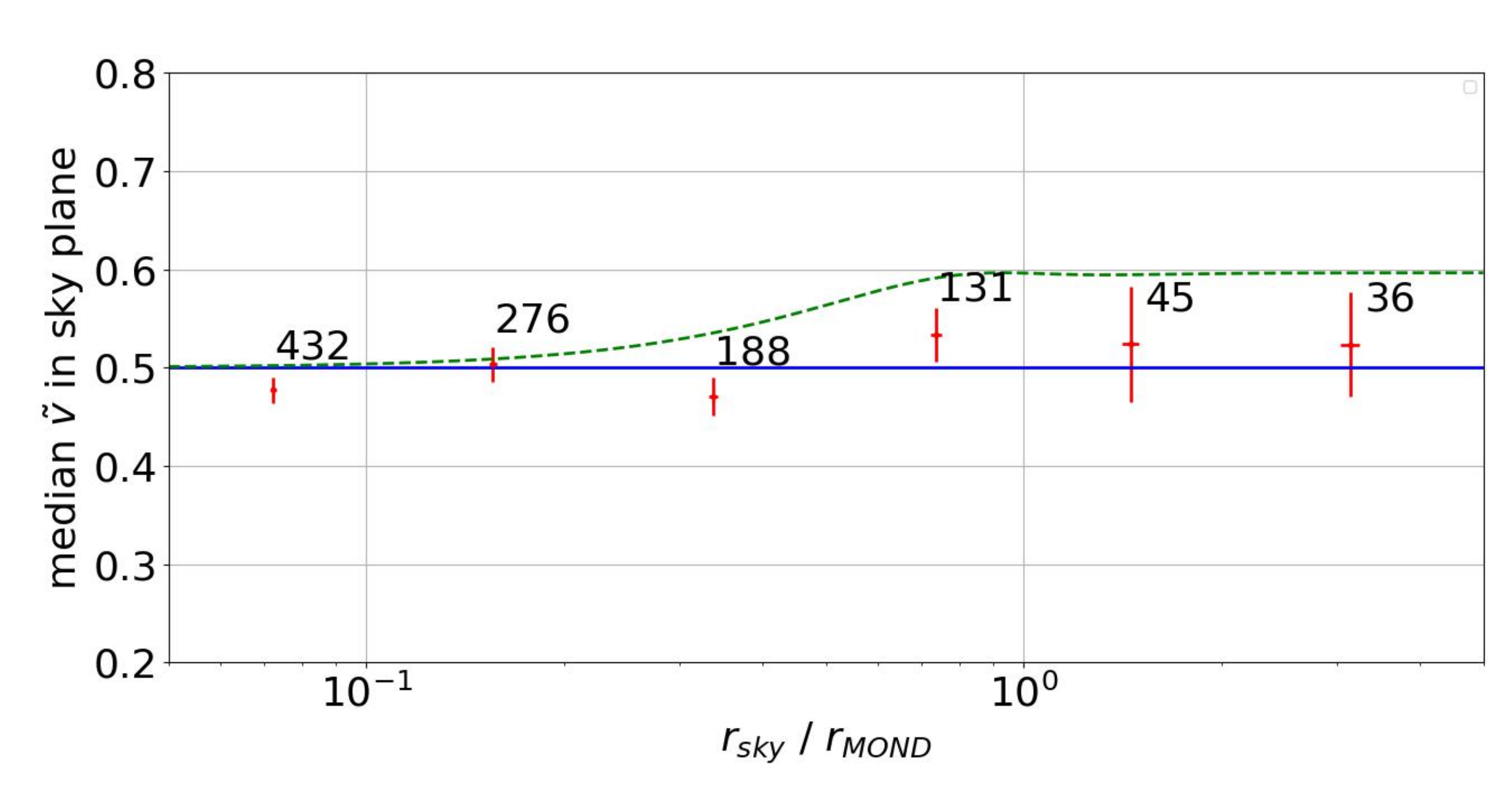}
    \hfill
    \includegraphics[width=0.49\linewidth]{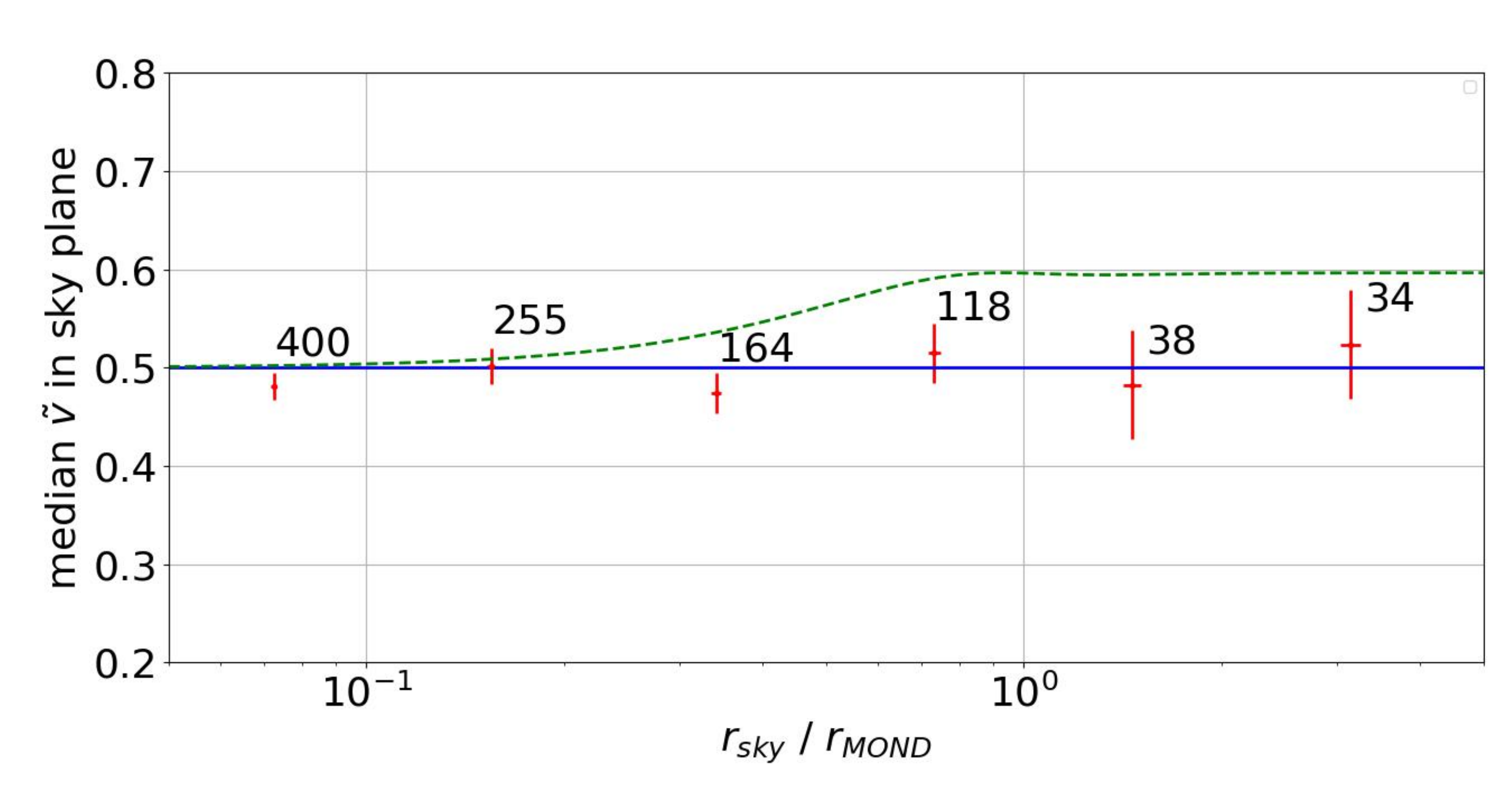}
    \caption{The distribution of $r_{\mathrm{sky}}/r_{\mathrm{MOND}}$ and $\widetilde{v}$ for a sample of WBs produced when using an input catalogue with parallax uncertainty below 1\% (catalogue C10; \emph{left}) and 2.5\% (catalogue C25; \emph{right}), following degrouping (Section~\ref{Degrouping}). We also show the approximate MOND expectation (Equation~\ref{v_ratio_MOND}) for each sample (dashed line) and the Newtonian expectation (continuous line). Allowing a larger parallax uncertainty leads to a larger sample prior to degrouping, improving the efficiency of the degrouping process and pushing the results closer to the flat Newtonian expectation. The sample totals in each bin are smaller on the right despite the greater number of pairs, because of the more thorough degrouping.}
    \label{Parallax_error_impact}
\end{figure*}

We illustrate this issue in Figure~\ref{Parallax_error_impact}, where the left panel has a stricter initial quality cut on the \emph{Gaia} trigonometric parallax ($S/N > 100$), while the right panel has a less strict quality cut ($S/N > 40$). This allows more CBs to be identified earlier in the process, allowing us to exclude triple systems that would otherwise masquerade as clean. Indeed, the left panel with stricter parallax cut has a slight MOND-like signal in the outer $r_{\mathrm{sky}}$ bins. This signal is diminished in the right panel with less strict parallax cut, which excludes just nine additional WBs across these two bins. We emphasize that these nine WBs are excluded on the basis of actually detected faint stars, albeit with poor parallax precision.

\subsubsection{Renormalized Unit Weight Error (RUWE) filter}
\label{RUWE}

The orbital acceleration of a CB is generally expected to be $3-4$ orders of magnitude larger than for the WB it is associated with \citep[see section~8.2 of][]{Banik_2018_Centauri}. This makes it quite possible that even in the short timeframe of \emph{Gaia} observations, there is appreciable astrometric acceleration of the contaminated star that we detect as part of a WB. It is not presently possible to search for this acceleration directly because astrometric time series are expected to be available only for \emph{Gaia} Data Release 4 (DR4), which will provide Keplerian orbital solutions for a large sample of binaries \citep{Badry_2024}. In the meantime, we can rely on the fact that astrometric acceleration would degrade the goodness of fit of the \emph{Gaia} astrometric solution, which only considers parallax and proper motion. The relevant catalogue quantity is called the Renormalized Unit Weight Error \citep[RUWE;][]{Lindegren_2018}.

A RUWE threshold of 1.25 has been suggested for DR3 as an upper limit for good astrometric solutions \citep{Penoyre_2022a, Penoyre_2022b, Castro-Ginard_2024}. Higher values such as 1.4 \citep{Andrew_2022} have been used specifically to select likely binary systems. The optimal single-source cutoff for \emph{Gaia}~DR3 varies between $1.15-1.37$ according to the sky position, with an average of about 1.25 \citep{Castro-Ginard_2024}. We therefore require RUWE $< 1.25$, though a more conservative cutoff at 1.15 might also be reasonable. However, our analysis shows that this makes little difference.

It is important to note that a RUWE cut will only filter out CBs with periods of months to years, but will be insensitive to CBs with both shorter and longer periods \citep[see][]{Penoyre_2022a,Penoyre_2022b,Badry_2025}. Therefore, a limit on RUWE may not detect CBs with periods of weeks or less, as the orbital separation is too small to resolve. For binaries with a mass ratio equal to their light ratio \citep[most commonly twin systems, where both components are roughly identical, see e.g.][]{Badry_2019_twin} there is no detectable astrometric motion \citep{Penoyre_2020}, biasing the luminosity-based mass estimate -- but with no astrometric signal at all beyond WB orbital motion \citep[see figure~8 of][]{Banik_2024_WBT}. A cut on RUWE is therefore complementary to a cut against stars that are over-luminous for their colour, which we discuss next.

\subsubsection{Hertzsprung-Russell (HR) filter}
\label{HR_filter}

Stars in certain regions of the Hertzsprung–Russell diagram (HRD) or colour–magnitude diagram (CMD) are less well suited for the WBT. A clear example is the white dwarf (WD) cooling track \citep[see figure~1 of][]{Belokurov_2020}, where the mass–luminosity relation differs significantly from that of main sequence (MS) stars \citep{Pecaut_2013}. Another complication arises from the presence of a `double main sequence' -- a track parallel to the MS but offset above it, such that stars appear roughly twice as luminous at the same colour \citep[see figure~2 of][]{Banik_2024_WBT}. This feature is thought to result from binaries with a mass ratio distribution that includes a Dirac delta function at $q=1$, i.e., equal-mass binaries \citep{Badry_2019_twin}. While a mass ratio of unity often implies a light ratio close to unity (since twin stars generally have similar brightness), this is not always the case. Even equal-mass binaries where one component is significantly brighter can exhibit detectable astrometric motion, complicating the interpretation.\footnote{The condition for the CB to have no astrometric motion is not actually $q = 1$ but instead for the mass ratio to equal the luminosity ratio. This is usually satisfied with two equally massive stars as they are usually equally bright, but other factors like metallicity can cause one star to be brighter than another equally massive star, in which case there would be some astrometric motion.}

A related population of interest comprises binaries in which one component is a compact object. These can contribute disproportionately to the astrometric signal and may bias WBT results, especially in hierarchical triple systems. A likely example is a main sequence star with a white dwarf companion. These systems can have high renormalised unit weight error (RUWE) values due to significant astrometric motion, yet exhibit little or no excess flux from the secondary and thus might escape detection in CMD-based filtering. However, these would be picked up both by the RUWE filter in Section \ref{RUWE} and by the $\widetilde{v}$ check in Section \ref{Flyby_cutoff_line}. 

\begin{figure*}
    \includegraphics[height=0.27\textheight]{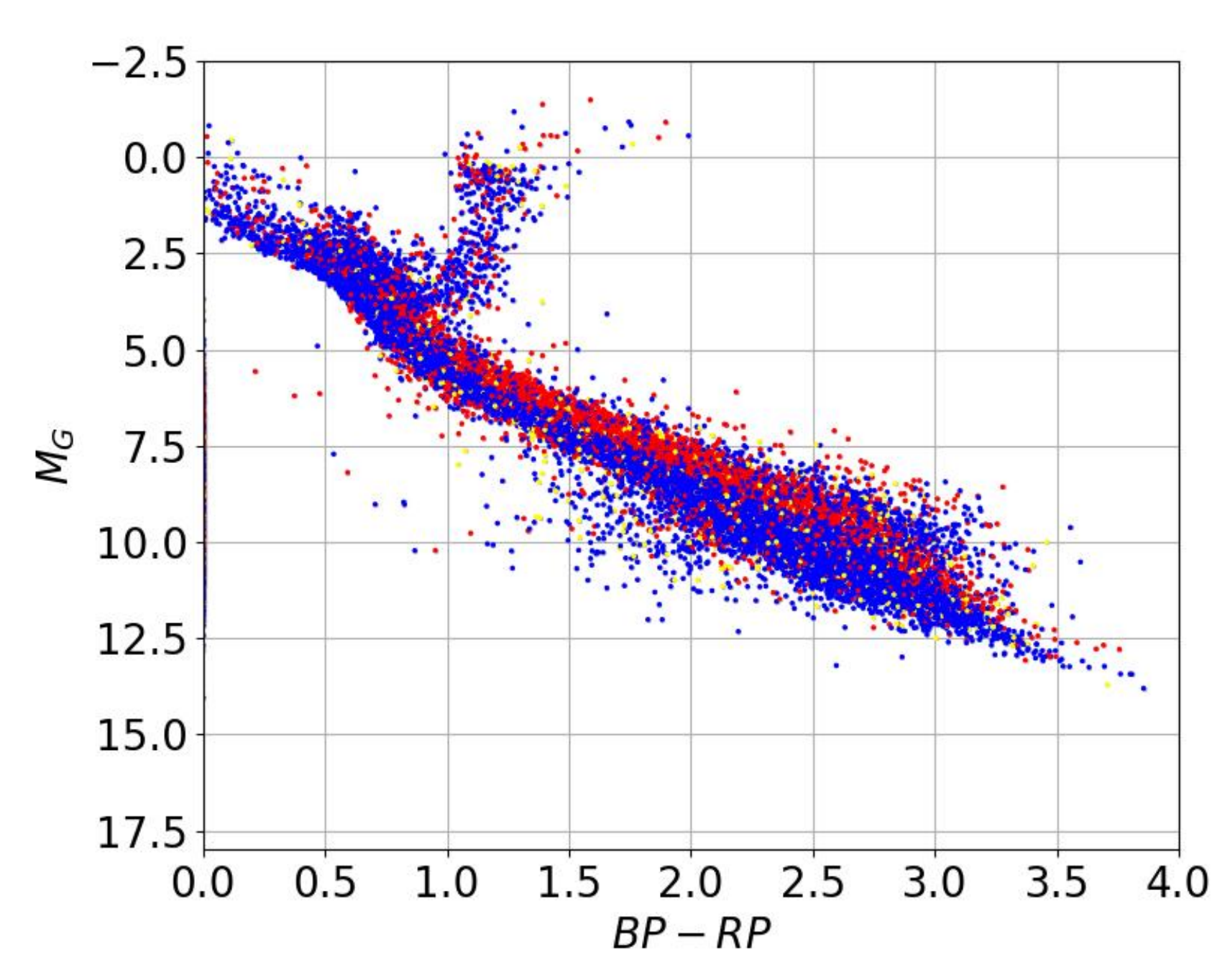}
    \hfill
    \includegraphics[height=0.27\textheight]{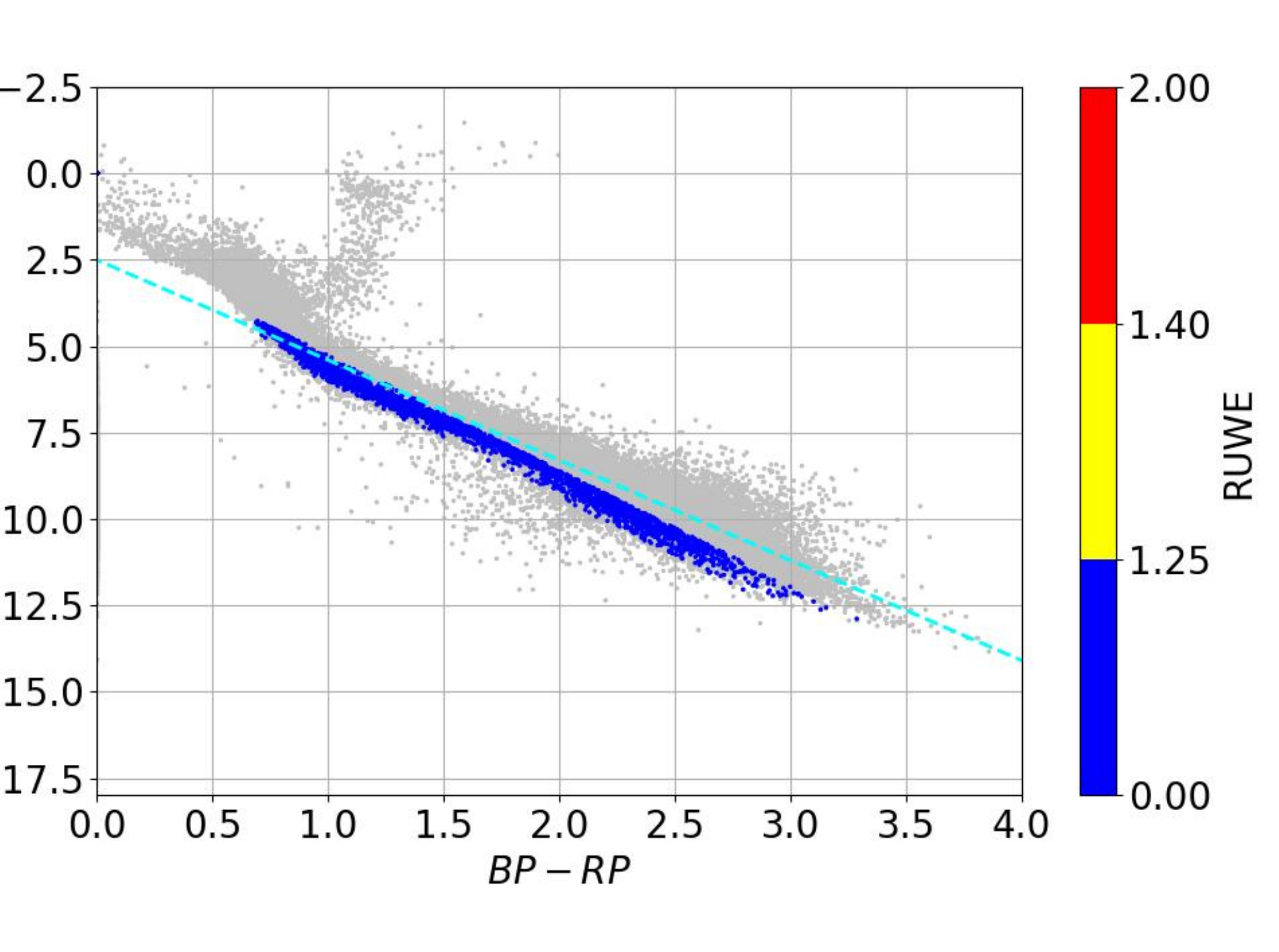}
\caption{Colour-magnitude diagram, with stars coloured by RUWE \citep[Section~\ref{RUWE}; see also][]{Belokurov_2020} as follows: blue ($<1.25$), yellow ($1.25-1.4$), and red ($>1.4$). Some 20k~grey points in the left panel are excluded from the final sample. The right panel restricts to stars with RUWE $<1.25$ truncated into a parallelogram on this diagram, as described in the text. It also implements a few kinematic quality cuts in our checklist, such as $\widetilde{v}$ < 2.5, though these have little effect on the overall appearance, amounting to $\approx 50$ pairs out of $\approx 900$. Notice how the RUWE cut preferentially removes stars which are overluminous for their colour, indicating a possible binary companion. The median colour-magnitude relation for Main Sequence stars used by \citet{Hartman_2022} is shown in cyan in the right panel.}
    \label{HRD-full}
\end{figure*}

To illustrate these issues, we construct our own HRD (Figure~\ref{HRD-full}). We would like to remove less suited stars to obtain a cleaner sample. We therefore define a parallelogram of high-quality stars following the MS, with $0.5 < BP - RP < 3.5$, a height of 0.8~mag, and $4 < M_G < 14$. This selection is illustrated in the right panel of Figure~\ref{HRD-full}, where deselected stars are shown in grey. \citet{Hernandez_2023} and \citet{Cookson_2024} also espouse the same technique.

A related idea is to use a ``Lobster Diagram'' to identify over-luminous components in a more analytical way \citep{Hartman_2022}. Although we do not follow their methodology, we note that they identify a reference value for K stars given by Equation~\ref{eq:kref_mg}, leading to the inclusion of stars stars from about $1.8 - 3.0$ on the \emph{Gaia} $BP - RP$ colour index, with a gradient of $2.9$ and an intercept of $+ 2.5$. This tracks the mid-point of the main sequence:
\begin{eqnarray}
    [M_G]_{\text{Kref}} ~=~ 2.9 \, (G_{\text{BP}} - G_{\text{RP}}) + 2.5.
    \label{eq:kref_mg}
\end{eqnarray}

It can be seen from Figure~\ref{HRD-full} that the parallel bars of MS stars continue in to about 0.6 or 0.7 on the colour index. In this paper, we set the inner limit to 0.5 and the outer limit to 3.5, covering stars of spectral type F to M (see Table~\ref{tab:spectral_types}). Because we use a simple trapezoid cut, we take the most effective simple cut across all four spectral types, requiring us to use a slightly steeper gradient of 3.3. In particular, we note that the K star part of the MS is somewhat bulbous in the MS turnoff region, where the stellar lifetime is similar to the Galactic age. We find we get a cleaner response if we keep the trapezoid to the lower edge of the bulge.

\begin{table}
    \centering
    \begin{tabular}{|c|c|}
        \hline
        \textbf{Spectral Type} & \textbf{$\boldsymbol{BP} - \boldsymbol{RP}$ Colour Index} \\
        \hline
        F & $0.5$ to $1.3$ \\
        G & $1.3$ to $1.8$ \\
        K & $1.8$ to $3.0$ \\
        M & $3.0$ to $5.5$ \\
        \hline
    \end{tabular}
    \caption{Spectral types corresponding to $BP-RP$ values.}
    \label{tab:spectral_types}
\end{table}

We note that every triple system erroneously included will positively bias $\widetilde{v}$. The RUWE and photometric cuts will exclude some of these triples, but will miss others (at shorter and longer periods than astrometry can capture, and at lower light ratios). Other observables (RV time series/scatter for short period systems, proper motion anomaly at longer periods) could flag more triple systems and thus further lower $\widetilde{v}$. However, many inner binaries are currently undetectable and may remain so, especially close to the 100 year mode of binary periods \citep[for the binary period distribution, see][]{Offner_2023}. Thus, we should always expect $\widetilde{v}$ to be a slight overestimate, at least while triples are the largest contaminant. Not identifying inner binaries with larger periods ($\ga 10$~years for RUWE) will be especially effective at boosting $\widetilde{v}$ at higher outer orbital separations (higher $r_{\rm sky}$) as any stable triple needs to have well-separated periods (at least an order of magnitude). In conclusion, we can expect that some proportion of systems in our sample are still undetected triples, and that detecting them (where possible) will further reduce the typical $\widetilde{v}$.

\subsubsection{Limiting the heliocentric distance}
\label{Limiting_dhel}

It is more difficult to identify a faint CB companion at a larger heliocentric distance $d_h$, both because this automatically makes the companion fainter, but also because it reduces the angular separation from the WB component it is orbiting. To mitigate the risk of including a WB contaminated by a faint third star bound to the system, we can impose an upper limit to $d_h$. Figure~4 of \citet{Hernandez_2023} nicely illustrates this issue, with many more WBs with anomalous kinematics identified when requiring $d_h < 175$~pc compared to a stricter limit of $d_h < 125$~pc. 

\begin{figure*}
    \centering
    \begin{subfigure}{0.48\textwidth}
        \centering
        \includegraphics[width=\linewidth]{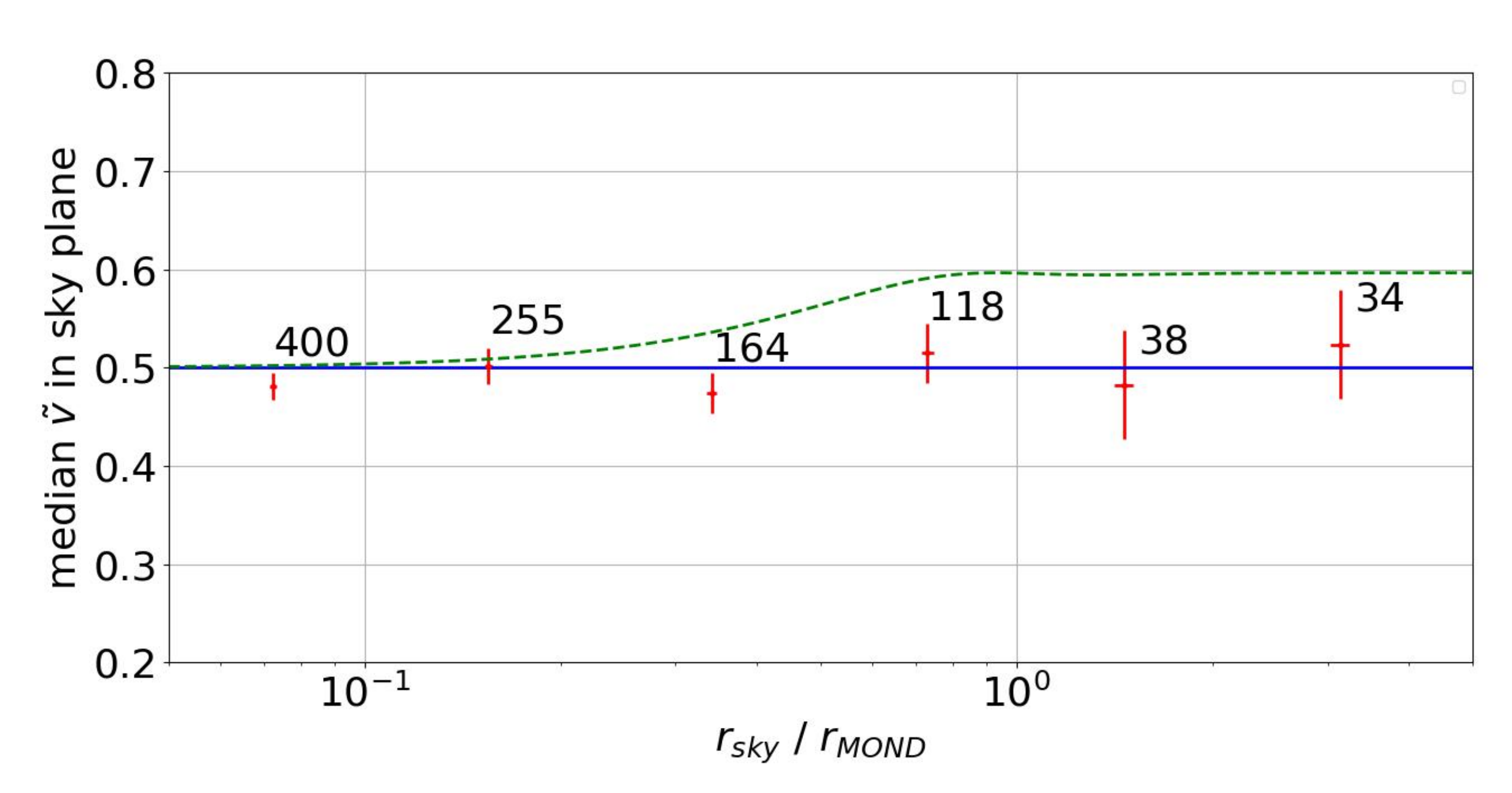}
        \caption{Catalogue C25 out to 130 pc. No \texttt{ipd\_frac\_multi\_peak} filter.}
        \label{fig:catA130}
    \end{subfigure}
    \hfill
    \begin{subfigure}{0.48\textwidth}
        \centering
        \includegraphics[width=\linewidth]{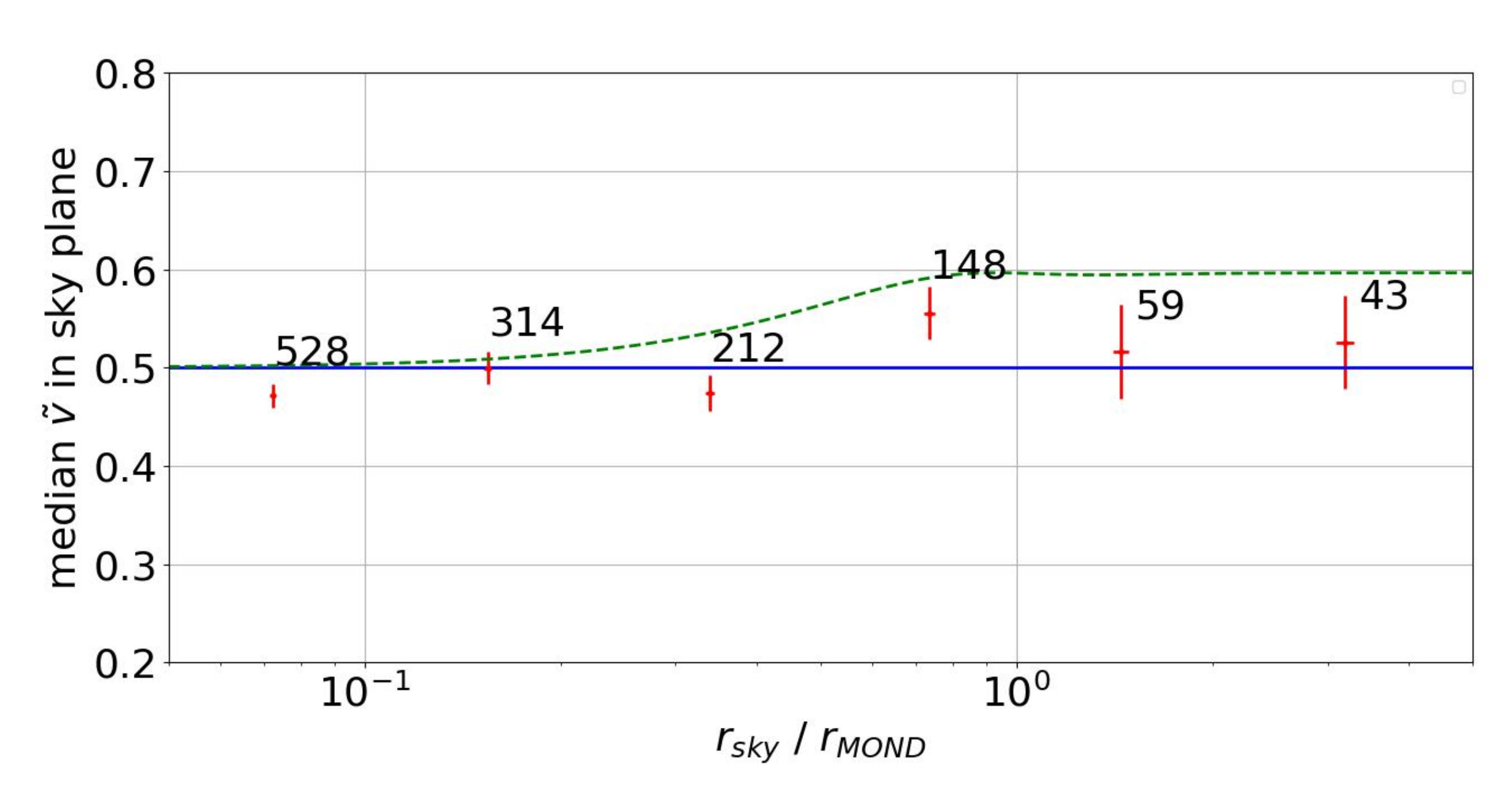}
        \caption{Catalogue C25 out to 150 pc. No \texttt{ipd\_frac\_multi\_peak} filter.}
        \label{fig:catA150}
    \end{subfigure}
    \caption{Comparison of Catalogue A with no \texttt{ipd\_frac\_multi\_peak} filter applied, for maximum distances of 130~pc (\emph{left}) and 150~pc (\emph{right}).}
    \label{fig:catA_comparison}
\end{figure*}

\citet{Cookson_2024} imposed an upper limit of 130~pc, which we adopt here. It is possible that CBs could contaminate the sample if going further out \citep[as done by][]{Banik_2024_WBT}. Here we show results for 130 pc (Figure \ref{fig:catA130}) and 150 pc (Figure \ref{fig:catA150}). We can mitigate some of the issues that arise when going out beyond 130 pc with extra quality cuts such as \texttt{ipd\_frac\_multi\_peak} \citep[see][]{Tokovinin_2023, Banik_2024_WBT}.

\begin{figure}
    \centering
    \includegraphics[width=\linewidth]{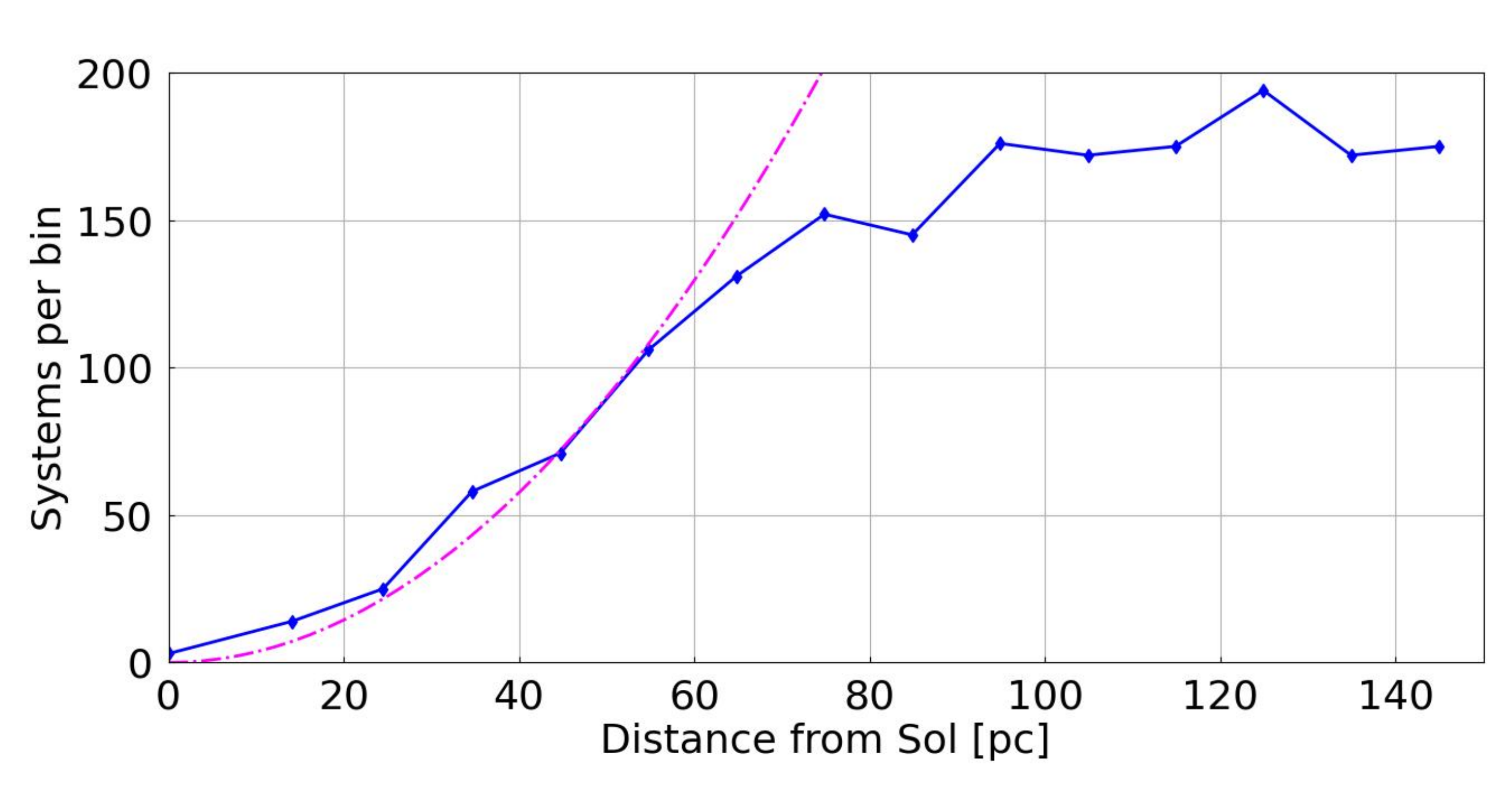}
    \caption{The number of WBs in the final sample in each $d_h$ bin. Perfect data should follow the parabolic expectation ($4\pi r^2 \rho_\star \Delta d_h$) shown in cyan, but the actual results flatten out at $d_h \ga 90$~pc, beyond which many of the WBs have insufficiently precise data for the WBT.}
    \label{fig:bin_count_with_d_h}
\end{figure}

The increasing difficulty of finding faint companions at larger $d_h$ is suggested by Figure~\ref{fig:bin_count_with_d_h}, which shows the distribution of $d_h$ for the WBs in our final sample. If we were able to detect all stars and obtain good data on them, we would expect a parabolic distribution of $d_h$ for a uniform density population. In reality, the number count per $d_h$ bin only rises out to $60-70$~pc and then stays approximately flat out to 150~pc in this sample. Given 150~pc is still very small compared to the thickness of the Galactic disc \citep{Juric_2008}, it must be difficult to detect and confidently characterize stars as uncontaminated with CBs beyond this distance. The loss of WBs from the sample despite a presumably uniform density of stars is partly because of the very strict quality cuts required for a star to be part of the WB sample, for instance the requirement for both components to have measured RVs. It is not necessary for a faint star to be characterized that well, since we just need to know of its existence in order to remove the WB it may be contaminating. It is therefore reasonable to use a somewhat larger maximum limit on $d_h$, but even so, it is clear that there are risks to going much beyond 130~pc. For instance, \citet{Banik_2024_WBT} argue in their section~3.2.3 that faint companions down to the bottom of the MS can be detected by \emph{Gaia} out to 200~pc, motivating those authors to go out to 250~pc. Regardless of the precise limit used, we expect any WBT to show that any claimed non-Newtonian effect is not distance-related before arguing that we should abandon such a well-established theory.

\subsection{Quality check: \texorpdfstring{$\texttt{ipd\_frac\_multi\_peak = 0}$}{ipd\_frac\_multi\_peak}}
\label{ipd_frac_multi_peak}

As the heliocentric distance increases, the risk of unresolved CB contamination grows. To mitigate this further, we can use the \emph{Gaia} parameter \texttt{ipd\_frac\_multi\_peak}. This statistic indicates the fraction of observations where the image parameter determination (IPD) found multiple peaks in the point spread function, suggesting that the source may not be a single star \citep{Tokovinin_2023, Banik_2024_WBT}. A non-zero value can indicate the presence of a marginally resolved companion that is occasionally detected depending on the observational conditions and orbital phase.

Requiring \texttt{ipd\_frac\_multi\_peak} $= 0$ for both stars in a WB provides an additional layer of quality control, helping to exclude systems where one component might be an undetected tight binary. This cut is particularly valuable when extending samples to larger distances ($d_h \ga 130$ pc), where direct detection of faint companions becomes more challenging. As shown in Section~\ref{Distance_ipd}, applying this filter to a sample out to 150~pc significantly cleans the sample, reversing a spurious MOND-like signal and restoring a strong preference for Newtonian dynamics. For the highest-purity samples, we recommend using this filter in conjunction with a conservative distance limit.

\section{Data analysis}
\label{Data_analysis}

\subsection{Relative velocity calculation and scaling}
\label{vtilde_calculation}

The importance of accurate error propagation in the WBT was highlighted in section~5.2 of \citet{Banik_2024_WBT}, where it was shown that the width of the $\widetilde{v}$ distribution can be broadened by $\widetilde{v}$ uncertainties preferentially at large separations, thereby mimicking a MOND signal. This section covers key considerations in propagating measurement errors.

\subsubsection{Mean heliocentric distance}
\label{Mean_heliocentric_distance}

The WBT suffers from various perspective effects \citep[for a detailed discussion, see][]{Shaya_2011}. One of these is that due to uncertainties in the trigonometric parallax, even a perfect proper motion measurement leads to some uncertainty in the relative velocity. This can be mitigated by placing the stars at the same heliocentric distance, which is the most likely scenario if the projected separation is small \citep[section~5.3 of][]{Banik_2019_line}. The best choice for the common distance is the inverse variance-weighted mean parallax distance to the stars in the WB \citep[as in][]{Banik_2024_WBT, Cookson_2024}. Using this as the basis for all subsequent calculations noticeably reduces any apparent non-Newtonian signal.

\subsubsection{Mass scaling}
\label{Mass_scaling}

The total mass $M$ of each WB is found by first estimating the mass of each star from its apparent magnitude and \emph{Gaia} trigonometric parallax, for instance using the method discussed in section~2.2 of \citet{Banik_2024_WBT}. The average $M$ is about $1.8 \, M_\odot$, slightly smaller than the $2 \, M_\odot$ assumed in \citet{Jiang_2010}. An accurate value of $M$ is required to calculate the Keplerian velocity and thus $\widetilde{v}$ (Equation~\ref{v_tilde}). It is also required to calculate $r_{_{\mathrm{M}}}$ and thus how far into the MOND regime a given system should be (Equation~\ref{r_M}). There are several ways to find the mass of a star from its absolute magnitude \citep[e.g.,][]{Hernandez_2023, Chae_2023, Chae_2024a}. The exact method is probably not that important as long as some attempt is made, since otherwise it becomes infeasible to test the gravity law for the above reasons. In addition, because we only care about how $\widetilde{v}$ varies with separation rather than the absolute value of $\widetilde{v}$, it does not really matter if the masses are biased low or high, as long as the bias does not change with separation. 

In this contribution, we find the mass of each star using the Final Luminosity, Age, and Mass Estimator (FLAME) package \citep{Pichon_2007}. FLAME masses are available in the \emph{Gaia} catalogue for about 19\% of the sample. For the remaining stars, \textsc{bynary} uses the distance $d_h$ to convert the \emph{Gaia} apparent magnitude into an absolute magnitude $M_G$ by subtracting the distance modulus $5 \, \log_{10} \left( d_h/ 10 \, \mathrm{pc} \right)$. The mass of the star in units of $M_\odot$ is then
\begin{eqnarray}
    M_\star ~=~ 10^{0.0725 \left(M_{G,\odot} - M_G \right)} \, ,
    \label{eq:mass_lum}
\end{eqnarray}
where $M_{G,\odot}$ is the absolute magnitude of the Sun in the \emph{Gaia} band \citep{Willmer_2018}. If the calculation gives $M_\star < 0.7 \, M_\odot$, we increase $M_\star$ by $0.05 \, M_\odot$ to achieve better agreement with FLAME results \citep{Hernandez_2023}.

\begin{figure}
    \centering
    \includegraphics[width=\linewidth]{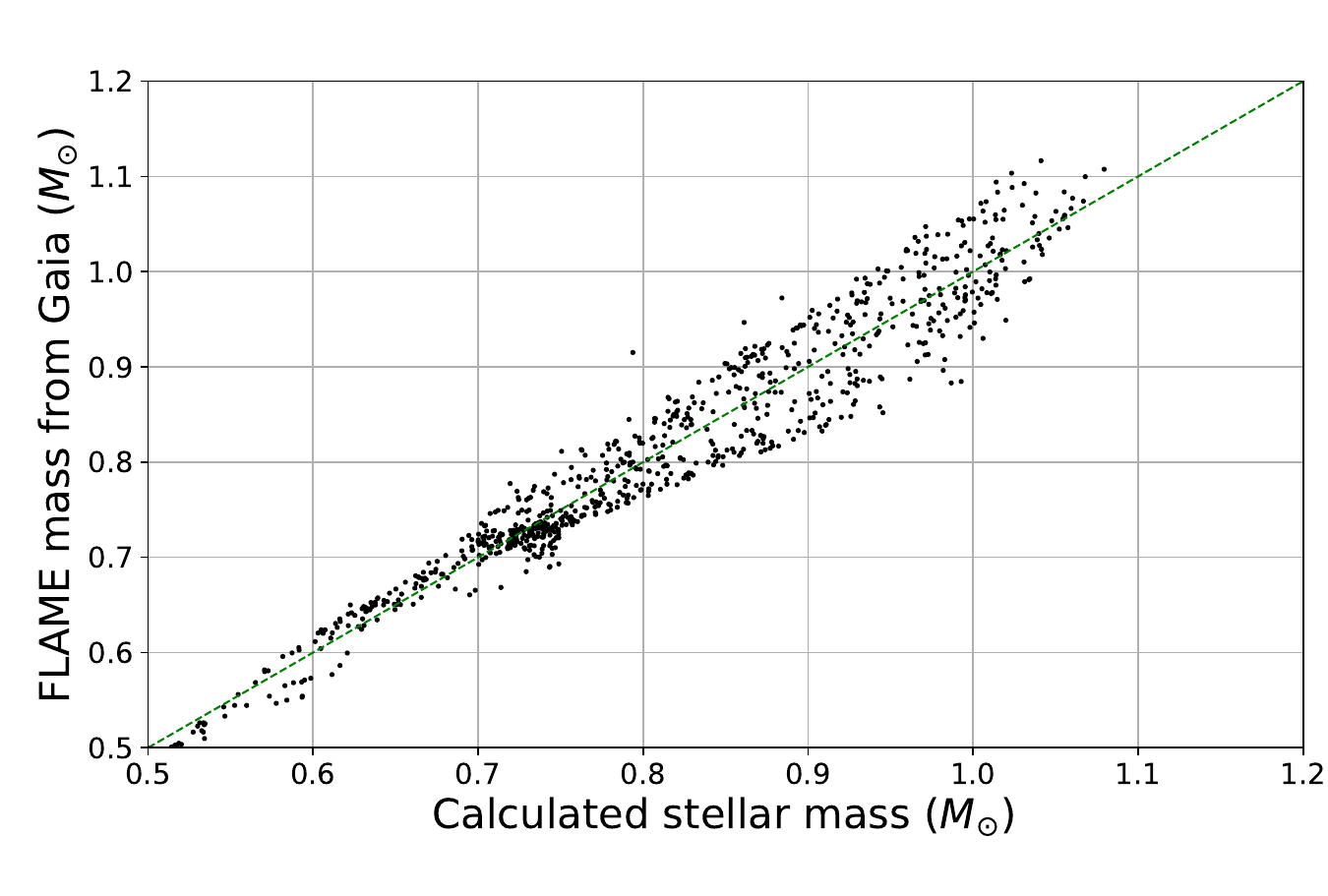}
    \caption{Calculated stellar mass $M_\star$ against the FLAME mass \citep{Pichon_2007} published in the \emph{Gaia} catalogue. When available, the latter provide an important check on the $M_\star$ estimates. These lie close to the dashed line of equality, indicating the estimates are reasonable.}
    \label{fig:FLAME_mass}
\end{figure}

The overall performance of this approach is shown in Figure~\ref{fig:FLAME_mass}, where the estimated masses are compared with those from FLAME \citep{Pichon_2007}. The differences are about $3-5\%$, which would affect the Newtonian $v_c$ by half as much. This is small compared to the predicted MOND enhancement of 20\%, justifying earlier optimism that estimating stellar masses accurately would not be a major challenge in the \emph{Gaia} era given accurate trigonometric parallaxes \citep{Banik_2018_Centauri}.

\subsubsection{Velocity representation}
\label{Velocity_representation}

The relative velocity in m/s between the stars in a WB can be quantified using $v_{\mathrm{sky}}$ or the projection of this along a particular line, which we call $v_{\mathrm{line}}$. The direction of this line can be chosen to optimize the accuracy of the WBT \citep{Banik_2019_line}, or it can simply be along the East-West and North-South directions for convenience, yielding two values per WB with some correlated uncertainties, e.g. from the heliocentric distance \citep{Hernandez_2019_WB}. While these are important first steps, the $\widetilde{v}$ parameter is far more useful for the WBT because it captures the expected Newtonian dependence of the relative velocity on the total mass and separation of the WB (Equation~\ref{v_tilde}). A simple way to implement the WBT is to plot some measure of the typical $\widetilde{v}$ against $r_{\mathrm{sky}}/r_{_{\mathrm{M}}}$ \citep[as in figure~11 of][]{Banik_2024_WBT}. Moreover, the use of $\widetilde{v}$ allows a much tighter cut of the form $\widetilde{v} \la 2$, which is far more effective at removing flybys and triple systems while retaining sensitivity to any genuine MOND signal \citep[e.g.,][]{Pittordis_2018, Pittordis_2019, Pittordis_2023, Banik_2018_Centauri, Banik_Zhao_2022, Banik_2024_WBT}.

\subsubsection{Scaling to the MOND radius}
\label{Scaling_to_r_M}

The WBT is often implemented by comparing populations of WBs with different $r_{\mathrm{sky}}$, but theoretically we expect a more consistent measure would actually be $r_{\mathrm{sky}}/r_{_{\mathrm{M}}}$ (Equation~\ref{r_M}). This distinction is not too crucial because $r_{_{\mathrm{M}}} \propto \sqrt{M}$, so the dynamic range in $r_{\mathrm{sky}}$ is typically much greater than in $r_{_{\mathrm{M}}}$. Even so, comparing WB subsamples with different $r_{\mathrm{sky}}$ would lead to unnecessary mixing of systems within and beyond their MOND radius, potentially diluting any MOND signal. We therefore recommend comparing (some measure of) the $\widetilde{v}$ distribution with $r_{\mathrm{sky}}/r_{_{\mathrm{M}}}$.

\subsubsection{Quantifying the velocity central tendency}
\label{Quantifying_typical_value}

While it is possible to consider the whole $\widetilde{v}$ distribution \citep[as done in][]{Pittordis_2023}, it is convenient to summarize the width of this distribution using a single statistic and then look for the expected step-like trend with $r_{\mathrm{sky}}/r_{_{\mathrm{M}}}$. Some obvious choices are the median, mean, and root mean square (RMS) value of $\widetilde{v}$. The mean suffers from a high degree of sensitivity to outliers due to flybys or CBs. This issue is exacerbated when using the RMS as it places even more weight on larger values. The median provides a more robust measure for quantifying the typical value in situations where there can be significant one-sided scatter, as arises for the WBT because $\widetilde{v} \geq 0$. If there is a genuine MOND signal, it should be apparent in any suitable statistic rising 20\% from low to high $r_{\mathrm{sky}}/r_{_{\mathrm{M}}}$, but the median is much less likely to do so in a Newtonian population of WBs due to spurious effects. A similar story arises with measurement of uncertainty. Here we use the Median Absolute Deviation (MAD) to measure the Gaussian-equivalent uncertainty $\sigma = 1.4826$~MAD.

\subsubsection{Spherical projection correction}
\label{Spherical_projection_correction}

The WBT is based on $v_{\mathrm{sky}}$ and will remain so for the foreseeable future \citep[see section~5.5 of][]{Banik_2024_WBT}. Since the WBT is best implemented by placing both stars at the same $d_h$, the main contribution to $\bm{v}_{\mathrm{sky}}$ is the product of $d_h$ with the difference in proper motions between the two stars. However, there are additional contributions known as perspective effects \citep{Shaya_2011}. These arise because the sky plane (with normal given by the line of sight) is not exactly the same for both stars given their angular separation on the sky. The curvature of the celestial sphere implies that systemic motion along the line of sight also contributes to $\bm{v}_{\mathrm{sky}}$. For instance, if $v_{\mathrm{sky}} = 0$ but the system as a whole is receding, then the stars will appear to approach each other on the sky \citep{Banik_2019_line}.

This issue was not taken into consideration in \citet{Hernandez_2019_WB}, leading to the suggestion that their claimed non-Newtonian signal might actually be due to perspective effects \citep{Badry_2019_geometry}. While perspective effects alone are not sufficient to explain the apparent deviation from Newtonian gravity \citep{Hernandez_2022}, it is still important to include the effect of the systemic RV on $v_{\mathrm{sky}}$. This requires knowledge of the RV for at least one star in each WB, since an accuracy of order 1~$\mathrm{km~s}^{-1}$ in the systemic RV is more than sufficient for this purpose. However, not correcting for this perspective effect altogether implicitly assumes that the systemic RV is 0, when in reality it might be many tens of $\mathrm{km~s}^{-1}$. This is often not negligible compared to $v_{\mathrm{sky}}$, even after multiplying by the angular WB separation. We note that WBs have a smaller angular size at larger $d_h$, reducing uncertainties from perspective effects \citep{Shaya_2011}.

\begin{table}
    \centering
    \begin{tabular}{ccc}
        \hline
        Bin & Median & Mean spherical \\
        no. & $r_{\mathrm{sky}}/r_{_{\mathrm{M}}}$ & correction to $\widetilde{v}$ \\ 
        \hline
        1 & 0.073 & 0.001 \\
        2 & 0.153 & 0.003 \\
        3 & 0.348 & 0.013 \\
        4 & 0.744 & 0.039 \\
        5 & 1.419 & 0.072 \\
        6 & 3.079 & 0.306 \\ \hline
    \end{tabular}
    \caption{Impact of the spherical projection correction in six different $r_{\mathrm{sky}}/r_{_{\mathrm{M}}}$ bins. For the WBs in each bin, the final column shows the mean absolute difference between the $\widetilde{v}$ of each WB with and without the spherical projection correction (Section~\ref{Spherical_projection_correction}). These corrections become important relative to the typical $\widetilde{v}$ of about 0.5 precisely when crossing the MOND radius.}
    \label{tab:sperical_corrections}
\end{table}

\begin{figure}
     \centering
     \includegraphics[width=1\linewidth]{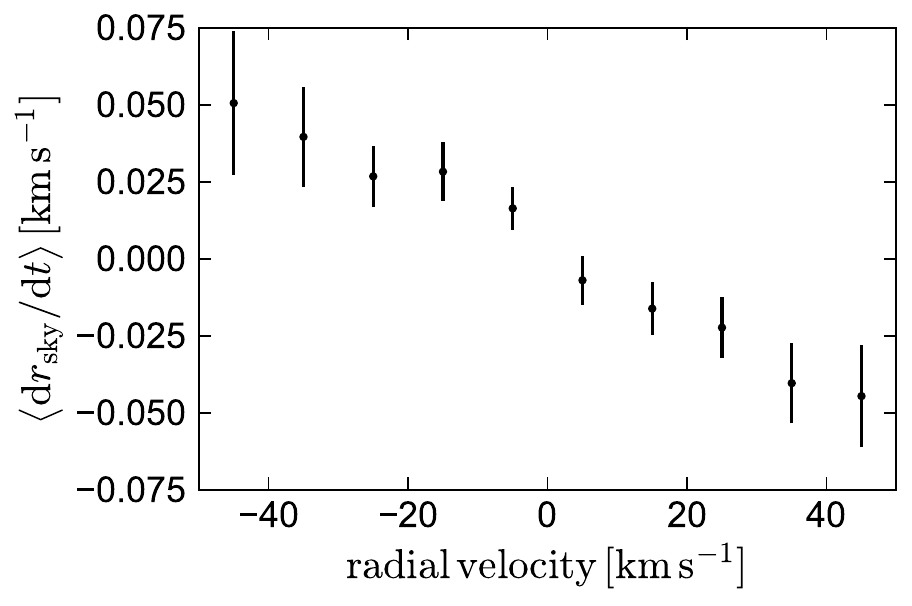}
     \caption{Spurius apparent orbital motion due to projection effects in the binary sample from \citet{Badry_2021}. We consider binaries with $r_{\rm sky} > 10$~kAU within 200~pc of the Sun. In each systemic RV bin, we calculate the apparent widening or shrinking of each binary from the components’ proper motions and plot the average value. Due to geometric effects, the components of binaries moving away from the Sun appear to be moving towards one another on average, while those of binaries moving toward the Sun appear to move apart. If not corrected for, this effect will manifest as a spurious relative proper motion, boosting the measured $\tilde{v}$ of the widest binaries.}
     \label{fig:spherical}
 \end{figure}

To quantify how important the spherical correction might be, we calculate the magnitude of the change in $\widetilde{v}$ for each WB in our sample if the correction is neglected. Table~\ref{tab:sperical_corrections} shows the mean value of these changes in $\widetilde{v}$ for the WBs in each of six $r_{\mathrm{sky}}/r_{_{\mathrm{M}}}$ bins. Since the median $\widetilde{v}$ is about 0.5 \citep{Banik_2024_WBT}, the spherical corrections to $\widetilde{v}$ are very small in the inner bins, but the corrections in the outer bins can be substantial. Coincidentally, the corrections become important when crossing the MOND radius, suggesting that mishandling of the spherical projection correction -- or neglecting it altogether -- could lead to an artificial MOND signal in underlying Newtonian data. In Section~\ref{sec:appendix_spher_proj}, we show how the MOND signal in the actual data is affected by not implementing the spherical projection correction. Figure~\ref{fig:spherical} shows the inverse relationship between the systemic RV and its spurious contribution to $\bm{v}_{\mathrm{sky}}$ along the line connecting the WB components.
 
To illustrate the effects of perspective effects on the apparent proper motions of real WBs, we consider the large WB sample constructed by \citet{Badry_2021}. We analyse all systems with projected separations $r_{\mathrm{sky}} > 10$~kAU, heliocentric distances $d_h < 200$\,pc, $<10\%$ chance alignment probabilities, and RVs reported for at least one component in \emph{Gaia} DR3. We calculate the apparent widening or shrinking of each binary, ${\rm d}r_{\rm sky}/{\rm d}t$, from the components’ proper motions. Finally, we bin the data by systemic RV and plot the average value in each RV bin. The results in Figure~\ref{fig:spherical} show a clear linear trend between ${\rm d}r_{\rm sky}/{\rm d}t$ and RV: WBs with positive RV appear to be getting tighter, while WBs with negative RVs appear to be widening. This illustrates that perspective effects clearly result in spurious relative proper motions that alter the measured $\widetilde{v}$ of WBs; these can largely be corrected for if the binaries’ systemic RVs are known.

\subsection{Measurement error propagation}
\label{sec:Error_propagation}

The WBT relies primarily on $v_{\mathrm{sky}}$, the sky-projected relative velocity between the stars in each WB. This requires good astrometric precision for a large sample of WBs, which has only recently become possible thanks to the \emph{Gaia} mission \citep{Gaia_2016}. In this section, we discuss how to obtain the uncertainty on $v_{\mathrm{sky}}$ and thus $\widetilde{v}$, and how uncertain it can be without much affecting the WBT.

\subsubsection{Velocity uncertainty calculation}
\label{sec:Velocity_uncertainty_calculation}

To quantify the uncertainty between the relative velocity of the stars in a WB, we propagate uncertainties in the astrometry through the Jacobian matrix \citep[following the approach explained in detail in appendix~A of][]{Cookson_2024}. It is also possible to use a Monte Carlo (MC) approach \citep{Banik_2024_WBT}, but results are expected to be similar given individual stars have quite small parallax and proper motion uncertainties, implying that the response of $\widetilde{v}$ to changes in the observational inputs should be approximately linear. Turning now to the induced uncertainty in the typical value of $\widetilde{v}$ in some population of WBs, a number of different methods are used in other papers. Except in bins with a large sample size $N$, the main error component in a population statistic like the median is due to intrinsic velocity dispersion, which arises from the different angles of the orbits to the sky plane and the different orbital phases along the generally eccentric orbits. Whatever mechanism is used to obtain an average measurement error, the calculation must include an allowance for intrinsic velocity dispersion. We find that $\sigma_{\mathrm{int}}$, the intrinsic width of the $\widetilde{v}$ distribution relative to the median value, is much larger than $E \left( \widetilde{v} \right)$, the rms of the individual $\widetilde{v}$ measurement errors of the WBs in each bin \citep[similarly to][]{Banik_2024_WBT}. This is no doubt related to the proper motions having a signal-to-noise ratio of about 4000. We therefore neglect $E \left( \widetilde{v} \right)$ and consider that the uncertainty of the median $\widetilde{v}$ within each bin is MAD/$\sqrt{N}$.

\subsubsection{Limiting the \texorpdfstring{$\widetilde{v}$}{scaled velocity} uncertainty}
\label{sec:v_tilde_uncertainty_cutoff}

The WBT is largely about whether the $\widetilde{v}$ distribution changes as a function of WB internal acceleration, which is to be expected in MOND but not in Newtonian gravity as it lacks a fundamental acceleration scale. Unfortunately, it is possible that a spurious MOND signal arises from systematic effects that scale with some other variable that strongly correlates with acceleration, with $r_{\mathrm{sky}}$ being an obvious example. $\widetilde{v}$ uncertainties are generally larger at wider separations because the same velocity uncertainty in m/s translates into a larger $\widetilde{v}$ uncertainty due to the lower Newtonian $v_c$ (Equation~\ref{v_tilde}). To avoid the median $\widetilde{v}$ being inflated by measurement uncertainties, these should be much smaller than the median $\widetilde{v}$ of $\approx 0.5$ \citep[e.g.,][]{Banik_2024_WBT}. This led those authors to conduct the WBT using only those WBs whose $\widetilde{v}$ is measured to an accuracy better than 0.1 in the $\widetilde{v}$ range of $0-2$ that is most critical to the WBT. They allowed a slightly higher $\widetilde{v}$ uncertainty at larger $\widetilde{v}$, but for simplicity and to improve the data quality further, we only consider WBs where the measurement uncertainty on $\widetilde{v}$ is below 0.1. In most cases, the uncertainties are far smaller.

\section{Results}
\label{Results}

\begin{figure}
    \includegraphics[width=\columnwidth]{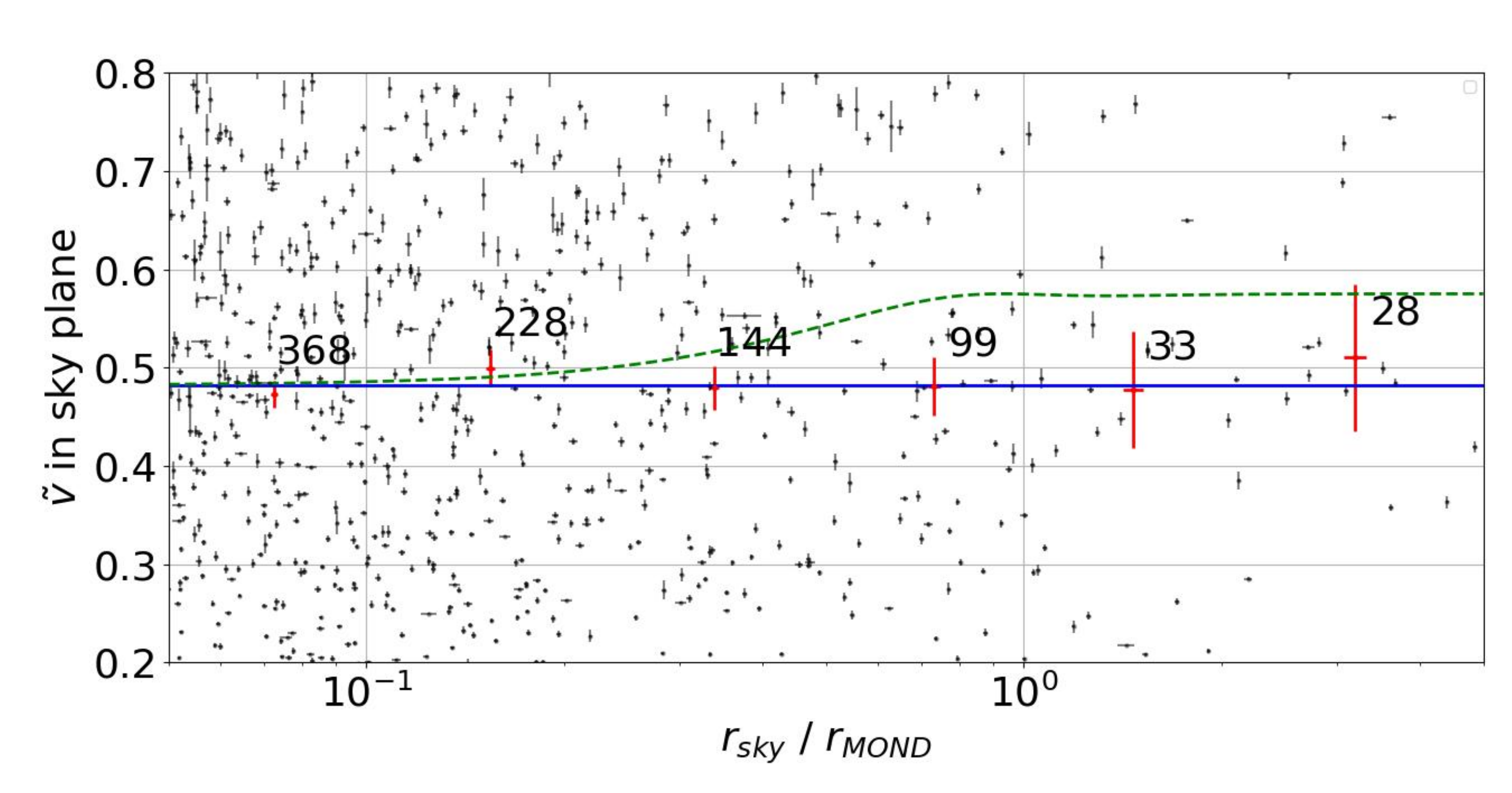}
    \caption{The distribution of $\widetilde{v}$ versus $r_{\mathrm{sky}}/r_{_{\mathrm{M}}}$ for WBs with $\widetilde{v} < 2.5$, showing also uncertainties along both axes (black points). The labelled crosses with uncertainties show the median $\widetilde{v}$ against the geometric mean of $r_{\mathrm{sky}}/r_{_{\mathrm{M}}}$, with the adjacent values showing the number of WBs in each corresponding bin. The Newtonian expectation is a horizontal line, which we illustrate here with a continuous line at 0.489. The MOND expectation is a step-like rise by 20\% at $r_{\mathrm{sky}} \approx r_{_{\mathrm{M}}}$ (Equation~\ref{v_ratio_MOND}). We show this using the dashed line with the same Newtonian normalization of 0.489.}
    \label{fig:v-tilde-r-rMOND}
\end{figure}

Our main result is the joint distribution of $\widetilde{v}$ and $r_{\mathrm{sky}}/r_{_{\mathrm{M}}}$ for a clean sample of WBs based on \emph{Gaia}~DR3 \citep{Gaia_2023}. We show this in Figure~\ref{fig:v-tilde-r-rMOND}, along with the median $\widetilde{v}$ in different bins of $r_{\mathrm{sky}}/r_{_{\mathrm{M}}}$. We also show a horizontal continuous line representing the Newtonian prediction at $\widetilde{v} = 0.489$ and a stepped dashed line showing the approximate MOND prediction (Equation~\ref{v_ratio_MOND}). Without more detailed modelling, both of these predictions can be scaled by an arbitrary factor, e.g. the Newtonian line could have been drawn at a different level, while the MOND line could rise from 0.7 to 0.84 rather than from 0.49 to 0.59. Even so, it is clear that the predicted 20\% step is not apparent in the data, which instead reveals only a flat trend to the median $\widetilde{v}$ values. This is in line with figure~11 of \citet{Banik_2024_WBT}, but here we have used different code to download the \emph{Gaia} data and prepare the WB sample, including a quite different approach to degrouping \citep[following][]{Cookson_2024}. We have also applied generally stricter quality cuts, leading to a sample size of 1421, more than the 280 in \citet{Hernandez_2012} and the 423 in \citet{Hernandez_2022}, but fewer than the 8611 WBs in \citet{Banik_2024_WBT}. Compared to this last work, the current study achieves a broader dynamic range in $r_{\mathrm{sky}}/r_{_{\mathrm{M}}}$, which goes down to 0.05 here, whereas their sample reaches down to about 0.2 \citep[addressing concerns raised in][]{Hernandez_2024_critical}. Our first two bins are at very small separations compared to the MOND radius, clarifying that the Newtonian normalization must be close to 0.5. Since this is also the median $\widetilde{v}$ at large separations,  we see that the data clearly prefers the Newtonian expectation over the Milgromian one.

Our conclusion assumes that the WB eccentricity distribution and the incidence of undetected CBs do not vary with the WB separation, since otherwise any such trends could either mimic or suppress a trend in the median $\widetilde{v}$ with our proxy for the WB acceleration. However, it is extremely unlikely that such trends not only precisely cancel out the predicted MOND signal and cause the results to look Newtonian, but also remain consistent with independent constraints \citep[see figures~15 and 16 of][]{Banik_2024_WBT}. In particular, the required changes to the eccentricity distribution are highly inconsistent with constraints from the angle between the projected WB separation and relative velocity \citep*{Hwang_2022}, while the CB fraction seems to be nearly independent of the WB separation \citep{Hartman_2022}. Therefore, trends in the median $\widetilde{v}$ with $r_{\mathrm{sky}}/r_{_{\mathrm{M}}}$ should be a reliable probe of the gravity law at the accelerations typical of galactic outskirts.

We use a standard $\chi^2$ approach to quantify the goodness of fit of the Newtonian and MOND models to the observed, binned median $\widetilde{v}$ with $r_{\mathrm{sky}}/r_{_{\mathrm{M}}}$. Since we do not perform detailed modelling of the WB population, we freely vary the median $\widetilde{v}$ in the Newtonian limit (the Newtonian normalization) to get the best overall fit. This means that although we have 6 binned data points, the $\chi^2$ values in Table~\ref{chi_sq_table} correspond to only 5 degrees of freedom. Once we have obtained $\chi^2$ in this way, we find the likelihood $P$ of a higher $\chi^2$ for 5 degrees of freedom. We then convert $P$ into an equivalent number $\chi$ of standard deviations for a single Gaussian variable with the following equation:
\begin{eqnarray}
    1 - \frac{1}{\sqrt{2 \mathrm{\pi}}} \int_{-\chi}^\chi \exp \left( -\frac{x^2}{2} \right) \, dx ~\equiv~ P \, .
    \label{Gaussian_equivalent_tension}
\end{eqnarray}
Figure~1 of \citet*{Asencio_2023} shows the relation between $P$ and the $\chi$ values obtained in this way, which we show in brackets in Table~\ref{chi_sq_table}. They summarize the overall goodness of fit to the run of median $\widetilde{v}$ with $r_{\mathrm{sky}}/r_{_{\mathrm{M}}}$ for each considered gravity law and upper limit to $\widetilde{v}$. It is clear that Newtonian gravity provides a significantly better fit than MOND in all cases. The $\Delta \chi^2$ of about $12-16$ implies a likelihood ratio of $\exp \left( \Delta \chi^2/2 \right) \approx 1500$.

\begin{table}
    \centering
    \caption{The $\chi^2$ between the indicated gravity model and the observed run of median $\widetilde{v}$ with respect to $r_{\mathrm{sky}}/r_{_{\mathrm{M}}}$, for different upper limits to $\widetilde{v}$. The $\chi^2$ values shown here are optimized with respect to the normalization in the Newtonian regime, since we refrain from any more advanced analysis of the WB population. Thus, the $\chi^2$ values correspond to 5 degrees of freedom (6 data points). The values in brackets give the likelihood of a higher $\chi^2$ as an equivalent tension for a single Gaussian variable (Equation~\ref{Gaussian_equivalent_tension}). The final column shows the preference for Newtonian gravity over MOND using the likelihood ratio $\exp \left( \Delta \chi^2/2 \right)$. All values shown here use catalogue C25 limited to $d_h < 130$~pc. The last row shows the impact of additionally requiring \texttt{ipd\_frac\_multi\_peak} = 0.}
    \begin{tabular}{llccl}
        \hline
        Quality & Sample & \multicolumn{2}{c}{Gravity law} & Newtonian \\
        filter & size & Newton & MOND &  preference\\ 
        \hline
        \multirow{2}{*}{$\widetilde{v} \leq 2.5$} & \multirow{2}{*}{1421} & 2.48 & 15.10 & 548 $\times$ \\
        & & ($0.28\sigma$) & ($2.58\sigma$) \\
        \multirow{2}{*}{$\widetilde{v} \leq 2$} & \multirow{2}{*}{1417} & 2.33 & 14.90 & 535 $\times$ \\
        & & ($0.25\sigma$) & ($2.55\sigma$) \\ 
        \multirow{2}{*}{$\widetilde{v} \leq 1.5$} & \multirow{2}{*}{1408} & 1.92 & 15.49 & 883 $\times$ \\
        & & ($0.18\sigma$) & ($2.58\sigma$) \\
        {$\widetilde{v} \leq 1.5$ \& } & \multirow{2}{*}{1216} & 1.46 & 16.09 & 1501 $\times$\\
        ipd m.p. = 0& & ($0.1\sigma$) & ($2.72\sigma$) \\
        \hline
    \end{tabular}
    \label{chi_sq_table}
\end{table}

Our main result is that Newtonian gravity is preferred over MOND by the WBT. We next discuss why this conclusion disagrees with some previous attempts at the WBT, but also why it agrees with other attempts.

\section{Applying the checklist to previous studies}
\label{Previous_studies}

This section evaluates all papers with a WB sample produced for the WBT, going back to its inception \citep*{Hernandez_2012}. Table~\ref{Review_table} applies the new checklist (Sections~\ref{Sample_selection} and \ref{Data_analysis}) to papers that implement the WBT using a largely new sample of WBs, though this can involve applying additional cuts to a previous sample. We review other literature related to the WBT in Appendix~\ref{Other_studies}, including predictions for the WBT and discussion on how to implement it, reviews of other studies, and reanalyses with minimal or no changes to the underlying WB sample. Since our focus is on preparing a suitable sample of WBs, these studies are not assessed in Table~\ref{Review_table}.

\begin{table*}
    \centering
    \caption{Summary of whether past studies that implemented the WBT using a new sample of WBs (including by significantly restricting a previous sample) correctly implemented the various steps required to obtain reliable results (see section link in first column). The rows in bold show a more general requirement, with the rows below showing the more detailed steps needed to meet this requirement. If most of the detailed requirements are satisfied and the ones not satisfied are relatively unimportant, we classify the study as passing the general requirement. The studies shown here are as follows: \citetalias{Hernandez_2012} \citep*{Hernandez_2012}, \citetalias{Hernandez_2019_WB} \citep{Hernandez_2019_WB}, \citetalias{Pittordis_2019} \citep{Pittordis_2019}, \citetalias{Hernandez_2022} \citep{Hernandez_2022}, \citetalias{Pittordis_2023} \citep{Pittordis_2023}, \citetalias{Hernandez_2023} \citep{Hernandez_2023}, \citetalias{Chae_2023} \citep{Chae_2023}, \citetalias{Chae_2024a} \citep{Chae_2024a}, \citetalias{Chae_2024b} \citep{Chae_2024b}, \citetalias{Banik_2024_WBT} \citep{Banik_2024_WBT}, and \citetalias{Cookson_2024} \citep{Cookson_2024}.}
    \begin{tabular}{|l|l|l|l|l|l|l|l|l|l|l|l|l|}
        \hline
        \textbf{Section} & \textbf{Test} & \textbf{\citetalias{Hernandez_2012}} & \textbf{\citetalias{Hernandez_2019_WB}} & \textbf{\citetalias{Pittordis_2019}} & \textbf{\citetalias{Hernandez_2022}} & \textbf{\citetalias{Pittordis_2023}} & \textbf{\citetalias{Hernandez_2023}} & \textbf{\citetalias{Chae_2023}} & \textbf{\citetalias{Chae_2024a}} & \textbf{\citetalias{Chae_2024b}} & \textbf{\citetalias{Banik_2024_WBT}} & \textbf{\citetalias{Cookson_2024}} 
        \\ \hline
        -- & Star catalogue (\emph{Gaia} if not stated) & Hip & DR2 & DR2 & eDR3 & eDR3 & DR3 & DR3 & DR3 & DR3 & DR3 & DR3 \\
        \textbf{\ref{Sample_selection}} & \underline{\textbf{Sample selection and quality cuts}} & & & & & & & & & & & \\
        \textbf{\ref{Flybys}} & \textbf{Flybys \& historical encounters} & \textbf{$\bm{\times}$} & \textbf{$\bm{\times}$} & \textbf{$\bm{\times}$} & \textbf{$\bm{\times}$} & \textbf{$\bm{\times}$} & \textbf{$\bm{\times}$}& \textbf{$\bm{\times}$}& \textbf{$\bm{\times}$}& \textbf{$\bm{\times}$}& \textbf{$\bm{\checkmark}$}& \textbf{$\bm{\times}$}  \\ 
        \ref{Flyby_cutoff_line} & Using a $\widetilde{v}$ cutoff (eg < 2.5) & $\times$ & $\times$ & $\times$ & $\times$ & $\times$ & $\times$ & $\times$ & $\times$ & $\times$ & $\checkmark$ & $\times$ \\
        \ref{Maximum_RV_difference} & Maximum RV difference (ie $< 10~\mathrm{km}^{-1}$) & $\times$ & $\checkmark$ & $\times$ & $\checkmark$ & $\times$ & $\checkmark$ & $\times$ & $\checkmark$ & $\checkmark$ & $\times$ & $\checkmark$ \\
        \ref{Limited_separation} & $r_{\mathrm{sky}} \la 30$~kAU & $\times$ & $\times$ & $\checkmark$ & $\checkmark$ & $\checkmark$ & $\checkmark$ & $\checkmark$ & $\checkmark$ & $\checkmark$ & $\checkmark$ & $\checkmark$ \\
        \hline
        -- & {Subtotal out of 3} & 0 & 1 & 1 & 2 & 2 & 2 & 1 & 2 & 2 & 2 & 2 \\ \hline
        \textbf{\ref{Close_binaries}} & \textbf{Close binaries} & \textbf{$\bm{\times}$} & \textbf{$\bm{\times}$} & \textbf{$\bm{\times}$} & \textbf{$\bm{\times}$} & \textbf{$\bm{\times}$} & \textbf{$\bm{\times}$}& \textbf{$\bm{\times}$}& \textbf{$\bm{\times}$}& \textbf{$\bm{\times}$}& \textbf{$\bm{\times}$}& \textbf{$\bm{\checkmark}$}  \\ 
        \ref{Degrouping} & Degrouping & $\times$ & N/A & $\checkmark$ & $\checkmark$ & $\checkmark$ & $\checkmark$ & $\checkmark$ & $\checkmark$ & $\checkmark$ & $\checkmark$ & $\checkmark$ \\
        \ref{Nulls} & Null values and high-noise stars & $\times$ & $\times$ & $\times$ & $\times$ & $\checkmark$ & $\times$ & $\checkmark$ & $\checkmark$ & $\checkmark$ & $\checkmark$ & $\checkmark$ \\
        \ref{RUWE} & RUWE $< 1.25 - 1.4$ & $\times$ & $\times$ & $\times$ & $\checkmark$ & $\checkmark$ & $\checkmark$ & $\times$ & $\times$ & $\checkmark$ & $\checkmark$ & $\checkmark$ \\
        \ref{HR_filter} & HR filter & $\times$ & $\times$ & $\times$ & $\checkmark$ & $\times$ & $\checkmark$ & $\times$ & $\times$ & $\times$ & $\times$ & $\checkmark$ \\
        \ref{Limiting_dhel} & Heliocentric distance limited to $\la 130 - 150$~pc & $\times$ & $\checkmark$ & $\times$ & $\checkmark$ & $\times$ & $\checkmark$ & $\checkmark$ & $\times$ & $\times$ & $\times$ & $\checkmark$ \\
        \ref{Limiting_dhel} & $\texttt{\textbf{ipd\_frac\_multi\_peak}}$\textbf{ = 0} & \textbf{$\bm{\times}$} & \textbf{$\bm{\times}$} & \textbf{$\bm{\times}$} & \textbf{$\bm{\times}$} & \textbf{$\bm{\times}$} & \textbf{$\bm{\times}$} & \textbf{$\bm{\times}$} & \textbf{$\bm{\times}$} & \textbf{$\bm{\times}$} & \textbf{$\bm{\checkmark}$} & \textbf{$\bm{\times}$} \\ \hline 
        -- & {Subtotal out of 6} & 0 & 1 & 1 & 4 & 3 & 4 & 3 & 2 & 3 & 4 & 4 \\ \hline
        \textbf{\ref{Data_analysis}} & \underline{\textbf{Data analysis}} & & & & & & & & & & & \\
        \textbf{\ref{vtilde_calculation}} & \textbf{Calculating $v_{\mathrm{sky}}$ and scaling to obtain $\widetilde{v}$} & \textbf{$\bm{\times}$} & \textbf{$\bm{\times}$} & \textbf{$\bm{\times}$} & \textbf{$\bm{\times}$} & \textbf{$\bm{\times}$} & \textbf{$\bm{\times}$} & \textbf{$\bm{\times}$} & \textbf{$\bm{\times}$} & \textbf{$\bm{\times}$} & \textbf{$\bm{\checkmark}$} & \textbf{$\bm{\times}$}  \\ 
        \ref{Mean_heliocentric_distance} & Mean heliocentric distance & $\times$ & $\times$ & $\checkmark$ & $\times$ & $\checkmark$ & $\times$ & $\checkmark$ & $\checkmark$ & $\checkmark$ & $\checkmark$ & $\checkmark$ \\
        \ref{Mass_scaling} & Mass scaling & $\times$ & $\times$ & $\times$ & $\times$ & $\checkmark$ & $\checkmark$ & $\checkmark$ & $\checkmark$ & $\checkmark$ & $\checkmark$ & $\times$ \\
        \ref{Velocity_representation} & $\widetilde{v}$ used instead of $v_{\mathrm{line}}$ or $v_{\mathrm{sky}}$ & $\times$ & $\times$ & $\checkmark$ & $\times$ & $\checkmark$ & $\times$ & $\times$ & $\checkmark$ & $\checkmark$ & $\checkmark$ & $\times$ \\
        \ref{Scaling_to_r_M} & $r_{\mathrm{sky}}/r_{_{\mathrm{M}}}$ used instead of $r_{\mathrm{sky}}$ & $\times$ & $\times$ & $\times$ & $\times$ & $\times$ & $\times$ & $\times$ & $\checkmark$ & $\checkmark$ & $\checkmark$ & $\times$ \\
        \ref{Quantifying_typical_value} & Using median for central tendency & $\times$ & $\times$ & $\times$ & $\times$ & N/A & $\times$ & $\checkmark$ & $\checkmark$ & $\checkmark$ & $\checkmark$ & $\times$ \\
        \ref{Spherical_projection_correction} & Spherical projection correction & $\times$ & $\times$ & $\times$ & $\checkmark$ & $\times$ & $\checkmark$ & $\times$ & $\times$ & $\times$ & $\checkmark$ & $\checkmark$ \\ \hline
        -- & {Subtotal out of 6} & 0 & 0 & 2 & 1 & 3 & 2 & 3 & 5 & 5 & 6 & 2 \\ \hline
        \textbf{\ref{sec:Error_propagation}} & \textbf{Measurement error propagation} & \textbf{$\times$} & \textbf{$\times$} & \textbf{$\times$} & \textbf{$\times$} & \textbf{$\times$} & \textbf{$\times$} & \textbf{$\times$} & \textbf{$\times$} & \textbf{$\checkmark$} & \textbf{$\checkmark$} & \textbf{$\times$} \\
        \ref{sec:Velocity_uncertainty_calculation} & Velocity uncertainty calculation & $\times$ & $\times$ & $\times$ & $\times$ & $\times$& $\times$ & $\checkmark$& $\checkmark$ & $\checkmark$ & $\checkmark$ & $\checkmark$ \\
        \ref{sec:v_tilde_uncertainty_cutoff} & $\widetilde{v}$ uncertainty limited to 0.1 & $\times$ & $\times$ & $\times$ & $\times$ & $\times$& $\times$ & $\times$& $\times$ & $\checkmark$ & $\checkmark$ & $\times$ \\ \hline
        -- & {Subtotal out of 2} & 0 & 0 & 0 & 0 & 0 & 0 & 1 & 1 & 2 & 2 & 1 \\ \hline
        -- & \textbf{Total score out of 17} & \textbf{0} & \textbf{2} & \textbf{5} & \textbf{7} & \textbf{8} & \textbf{8} & \textbf{8} & \textbf{10} & \textbf{12} & \textbf{14} & \textbf{10} \\ \hline
        -- & \textbf{Still Valid? (Yes/No/Inconclusive)} & No & No & Yes & No & Yes & No & No & No & Inc & Yes & Inc \\ \hline
    \end{tabular}
    \label{Review_table}
\end{table*}

Between the first study in 2012 and 2018, researchers were looking for a velocity curve flattening reminiscent of galaxy rotation curves, i.e., they were testing if the typical $v_{\mathrm{sky}}$ becomes independent of $r_{\mathrm{sky}}$ at large radii. But by 2018, it was understood that the Galactic EFE plays an important role such that a Keplerian decline is also expected beyond $r_{_{\mathrm{M}}}$, with MOND in fact predicting only a 20\% velocity uplift over the Newtonian $v_c$ \citep{Pittordis_2018, Banik_2018_EFE, Banik_2018_Centauri}. This leaves us with the conclusion that all earlier studies purportedly showing flattening were in fact using samples dominated by spurious velocities from CBs and flybys, or perhaps hitting a noise floor due to uncertainties in $v_{\mathrm{sky}}$.

\subsection{Wide binaries as a critical test of classical gravity \citep*{Hernandez_2012}}

This landmark study set the stage for the WBT, highlighting that since $r_{_{\mathrm{M}}} = 7$~kAU for a Sun-like star, MOND predicts anomalous behaviour in WBs. This had already been discussed earlier in the context of Proxima Centauri, which is 13~kAU from $\alpha$~Centauri A and B \citep{Beech_2009, Beech_2011}. It is possible to test MOND using just this single system if we can directly measure the orbital acceleration, but this would require $\mu$as astrometric precision over many years \citep{Banik_2019_Proxima}. The main advance of \citet*{Hernandez_2012} was in proposing a statistical analysis of many WBs using only their measured separation and relative velocity, which is much more practical. The authors then implemented the WBT using available data from the \emph{Hipparcos} space-based astrometry mission \citep{Perryman_1997} and the Sloan Digital Sky Survey \citep[SDSS;][]{SDSS}. Their analysis showed a flat turn-off at 1~$\mathrm{km~s}^{-1}$ ``suggestive of a breakdown of Kepler's Third Law'' beyond $r_{_{\mathrm{M}}}$. The data and techniques required for the WBT were not available in 2012, limiting the extent to which the study could meet the later-defined criteria. Further developments to include the EFE \citep[including][whose finding was accepted by other authors like \citealt{Hernandez_2023}]{Pittordis_2019} mean that a flat turn-off is no longer expected.

\subsection{Challenging a Newtonian prediction through \emph{Gaia} wide binaries \citep{Hernandez_2019_WB}}
\label{sec:Hernandez_2019_WB}

This study implemented the WBT using sources in \emph{Gaia}~DR2 \citep{Gaia_2018} cross-matched to \emph{Hipparcos}, heavily relying on the catalogue of \citet{Shaya_2011}. The authors found that the RMS $v_{\mathrm{line}}$ remains flat with $r_{\mathrm{sky}}$, arguing that ``this result challenges Newtonian gravity at low accelerations.''

The results of our assessment are shown in Table~\ref{Review_table}. The main issues include calculating MOND radii using a fixed mass of $1 \, M_\odot$ for all WBs, considering only $r_{\mathrm{sky}}$ rather than $r_{\mathrm{sky}}/r_{_{\mathrm{M}}}$, not using $\widetilde{v}$ to normalize velocities to the Newtonian expectation, not using any HR filter, and not using the RUWE statistic, which was not yet available. The quantitative results are discounted for the same reasons as \citet{Hernandez_2012} and were superseded by subsequent studies.

\subsection{Testing modified gravity with wide binaries in \emph{Gaia}~DR2 \citep{Pittordis_2019}}
\label{sec:Pittordis_2019}

This study also used \emph{Gaia}~DR2 \citep{Gaia_2018} to make two clear points:
\begin{enumerate}
    \item By considering the mode of the $\widetilde{v}$ distribution as a function of $r_{\mathrm{sky}}$, the study conclusively rejected MOND-like theories without the Galactic EFE \citep[see figure~11 of][]{Pittordis_2019}. Later studies are all in agreement with this important conclusion: WBs with $r_{\mathrm{sky}} > r_{_{\mathrm{M}}}$ also have a Keplerian decline in their relative velocity. This shifted the focus of the WBT to careful comparison of the normalization in this regime with that in the Newtonian regime, to look for the predicted 20\% excess in MOND.
    \item There is an extended tail to the $\widetilde{v}$ distribution, which reaches values much higher than can plausibly arise due to genuine WB orbital motion. This explains why some authors get results showing rather high $\widetilde{v}$, especially if using the RMS to quantify its typical value. The reasons for the extended tail were not well understood at that time $-$ we will return to this later (see Appendix~\ref{sec:Clarke_2020}).
\end{enumerate}

In terms of our checklist, there is no upper limit to $\widetilde{v}$, though the study lays the foundations for such a limit by showing an extended tail that needs to be excluded. There is no RV cut because RVs were not available in sufficient numbers until \emph{Gaia}~DR3. This prevents spherical projection corrections. On the positive side, the study implements a Galactic latitude limit of $\left| b \right| < 15^{\circ}$ and a limit to the projected separation of $r_{\mathrm{sky}} < 20$~kAU. With regards to CBs, the study does degrouping but fails most of the other checks. However, the use of a faint star catalogue to try and remove CBs is potentially more rigorous than the method used here.

The study concluded that MOND-like theories that include the expected Galactic EFE could not be decisively tested at that time. However, it demonstrated significant progress towards this goal.

\subsection{Internal kinematics of \emph{Gaia}~eDR3 wide binaries \citep*{Hernandez_2022}}
\label{sec:Hernandez_2022}

This study executed the WBT using \emph{Gaia} early Data Release 3 \citep[\emph{Gaia}~eDR3;][]{Gaia_2021} using a modified form of appendix~D in \citet[][]{Badry_2018}, updated mostly for eDR3. The results were found to be ``at odds not only with Newtonian expectations, but also with MOND predictions'' once the Galactic EFE is considered. This is because the relative velocity flattens out at $\approx 0.5$~$\mathrm{km~s}^{-1}$ instead of showing a Keplerian decline with an uplift of 20\% compared to the Newtonian normalization, demonstrating that the data were not reduced well enough to distinguish between Newtonian gravity and MOND with the Galactic EFE. Considering our quality checklist (Sections~\ref{Sample_selection} and \ref{Data_analysis}), possible reasons are that the study:
\begin{enumerate}
    \item Lacked any upper limit to the $\widetilde{v}$ distribution in the analysis;
    \item Excluded high-noise stars prior to the degrouping stage \citep{Cookson_2024};
    \item Calculated the MOND radius assuming every star is like the Sun ($M = 2 \, M_\odot$) rather than working out $r_{_{\mathrm{M}}}$ separately for each WB;
    \item Used separate trigonometric parallax distances to the stars in each WB rather than a common distance, which would have minimized perspective effects \citep{Banik_2019_line};
    \item Used the less robust RMS statistic rather than the median;
    \item Neglected the intrinsic width of the relative velocity distribution.
\end{enumerate}

These challenges mean that the study was superseded by subsequent studies for much the same reasons as \citet{Hernandez_2012} and \citet{Hernandez_2019_WB}. On the other hand, \citet*{Hernandez_2022} was helpful in pointing out that the extended tail to the $\widetilde{v}$ distribution evident in \citet{Pittordis_2019} has a declining rather than rising form, making it unlikely to arise from chance alignments or flybys. However, by then it was already clear that the most plausible interpretation is undetected CBs \citep{Clarke_2020}.

\begin{figure}
    \centering
    \includegraphics[width=\linewidth]{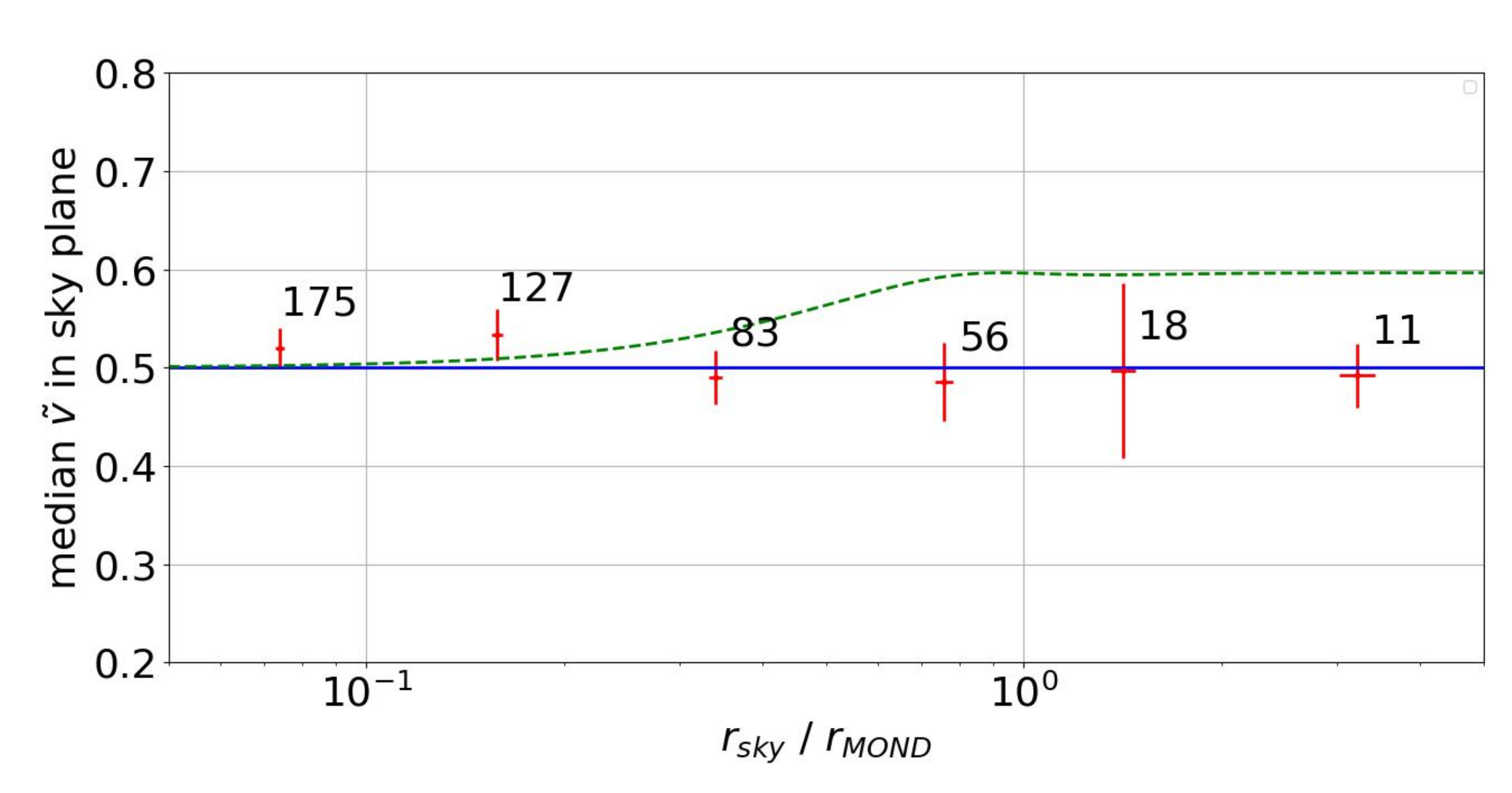}
    \caption{Similar to our main result (Figure~\ref{fig:v-tilde-r-rMOND}), but now using the sample from \citet*{Hernandez_2022} restricted further according to our quality checklist.}
    \label{fig:Hernandez_2022}
\end{figure}

The importance of our quality checklist is illustrated in Figure~\ref{fig:Hernandez_2022}, where we show the WB sample of \citet*{Hernandez_2022} but with some further systems removed to make it compliant with the checklist. Those authors claim to find a non-Newtonian signal in the data, but applying our quality checklist leads to a result that is much more in line with Newtonian expectations than with MOND.

\subsection{Internal kinematics of \emph{Gaia}~DR3 wide binaries: anomalous behaviour in the low acceleration regime \citep{Hernandez_2023}}

The overall approach of this study was similar to previous studies by the same author, but using the newer \emph{Gaia}~DR3 \citep{Gaia_2023}. Methodological concerns with \citet{Hernandez_2023} are similar to those discussed in Section~\ref{sec:Hernandez_2022}. A significant improvement is more careful estimates of the mass of each star, though this information is not used as much as it could be -- the main analyses focus on relative velocity against projected separation. However, the $\widetilde{v}$ distribution of the data is shown and compared with Newtonian and Milgromian expectations (albeit neglecting any contamination). This led to a slight increase in the quality score, but the other checklist points invalidate the main finding of the paper, namely that ``a non-Newtonian low acceleration phenomenology is thus confirmed.'' Significant improvements to the quality controls and analysis would be required to draw such a conclusion. The results of this study have been superseded for similar reasons as earlier studies by the same author.

\subsection{Wide Binaries from \emph{Gaia}~EDR3: preference for GR over MOND? \citep{Pittordis_2023}}
\label{sec:Pittordis_2023}

The authors slightly modify the techniques in their 2019 study (Section~\ref{sec:Pittordis_2019}), hoping that the improved quality of data in \emph{Gaia}~eDR3 will provide a more definitive answer for the WBT. The study was the first to do the WBT using a comparison of the full $\widetilde{v}$ distribution. While its authors openly acknowledge some issues with modelling flybys and the undetected CB population, they find a clear preference for Newtonian gravity over MOND. 

This study provided the first hint that the WBT might end up favouring GR over MOND. However, it also contained a few deficiencies, as summarized in Table~\ref{Review_table}. Perhaps the most important was the lack of any HR filters or spherical projection corrections, the latter being infeasible due to the very small number of available RVs. The study considered the full $\widetilde{v}$ distribution rather than a summary statistic like the median or RMS. While one can argue that this makes the study even better than if following our checklist, results obtained from fitting the full $\widetilde{v}$ distribution become more model-dependent. Assessing the forward WB model -- including flybys and CBs -- goes beyond the scope of our contribution.

\subsection{Breakdown of the Newton–Einstein Standard Gravity at Low Acceleration in Internal Dynamics of Wide Binary Stars \citep{Chae_2023}}

This study executed its version of the WBT using a sample of 26,615 WBs from \citet*{Badry_2021}. The WBs were chosen to be within 80~pc or within 200~pc, with similar results reported in both cases. \citet{Chae_2023} claimed a $10\sigma$ deviation from Newtonian dynamics. The upper limit to the WB separation is $10^{4.5}$~AU or 32~kAU, which is quite reasonable. However, a number of issues with the sample selection are evident, the most important being the lack of any $\widetilde{v}$ quality cut. The discussion mentions using RUWE $<1.2$, but not for the main analysis. Neither this study nor the underlying WB sample \citep{Badry_2021} appear to use filters based on $\Delta \mathrm{RV}$. We also consider that \citet{Chae_2023} did not adequately mitigate the impact of flybys and historical encounters, since these inflate relative velocities at large separations $-$ as indeed reported in this study. It only takes a modest amount of contamination to skew the whole result \citep[see figure~11 of][]{Banik_2024_WBT}.

Another kind of contamination is genuine WBs whose $\widetilde{v}$ is too uncertain for the WBT (Section~\ref{sec:v_tilde_uncertainty_cutoff}). If a significant number of WBs in some $r_{\mathrm{sky}}/r_{_{\mathrm{M}}}$ (or acceleration) bin have a large $\widetilde{v}$ uncertainty, the median $\widetilde{v}$ for that bin will be inflated as intrinsic dispersion and measurement errors combine to broaden the $\widetilde{v}$ distribution. This issue is known to be a serious problem for the \citet{Chae_2023} analysis, as discussed in detail in section~5.3 of \citet{Banik_2024_WBT}. In particular, those authors showed that an appropriate cut on the $\widetilde{v}$ uncertainty that removes just 20\% of the WBs completely eliminates the apparent rising trend in median $\widetilde{v}$ with respect to $r_{\mathrm{sky}}/r_{_{\mathrm{M}}}$. Prior to this cut, the result departs substantially from the flat Newtonian expectation and instead resembles the expected behaviour in MOND, though flat behaviour beyond the MOND radius is still somewhat unclear. This was not discussed in \citet{Chae_2023} because it used a much smaller number of bins, making it unclear if the relative velocities follow a Keplerian decline in the quasi-Newtonian regime.

Other issues with \citet{Chae_2023} are the lack of a cut on the HRD, with the parallel band of equal-brightness stars clearly evident in all four panels of its figure~1. This is related to the twin population of WBs with exactly equal mass stars \citep{Badry_2019_twin}. These issues prompt us to consider \citet{Chae_2023} as not adequately accounting for CBs, especially given the analysis assumes that CBs play only a minor role. We therefore do not consider the sample to have adopted all the practices recommended here.

\subsection{Robust Evidence for the Breakdown of Standard Gravity at Low Acceleration from Statistically Pure Binaries Free of Hidden Companions \citep{Chae_2024a}}

This study builds on the earlier analysis of \citet{Chae_2023} and makes some improvements. It implements a strict quality cut based on $\Delta \mathrm{RV}$ and clearly specifies that RUWE must be $<1.2$, though unfortunately the RUWE cut was not implemented in the main analysis because of an oversight \citep{Chae_2024a_erratum}. The relative velocity calculation assigns the same heliocentric distance to both stars, this being the ``error-weighted mean'' (presumably the inverse variance weighted mean) of the individual parallax distances. \citet{Chae_2024a} uses $\widetilde{v}$ as a key component of the analysis and quantifies its dependence on $r_{\mathrm{sky}}/r_{_{\mathrm{M}}}$, though no upper limit is imposed on $\widetilde{v}$ as recommended in our checklist (Section~\ref{Flyby_cutoff_line}). Even if other cuts are handled appropriately, using an insufficiently strict upper limit to $\widetilde{v}$ of 5 rather than $\approx 2$ can lead to a MOND-like signal in the former case that disappears in the latter case, which cannot happen with a genuine MOND signal because genuine WBs should not have $\widetilde{v} > 2$ \citep[see figure~11 of][]{Banik_2024_WBT}.

Other issues with the analysis of \citet{Chae_2024a} are the use of a somewhat high limit to $d_h$ of 200~pc and the lack of a suitable HRD cut, though this is used to some extent in the underlying catalogue \citep{Badry_2021}. Moreover, equations 6--9 of \citet{Chae_2024a} clearly demonstrate that the difference in proper motion is multiplied by the systemic distance to get the projected relative velocity, so there is no spherical projection correction. It has previously been demonstrated that this can lead to the inference of anomalously large orbital velocities at large separations \citep{Badry_2019_geometry}. In Section~\ref{sec:appendix_spher_proj}, we demonstrate the importance of this issue for our own WB sample. Given the above, there are still too many issues with the sample selection and data analysis for the conclusions to be considered safe.

\subsection{Measurements of the Low-Acceleration Gravitational Anomaly from the Normalized Velocity Profile of \emph{Gaia} Wide Binary Stars and Statistical Testing of Newtonian and Milgromian Theories \citep{Chae_2024b}}

This study reuses the WB sample considered in \citet{Chae_2024a} with appropriate RUWE cut and makes some further improvements to the analysis, though it ultimately again reaches the earlier conclusion that the data prefers MOND over Newtonian dynamics. 

Applying the checklist, we see that it implements only a minimal HRD cut rather than the much tighter cut we recommend. Although it does not explicitly require $\widetilde{v} < 2.5$, the main sample nonetheless lacks systems outside this range. A potentially serious problem is that its equation~6 demonstrates that no spherical correction has been applied despite its importance to the WBT \citep[see our Section~\ref{Spherical_projection_correction} and][]{Badry_2019_geometry}. Following the same idea as section~2.3.1 of \citet{Banik_2024_WBT}, we can estimate the possible impact of systemic RV on $\widetilde{v}$ by considering a WB 50~pc away with a total mass of $1.5 \, M_\odot$ and $r_{\mathrm{sky}} = 10$~kAU. If the WB has a quite plausible systemic RV of 50~$\mathrm{km~s}^{-1}$, then the impact on $\widetilde{v}$ could be up to 0.13, depending on the angle between the projected separation and relative velocity. The effects are larger for less massive and less distant WBs with larger separations, since perspective effects scale with the angular separation of the WB. Because only a small number of WBs with incorrectly estimated $\widetilde{v}$ can skew the WBT \citep[see section~5.2 of][]{Banik_2024_WBT}, the lack of spherical projection corrections could be an important deficiency of \citet{Chae_2024b}, as demonstrated in Section~\ref{sec:appendix_spher_proj} of this paper.

\subsection{Strong constraints on the gravitational law from \emph{Gaia}~DR3 wide binaries \citep{Banik_2024_WBT}}

This study compared a large sample of 8611 WBs with a forward model for the WB population, undetected CBs, and line of sight contamination (see their figure~10 for how each of these affects the model prediction). The authors introduced the $\alpha_{\mathrm{grav}}$ parameter to interpolate between the predictions of Newtonian gravity (0) and MOND (1), but also allowed values somewhat outside this range. The main conclusion of the study was that $\alpha_{\mathrm{grav}} = -0.021^{+0.065}_{-0.045}$, excluding MOND at $16\sigma$ confidence but well within $1\sigma$ of the Newtonian expectation. The authors argued that the very close agreement with the Newtonian expectation of 0 would be a remarkable coincidence if MOND were actually correct but there were systematic errors in some aspect(s) of the analysis.

Nearly all the steps identified in the quality checklist were implemented by this study. A second faint-star catalogue was used to remove CBs as far as possible, avoiding the need to consider nulls and high-noise stars. This is probably the preferred method because it avoids using the same catalogue for both the main WB sample and to remove triples. The authors also used the $\texttt{ipd\_frac\_multi\_peak}$ parameter to further remove stars which sometimes appear to not be a single source. While the main purpose of the study was to construct a detailed forward model of the WB population including undetected CBs and chance alignments, here we only assess their sample selection and the basic median $\widetilde{v}$ analysis presented in their figure~11.

The use of a forward model meant their analysis could handle a significant level of CB contamination, which was undoubtedly present in their sample. This is partly due to the lack of the HRD cut (apart from a cut to exclude WDs, which removed only a very small number of WBs). Moreover, the sample went out to 250~pc, much higher than our recommended limit of 150~pc with an \texttt{ipd\_frac\_multi\_peak} filter. Another possible source of contamination is that the study imposed no cut on $\Delta \mathrm{RV}$ in systems where only one star in a WB had known RV. Knowing the RV of just one star in a WB is sufficient for the spherical projection correction, but not for our recommended quality cut related to $\Delta \mathrm{RV}$.

Our overall assessment of the \citet{Banik_2024_WBT} WB sample is that although most aspects are handled appropriately, their quality cuts fail to adequately remove CBs due to the inclusion of WBs out to 250~pc and the lack of an HRD cut. We identified no issues with the way in which the authors computed $r_{\mathrm{sky}}/r_{_{\mathrm{M}}}$ and $\widetilde{v}$ for their sample. Their calculation of $\widetilde{v}$ uncertainties included astrometric uncertainties, perspective effects, and mass uncertainties. We therefore have no issues with their conclusion that the flat median $\widetilde{v}$ with respect to $r_{\mathrm{sky}}/r_{_{\mathrm{M}}}$ favours Newtonian gravity over MOND (see their figure~11).

\subsection{Influence of selection criteria on the interpretation of rotational behaviour of wide-binary star systems \citep{Cookson_2024}}

This study focused on a hitherto overlooked issue with the removal of CBs in some prior studies (Section~\ref{Nulls}). Addressing  this issue led to a ``complete elimination of any apparent non-Newtonian motion''. The study focused on the removal of CBs, thereby scoring well against our criteria in this regard. However, the study is not ideal as an implementation of the WBT all by itself. The main issue is the use of $v_{\mathrm{line}}$ rather than $\widetilde{v}$ and $r_{\mathrm{sky}}$ rather than $r_{\mathrm{sky}}/r_{_{\mathrm{M}}}$, which makes it difficult to look for non-Newtonian behaviour because the Newtonian $v_c$ depends on mass (which might have a different distribution at larger separations) and the MOND regime starts at different separations for WBs of different mass, respectively. The lack of any calculation of $v_{\mathrm{sky}}$ (let alone $\widetilde{v}$) also meant that the study imposed no upper limit to $\widetilde{v}$. The study also used the less robust RMS statistic rather than the median. We handle these issues much more carefully in this contribution.

\section{Discussion}
\label{Discussion}

The chronologically ordered Table~\ref{Review_table} shows that from the inception of the idea behind the WBT \citep*{Hernandez_2012}, there were a series of advances in the theory \citep{Banik_2018_Centauri, Pittordis_2019, Belokurov_2020, Clarke_2020} and data, in particular from \emph{Hipparcos}, SDSS, \emph{Gaia}~DR2, \emph{Gaia}~eDR3, and most recently \emph{Gaia}~DR3. This led to earlier results on the WBT successively being superseded as theory and data advanced \citep{Hernandez_2012, Hernandez_2019_WB, Pittordis_2019, Hernandez_2022}. We are now in a situation where a number of papers have been published either denying non-Newtonian anomalies \citep{Banik_2024_WBT, Cookson_2024} or claiming the presence of such anomalies \citep{Chae_2024b}.

It can be seen that as the total scores increase across Table~\ref{Review_table}, the claimed deviations from Keplerian motion on kAU scales gradually become smaller. \citet*{Hernandez_2012}, \citet{Hernandez_2019_WB}, and \citet*{Hernandez_2022} all suggest that WBs reach a velocity floor of about 0.5~$\mathrm{km~s}^{-1}$ within errors, reminiscent of flat galaxy rotation curves. \citet*{Hernandez_2012} reported that the flat level is about 1~$\mathrm{km~s}^{-1}$, but the study scored zero in the checklist. The score had risen to 7 a decade later \citep*{Hernandez_2022}, but the flat level had dropped to about 0.5~$\mathrm{km~s}^{-1}$.

\citet{Pittordis_2019} took an important step by invalidating the notion that WB orbital velocities remain flat beyond some separation, and thus versions of MOND without the EFE. Although their analysis only scored 5, their later study \citep{Pittordis_2023} scored 8 and tentatively claimed a problem for MOND even with the EFE. Their cautious wording is somewhat at odds with their quantitative results (Section~\ref{sec:Pittordis_2023}), but this is reasonable for the first study which appeared sensitive enough to discriminate between Newtonian gravity and MOND with the Galactic EFE. It is also reasonable given the many steps in our quality checklist which those authors did not follow, though of course this could not have been known at the time given the limited understanding then.

Other studies claimed to find evidence for a $10-20\%$ uplift to the Newtonian $v_c$ for separations $\ga 0.03$~pc \citep{Chae_2023, Chae_2024a, Chae_2024b}. These studies receive scores of $8-11$ against our quality checklist, thus implementing many of the quality controls necessary to reliably implement the WBT. However, they also suffer from important deficiencies, for instance the lack of spherical projection corrections (Section~\ref{sec:appendix_spher_proj}).

These studies are contradicted by \citet{Banik_2024_WBT} and \citet{Cookson_2024}, which receive scores of 14 and 10, respectively. The latter study highlighted an important issue with some earlier studies regarding the removal of CBs. Bearing this in mind and incorporating insights from earlier studies, the current investigation ensures full compliance with the quality checklist we introduced earlier. This is the most thorough implementation of the WBT to date, and it also finds against MOND (Figure~\ref{fig:v-tilde-r-rMOND}).

\subsection{Excluding low Galactic latitudes}
\label{Excluding_low_b}

\begin{figure}
    \centering
    \includegraphics[width=\linewidth]{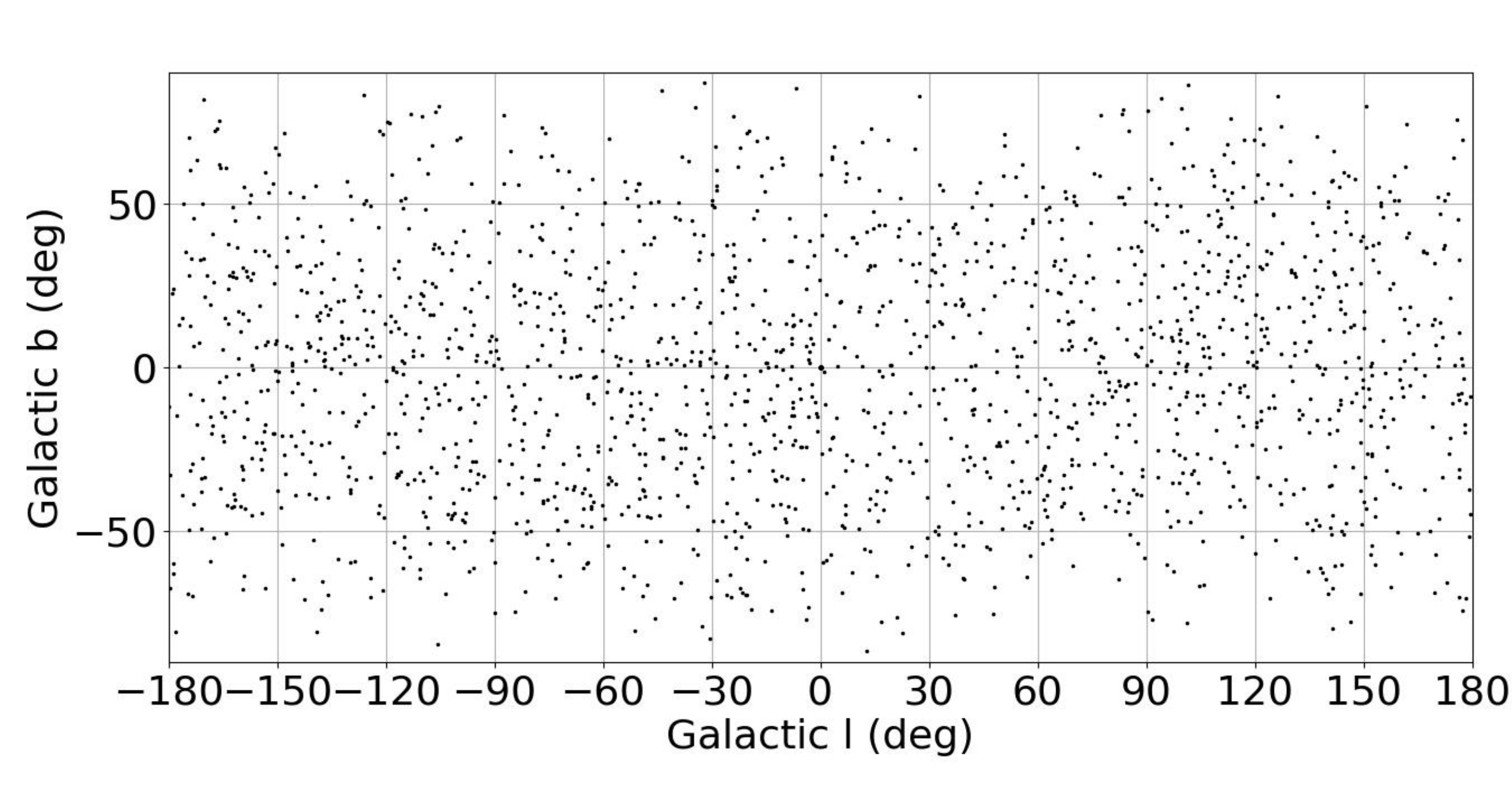}
    \caption{Distribution of sky positions in Galactic coordinates for our WB sample.}
    \label{fig:skymap-b}
\end{figure}

A number of authors have excluded low Galactic latitudes to reduce the possibility of chance alignments \citep{Pittordis_2019, Cookson_2024}. There is a related issue that stars could have a nearby faint star in projection, which could perturb the astrometry even if the faint star is much more distant. A cut on the Galactic latitude $\left| b \right|$ could also help here by excluding crowded fields of view. However, tests have shown that a limit on $\widetilde{v}$ is effective at avoiding all these issues. We have therefore excluded this test from the checklist. We show the sky distribution of our sample in Figure~\ref{fig:skymap-b}. The stellar density appears fairly constant for $\left| b \right| \la 50^\circ$, rather than heavily clustered around $b = 0$, showing that degrouping has successfully removed triple and higher-order systems, comoving star clusters, etc.

\subsection{Effects of spherical projection corrections and the HRD cut}
\label{sec:appendix_spher_proj}

\begin{figure}
    \centering
    \begin{subfigure}{\linewidth}
        \centering
        \includegraphics[width=\linewidth]{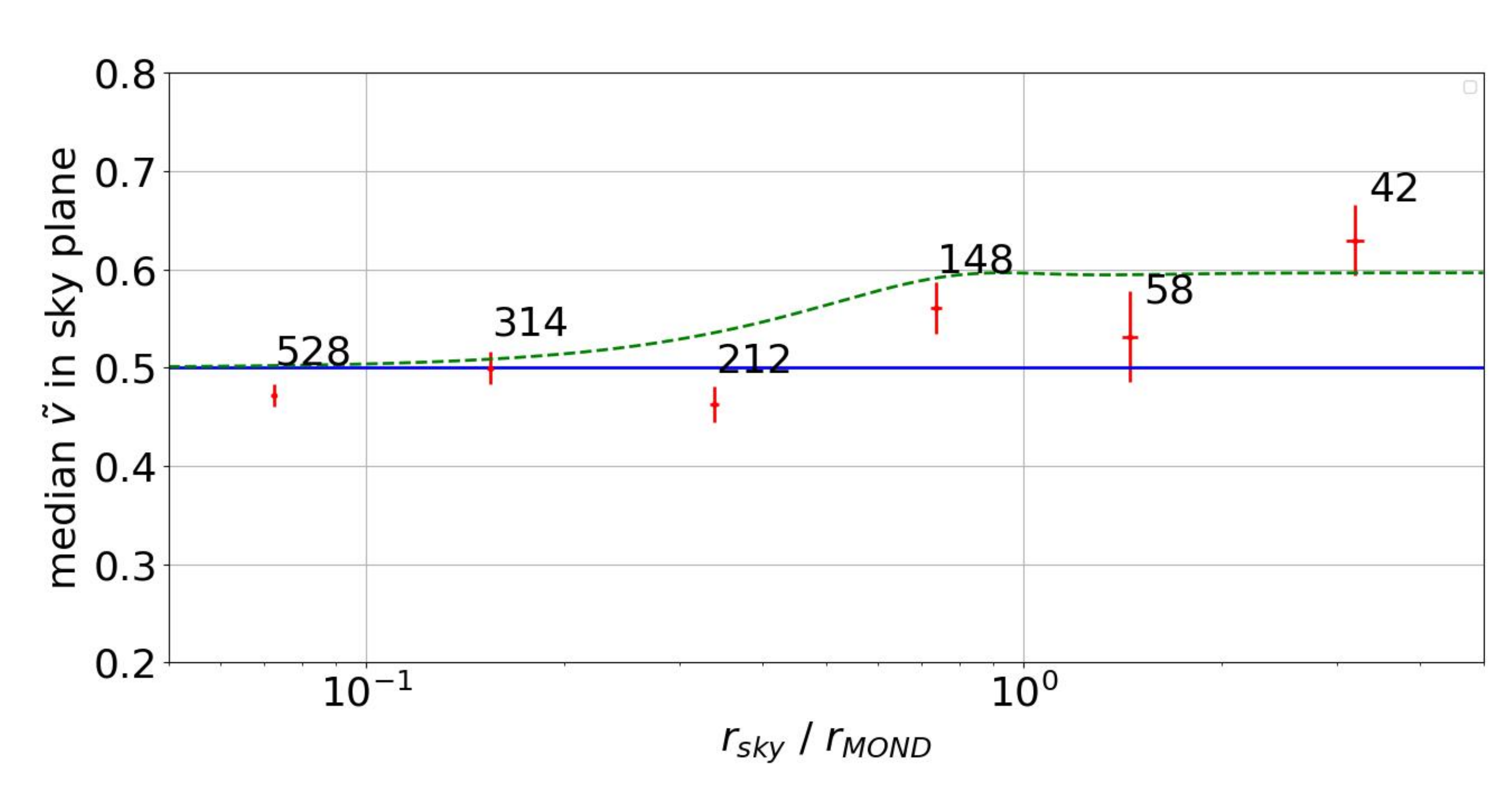}
        \caption{Omitting the spherical projection correction.}
        \label{fig:no_SC}
    \end{subfigure}

    \vspace{1em}

    \begin{subfigure}{\linewidth}
        \centering
        \includegraphics[width=\linewidth]{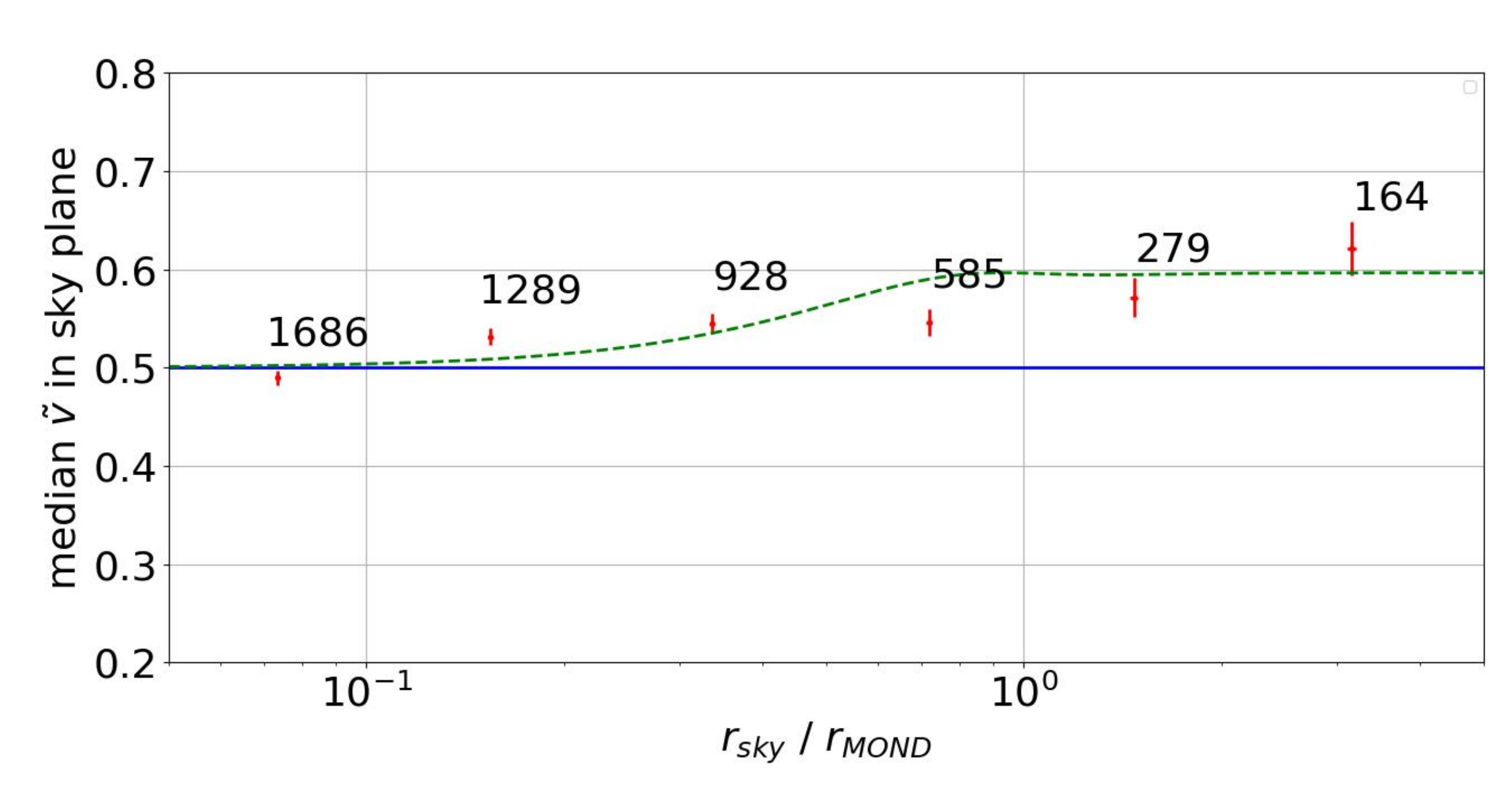}
        \caption{Omitting the HRD cut.}
        \label{fig:no_HRD_cut}
    \end{subfigure}

    \vspace{1em}

    \begin{subfigure}{\linewidth}
        \centering
        \includegraphics[width=\linewidth]{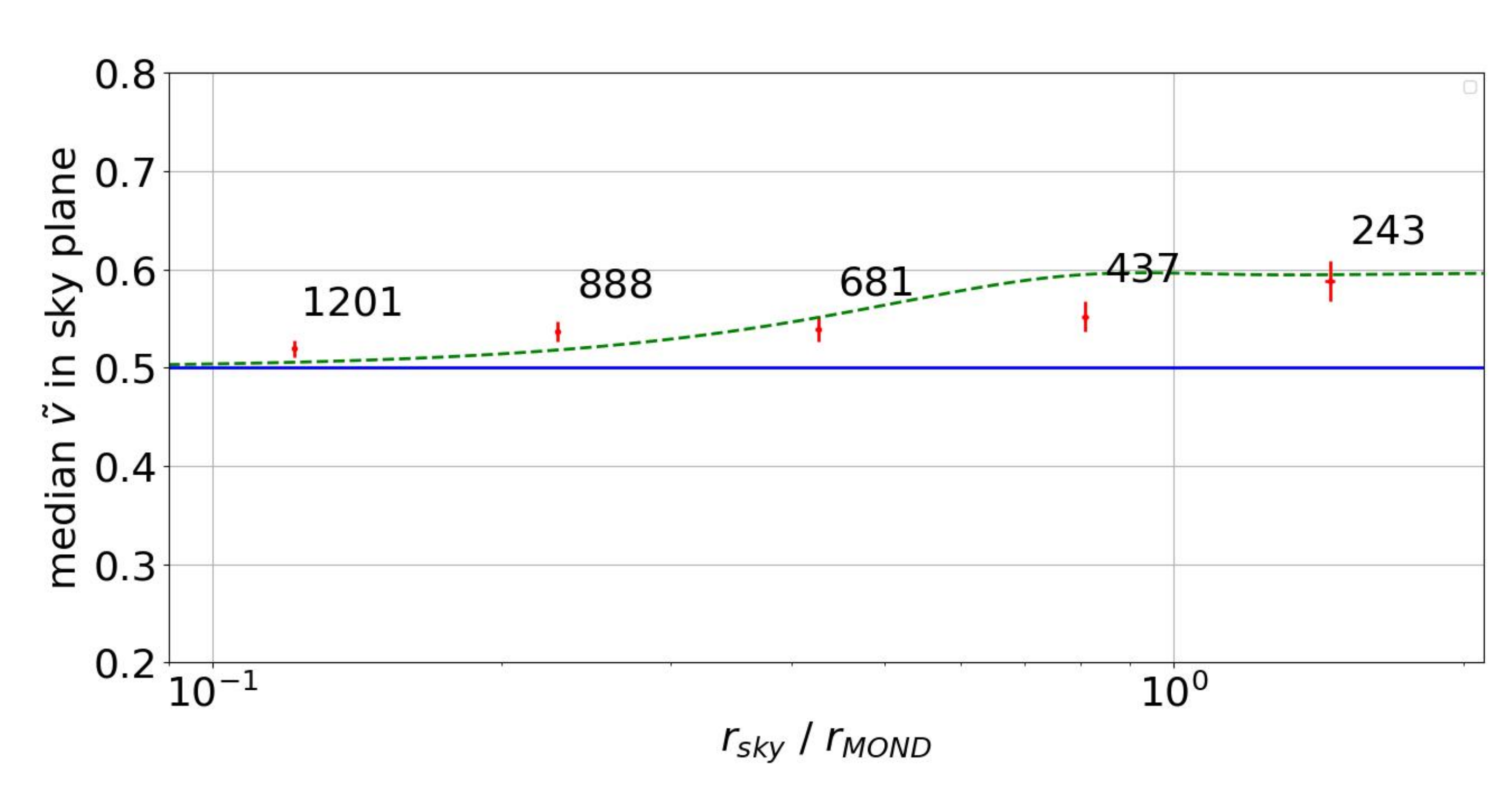}
        \caption{Omitting both corrections and applying smoothing.}
        \label{fig:smoothing}
    \end{subfigure}

    \caption{Three variations of our main result (Figure~\ref{fig:v-tilde-r-rMOND}). a) No spherical projection correction. b) No HRD cut. c) Combining the above changes and applying smoothing. These changes can cause the results to resemble the MOND prediction (dashed line).}
    \label{fig:appendix}
\end{figure}

Although the angular separation of each WB is small, it is still important to account for spherical projection corrections (Section~\ref{Spherical_projection_correction}). Not doing so leads to increasingly larger errors in $v_{\mathrm{sky}}$ at larger separations, potentially creating a spurious MOND signal \citep{Badry_2019_geometry}. To illustrate this effect for our sample, we omit the spherical projection correction in the top panel of Figure~\ref{fig:appendix}. The effect on the median $\widetilde{v}$ is very minor ($\la 0.01$) at smaller radii, but in an unfortunate coincidence, it rapidly escalates to an uplift by about $0.15$ as we cross the MOND radius (c.f. Table~\ref{tab:sperical_corrections}).

The typical absolute difference between $\widetilde{v}$ with and without SC is a distinct issue to how much the median $\widetilde{v}$ is boosted, which is a second-order effect since SC can affect $\widetilde{v}$ either way. Comparing to Figure \ref{fig:v-tilde-r-rMOND} shows that the boost is about 0.15.

In the middle panel of Figure~\ref{fig:appendix}, we relax the HRD cut (Section~\ref{HR_filter}). Again, the results are appreciably different to our main analysis (Figure~\ref{fig:v-tilde-r-rMOND}). On top of the above-mentioned coincidence that spherical projection corrections become important beyond $r_{_{\mathrm{M}}}$, the effect of unseen third stars also becomes more apparent at approximately the same point. The left panel of Figure~\ref{HRD-full} shows that the RUWE value is raised above the MS in a pattern corresponding to the parallel bar of equal brightness double stars \citep{Badry_2019_twin}, which will have an underestimated mass \citep{Clarke_2020}. However, a CMD cut is safer because RUWE is not a perfect indicator. This is partly because RUWE is sensitive to CBs only to the extent that they introduce extra astrometric acceleration due to a mismatch between the photocentre and barycentre \citep{Penoyre_2022a, Banik_2024_WBT}. An identical pair of stars in a binary has no such mismatch, being problematic for the WBT only due to underestimated mass thanks to the steep mass-luminosity relation \citep{Pecaut_2013}.

In the bottom panel of Figure~\ref{fig:appendix}, we ignore both corrections and we alter the bin positions in $r_{\mathrm{sky}}/r_{_{\mathrm{M}}}$ to obtain a smoother curve. These adjustments create an illusion of MONDian kinematics, especially given that we do not know the exact form of the interpolation between the Newtonian and quasi-Newtonian regimes. It is therefore clear that a MOND-like signal can be generated quite by chance if we are not careful with our quality cuts and data analysis, here exemplified through the HRD cut and spherical correction, respectively.

\subsection{Distance and ipd multi-peak fraction}
\label{Distance_ipd}

\begin{figure}
    \centering

    \begin{subfigure}{\linewidth}
        \centering
        \includegraphics[width=\linewidth]{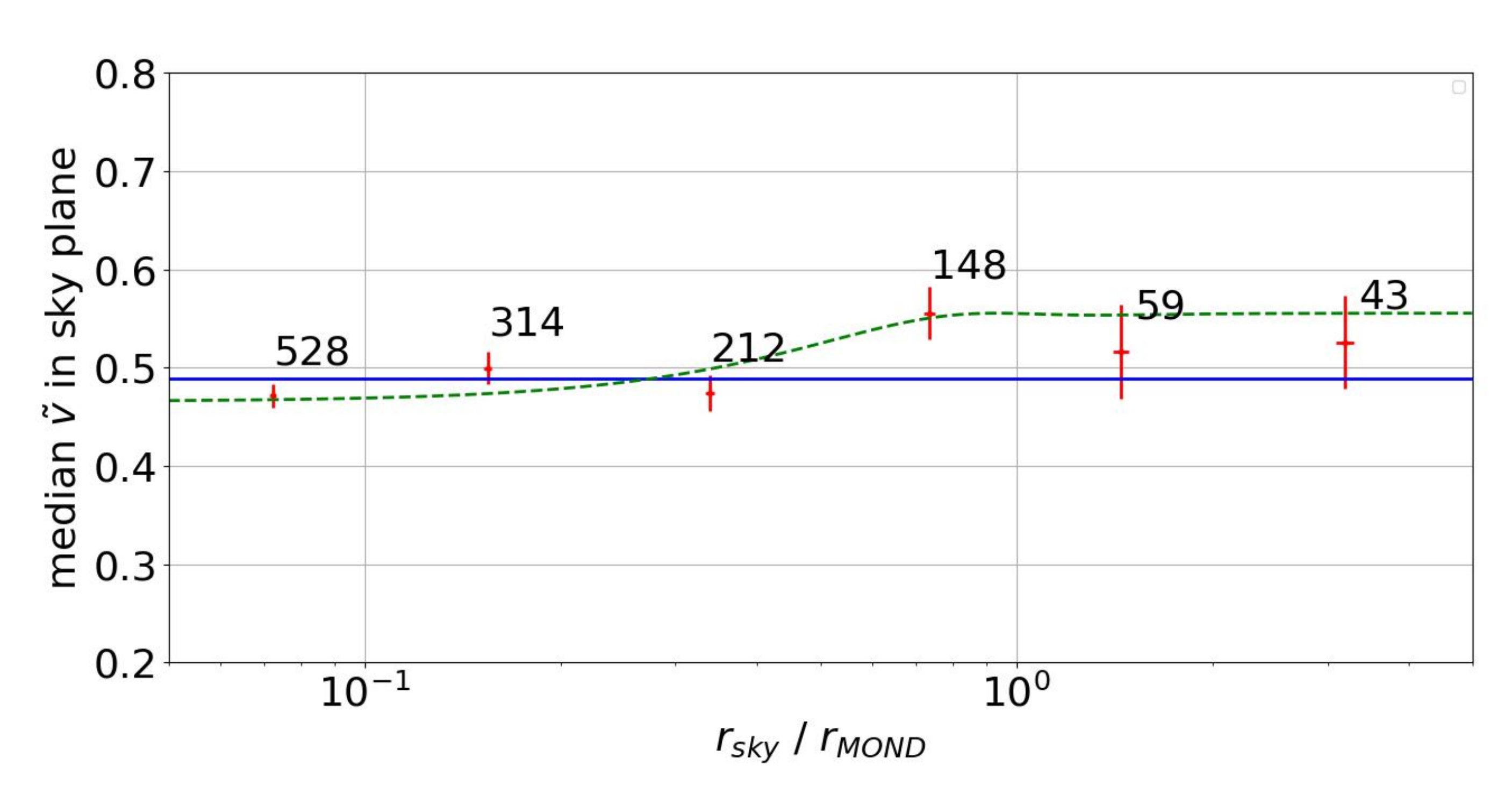}
        \caption{Results for $d_h < 150$~pc. The best MOND fit (dashed line) is $12\times$ more likely than the best Newtonian fit (solid line)}
        \label{fig:p40_d150_ipdmp500_r05_05_MOND}
    \end{subfigure}


    \begin{subfigure}{\linewidth}
        \centering
        \includegraphics[width=\linewidth]{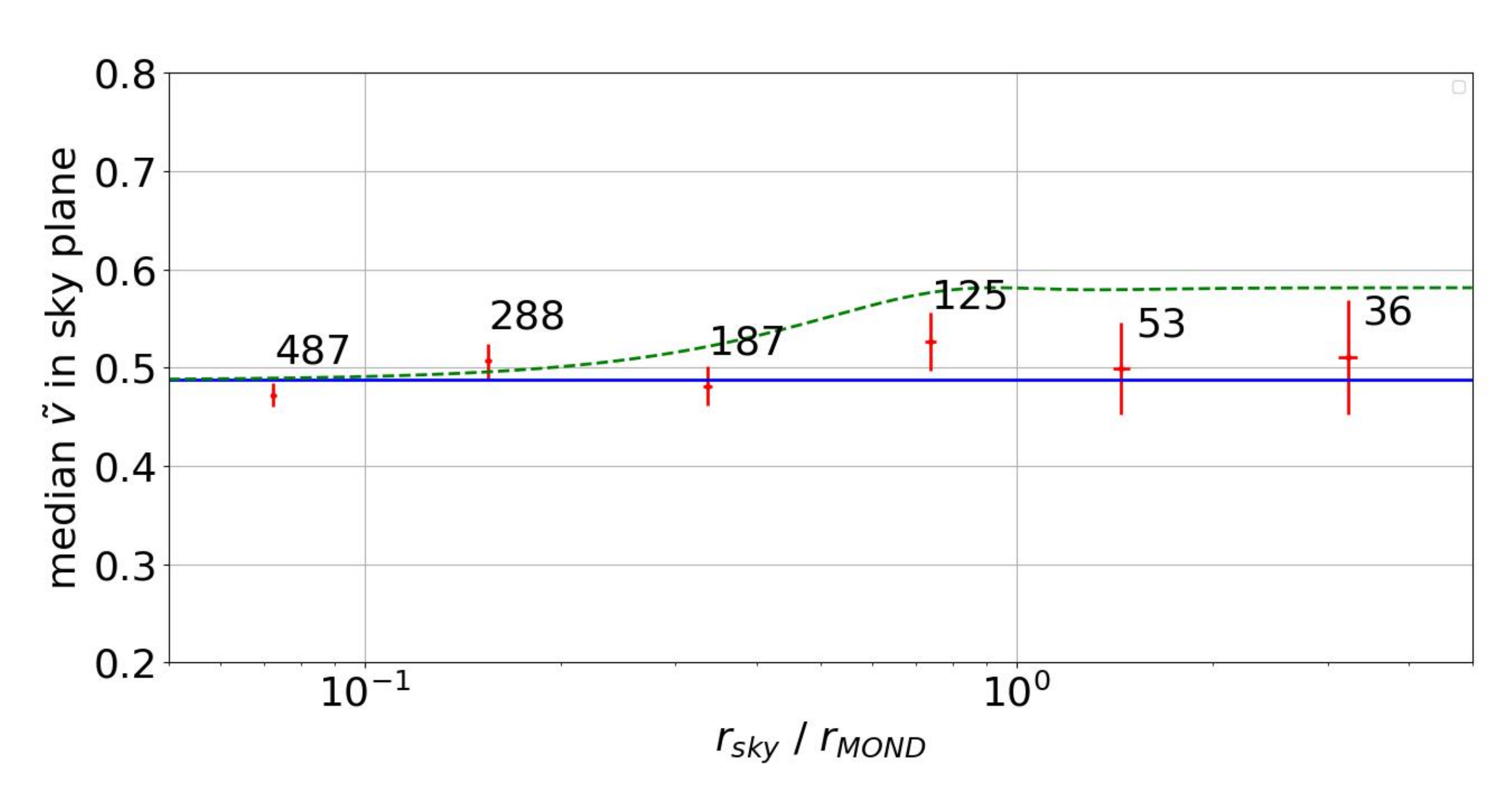}
        \caption{Results for $d_h < 150$~pc, but also requiring \texttt{ipd\_frac\_multi\_peak} = 0. The best Newtonian model is now $88\times$ more likely than the best MOND model.}
        \label{fig:p40_d150_ipdmp0_r05_05}
    \end{subfigure}


    \begin{subfigure}{\linewidth}
        \centering
        \includegraphics[width=\linewidth]{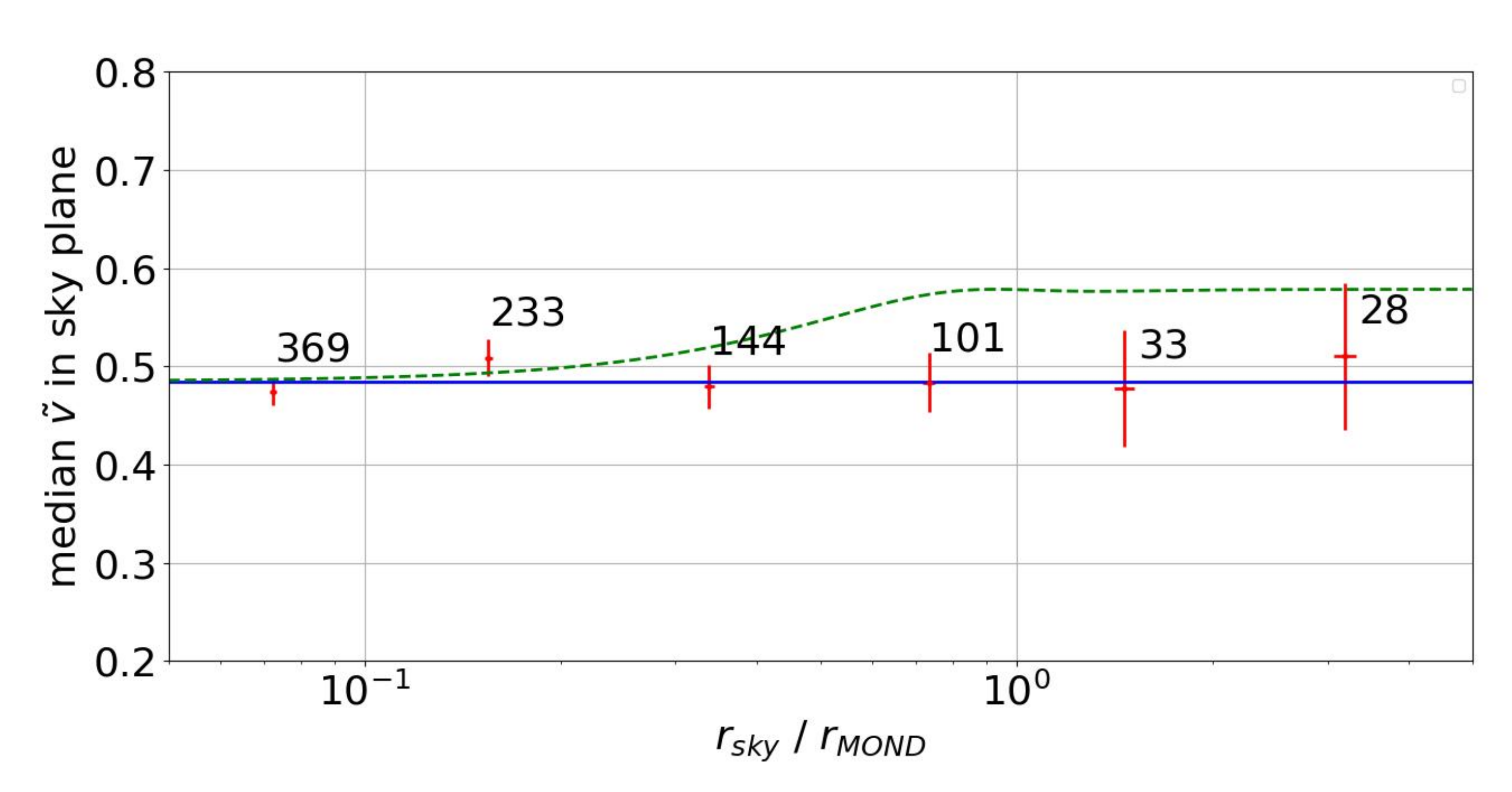}
        \caption{Results for $d_h < 130$~pc as in our main analysis, but also requiring \texttt{ipd\_frac\_multi\_peak} = 0. The best Newtonian model is now $1500\times$ more likely than the best MOND model.}
        \label{fig:p40_d130_ipdmp0_r05_05}
    \end{subfigure}

    \caption{Exploration of the fitting results for different limits to $d_h$, showing that we can get a fake MOND signal if we extend our sample out to 150~pc. This can be done reliably only if we also impose a quality cut on \texttt{ipd\_frac\_multi\_peak}, which recovers the preference for Newton.}
    \label{fig:ipd_explorations}
\end{figure}

The strong preference for Newton in our main analysis should remain robust with respect to the maximum allowed $d_h$. We therefore extend our sample out to 150~pc in Figure~\ref{fig:p40_d150_ipdmp500_r05_05_MOND}. MOND is now $12\times$ more likely than Newton. However, extending our sample makes it more vulnerable to contamination by CBs, which are harder to detect further out because they become fainter and have a smaller angular separation. To compensate for this, we introduce the additional requirement that \texttt{ipd\_frac\_multi\_peak} = 0 \citep[section~2.4.3 of][]{Banik_2024_WBT}. This is because a non-zero value indicates that one of the stars in a WB is sometimes resolved into multiple sources by \emph{Gaia}, suggesting that it might have a marginally detected CB companion that is sometimes discernible and sometimes not, depending on the observational conditions and possibly the CB orbital phase. In Figure~\ref{fig:p40_d150_ipdmp0_r05_05}, we show results out to 150~pc using only stars with \texttt{ipd\_frac\_multi\_peak} = 0, which only modestly reduces the sample size. The best-fitting Newtonian model is now $88\times$ more likely than the best-fitting MOND model. Finally, in Figure~\ref{fig:p40_d130_ipdmp0_r05_05}, we limit the sample to $d_h < 130$~pc as in our main analysis, but keep the requirement that \texttt{ipd\_frac\_multi\_peak} = 0. Newtonian dynamics is now $1500\times$ more likely than MOND.

\subsection{Broader implications}

Once the Galactic EFE is included, the rotation curve should follow a Keplerian decline with increasing separation, but the normalization should be 20\% larger in the MOND regime than in the Newtonian regime \citep{Banik_2018_EFE}, a finding that is now widely accepted. Indeed, \citet{Pittordis_2019} conclusively rule out MOND without the EFE. This invalidates studies which claim to show that WBs follow isolated MOND expectations, as if the Galactic EFE is not there \citep{Hernandez_2012, Hernandez_2019_WB, Hernandez_2022}. The reported flattening of the WB relative velocity with increasing separation could be interpreted as a noise floor, perhaps due to considering WBs with too large a separation such that chance alignments and tides from other stars become relevant \citep{Banik_2019_line, Hernandez_2019_WB}.

Several studies advance the idea of an upper limit to $\widetilde{v}$ \citep{Banik_2018_EFE, Pittordis_2018, Pittordis_2019, Pittordis_2023, Banik_2024_WBT}. We fully implement this here and in the \textsc{bynary} software on \textsc{github}, which is available for download. Since it is not necessary to consider $\widetilde{v} > 2.5$ (or even 2) to find a genuine MOND signal, doing so unnecessarily risks allowing contaminating effects that undermine the WBT \citep[see figure~11 of][]{Banik_2024_WBT}. Our analysis therefore supersedes all studies that fail to require $\widetilde{v} < 2.5$, unless of course they obtain a sample which satisfies this condition without requiring it \citep{Chae_2024b}. Similarly, the uncertainty on $\widetilde{v}$ needs to be sufficiently small to avoid biasing the result \citep[as discussed in detail in section~5.2 of][]{Banik_2024_WBT}. We find that this cut is nearly always satisfied if the other steps in our quality checklist are followed. Spherical projection corrections also need to be accounted for \citep{Shaya_2011, Badry_2019_geometry} -- their importance to our WB sample is demonstrated in Section~\ref{sec:appendix_spher_proj}. \citet{Chae_2024b} do not include this correction (see their equation~6).

As there are no further studies purporting to show a MONDian response in local WBs, we are drawn to the conclusion that if there is a low-acceleration departure from Newtonian dynamics, it must also depend on other factors besides the acceleration. This is in line with the historical development of MOND as a theory of galactic dynamics \citep{Milgrom_1983} rather than dynamics in the Solar System, where MOND has recently encountered difficulties \citep{Desmond_2024, Vokrouhlicky_2024}. Perhaps MOND is only effective on kpc scales, or if the density is sufficiently low -- after all, WBs have a higher density than the Galaxy as a whole \citep[see section~5.4 of][]{Banik_2024_WBT}. The MONDian departure from Newtonian dynamics might also be dependent on the potential energy, a particularly interesting idea that may be related to a deeper quantum gravity theory.

\section{Conclusions}
\label{Conclusions}

An important goal of astrophysics is to test if the laws of physics validated in laboratory environments remain applicable in the much more extreme environments accessible when observing on much larger scales. In particular, the Newtonian inverse square law of gravity known to work in the Solar System appears to break down in galaxies, which follow a tight non-Newtonian relation between their visible mass distribution and the kinematically inferred acceleration \citep{Famaey_McGaugh_2012, Lelli_2017, Li_2018, McGaugh_2020, Desmond_2023, Stiskalek_2023}. This may be due to the onset of MOND when the acceleration becomes $\la a_{_0}$. This possibility can be tested using the dynamics of WBs in the Solar neighbourhood, which should orbit each other 20\% faster than the Newtonian expectation if their separation exceeds $r_{_{\mathrm{M}}}$ (Equation~\ref{r_M}). Testing gravity in this way has proven controversial, with some attempts at the WBT finding the predicted MOND signal at high confidence, even as other studies published at almost the same time using similar datasets rule out the MOND signal at similarly high confidence.

This study introduces a quality checklist to ensure consistency and accuracy in studies of WBs that aim to conduct the WBT, covering quality cuts that should be applied to select an appropriate WB sample (Section~\ref{Sample_selection}) and how the data on this sample should be analysed (Section~\ref{Data_analysis}). The use of standardized procedures for managing measurement uncertainties and reducing the contaminating effects of CBs, flybys, and historical encounters will improve the reliability of comparisons between GR and MOND.

We follow these procedures to prepare a sample of WBs. We find that the results favour Newtonian gravity over MOND (Figure~\ref{fig:v-tilde-r-rMOND}). A standard $\chi^2$ analysis indicates that Newtonian dynamics is some $1500\times$ more likely than MOND given the observed flat trend in the median $\widetilde{v}$ (Equation~\ref{v_tilde}) with respect to $r_{\mathrm{sky}}/r_{_{\mathrm{M}}}$, our proxy for the Newtonian internal orbital acceleration of each WB (Table~\ref{chi_sq_table}).

We next attempt to disentangle why our result disagrees with earlier studies that claimed to find a MOND signal in the data, but also why we agree with earlier studies that claimed no such signal exists. This leads us to review whether existing attempts at the WBT perform the steps outlined in our quality checklist (Table~\ref{Review_table}). We find that the quality of studies has generally improved over the years, especially after the release of \emph{Gaia} data. By reviewing each study, we are able to identify significant departures from the standard checklist in nearly every case, with even the most complete assessment so far suffering from a few issues. This quality checklist is readily able to identify issues with previous studies that claimed to find a MOND signal. The most compelling and most recent such study \citep{Chae_2024b} overlooks spherical projection and does not include an HRD cut, casting doubt on its preference for MOND. We demonstrate in Section~\ref{sec:appendix_spher_proj} that these issues combined with careful choices of bin positions can lead to a spurious MOND signal in the data.

During the final stages of the refereeing process for this work, several new studies were published concerning the WBT \citep{Pittordis_2025, Saglia_2025, Saad_2025, Yoon_2025, Makarov_2026}. While a full assessment using the quality checklist proposed here is beyond the present scope, we note that these recent contributions continue the active development of methodologies in this field. The framework established here provides a foundation for evaluating these and future studies. We note that some of these studies implement the WBT in 3D, which would require additional quality controls on data along the line of sight.

A more general issue with the WBT is that since difficulties with the data are more likely to arise at large separations and since relative velocities cannot be negative, unaccounted systematics are more likely to inflate the median $\widetilde{v}$ at large $r_{\mathrm{sky}}/r_{_{\mathrm{M}}}$, thereby mimicking a MOND signal \citep[see figure~11 of][]{Banik_2024_WBT}. It seems unlikely to have the converse situation where a genuine MOND signal is precisely countered by systematic uncertainties, for instance trends in the CB population or the eccentricity distribution (see their figures~15 and 16). A clear demonstration of MOND would require a sample selected and analysed according to the checklist to follow a step-like trend in the median $\widetilde{v}$ with respect to $r_{\mathrm{sky}}/r_{_{\mathrm{M}}}$ (dashed line on Figure~\ref{fig:v-tilde-r-rMOND}). Importantly, there should be a flat trend also beyond $r_{_{\mathrm{M}}}$, not merely a deviation from a flat trend upon crossing $r_{_{\mathrm{M}}}$. This is because systematic effects can coincidentally become significant upon crossing the MOND radius, given the transition is in any case somewhat blurred by projection effects and other uncertainties. However, systematics would generally continue becoming more important at larger separations, leading to a continuous rising trend in the median $\widetilde{v}$. It is unlikely for systematics to create the MOND-predicted step-like behaviour in the median $\widetilde{v}$ as a function of internal WB acceleration.

Our attempt to find the MOND signal in this way shows instead that the MOND prediction is very likely incorrect. We also identify problems with all attempts at the WBT so far, perhaps explaining why some studies have claimed that the WBT decisively favours MOND. We hope that our quality checklist serves as a useful guide on how any future attempt at the WBT should prepare an appropriate sample and analyse it.

\section*{Acknowledgements}

IB acknowledges support from Royal Society University Research Fellowship grant 211046. ZP acknowledges support from the European Research Council (ERC) grant 101002511. We thank the referee for their generous comments and suggestions.

\section*{Data Availability}

The sample of WBs analysed in this study will be provided upon reasonable request to the lead author. All the data here was downloaded from the \emph{Gaia} collaboration.\footnote{\url{https://gea.esac.esa.int/archive/}} It needs a login. The \textsc{python} code \textsc{bynary} is available for download and use by interested parties.\footnote{\url{https://github.com/SteveBz/Bynary}} Currently the software has only been installed on the \textsc{ubuntu} family of operating systems. Requests for support to install on other \textsc{linux} operating systems or other operating systems would be welcome. Please email the lead author with any questions.

\bibliographystyle{mnras}
\bibliography{WBQA_bbl}


\begin{appendix}

\section{Other studies advancing the WBT}
\label{Other_studies}

The WBT relies on a significant amount of groundwork in gathering the relevant observations of millions of stars and identifying WBs. It also relies on accurate theoretical predictions of their expected behaviour in Newtonian and Milgromian gravity. In this section, we review several studies that are of significant importance to the WBT, even though they did not implement it.

\subsection{Testing modified gravity theories via wide binaries and \emph{Gaia} \citep{Pittordis_2018}}

This purely theoretical study introduced the $\widetilde{v}$ parameter and investigated its distribution in a wide variety of gravity theories, including MOND without the Galactic EFE. An important conclusion was that although the $\widetilde{v}$ distribution is affected by both the eccentricity distribution and the gravity law, the effects are quite distinct, so a detailed analysis should be able to break the degeneracy. While the consideration of MOND without the EFE is interesting due to the predicted dramatic effects, the EFE is an integral part of MOND \citep{Milgrom_1986}. Unfortunately, \citet{Pittordis_2018} were unable to accurately calculate the predicted effect of MOND including a realistic Galactic EFE, though they estimated an enhancement to the orbital velocities of about $4-8\%$.

\subsection{Testing gravity with wide binary stars like $\alpha$ Centauri \citep{Banik_2018_Centauri}}

This study was the first to present detailed analytic and numerical calculations of the predicted MOND signal in local WBs. Prior to its publication, predictions were often obtained using an approach similar to equation~59 of \citet{Famaey_McGaugh_2012}, which can give incorrect results, even yielding the prediction that Milgromian WBs orbit slower than the Newtonian prediction for some choices of the MOND interpolating function \citep[see figure~4 of][]{Pittordis_2018}. The work of \citet{Banik_2018_Centauri} demonstrated that local WBs at large separations should orbit about 20\% faster than the Newtonian expectation. It was also demonstrated that the $\widetilde{v}$ distribution responds quite differently to changes in the eccentricity distribution and in the gravity law \citep[see figure~3 of][]{Banik_2018_Centauri}. The study went on to forecast that only a few hundred WBs would be required to implement the WBT (see their figure~10), albeit in an idealised scenario without measurement errors or astrophysical systematics -- which the authors considered at length (see their section~8.1). Since the study was published prior to suitable \emph{Gaia} data, no attempt was made to analyse actual data on WBs, making the study an important source of \emph{a priori} predictions. Other authors have subsequently accepted the predictions, including the predicted 20\% velocity excess in MOND and the importance of CBs.

\subsection{A new line on the wide binary test of gravity \citep{Banik_2019_line}}

The main purpose of this study was to consider various techniques for implementing the WBT, depending on precisely which information would ultimately be available in the \emph{Gaia} data. A key point in their section~2.3 was that both stars in a WB should be placed at the same heliocentric distance (Section~\ref{Mean_heliocentric_distance}). The study tried out the new techniques on the WB sample used in \citet{Hernandez_2019_WB}, finding no evidence for a departure from Newtonian gravity of the form expected in MOND, but at the same time finding no strong evidence against MOND either. Since \citet{Hernandez_2019_WB} did claim a preference for MOND (Section~\ref{sec:Hernandez_2019_WB}), \citet{Banik_2019_line} considered why their conclusions were different. One reason was the low importance attached to WBs with $r_{\mathrm{sky}} \gg 50$~kAU, which are less suitable for the WBT due to the possible impact of other stars (Section~\ref{Limited_separation}). The other reason was the use of an outlier rejection technique to better focus on the main part of the $v_{\mathrm{line}}$ distribution. While the modern techniques were then in their infancy, this was the first hint that the WB relative velocity distribution has an extended tail even in the \emph{Gaia} era, suggesting that it cannot be explained away by measurement errors.

\subsection{Unresolved stellar companions with \emph{Gaia} DR2 astrometry \citep{Belokurov_2020}}

This study discussed how the RUWE statistic could be helpful in reducing contamination from undetected CBs (Sections~\ref{RUWE} and ~\ref{HR_filter}). The basic principle is that an undetected CB would cause an astrometric acceleration of its companion. This would typically be thousands of times greater than the WB orbital acceleration, but it might still be undetectable in a narrow range of parameters while still impacting the measured $\widetilde{v}$ of the WB \citep*{Manchanda_2023}. Mitigating and/or modelling CB contamination is nowadays considered a vital aspect of the WBT, with the RUWE statistic thought to be a good way of achieving the former. The principle has been accepted by subsequent authors  and is further supported by dedicated surveys of higher-order stellar multiplicity, which find triple fractions consistent with the $\approx$50\% required to explain the high-$\widetilde{v}$ tail \citep{Riddle_2015}.

\subsection{The distribution of relative proper motions of wide binaries in \emph{Gaia}~DR2: MOND or multiplicity? \citep{Clarke_2020}}
\label{sec:Clarke_2020}

This study proposed that the extended tail to the $\widetilde{v}$ distribution evident in the studies of \citet{Hernandez_2019_WB} and \citet{Pittordis_2019} might be ``a consequence of the higher order multiplicity of the WB sample'', i.e., due to undetected CBs. By fitting the observed tail, \citet{Clarke_2020} suggested that the triple fraction is around 50\% \citep[similar to that found by][]{Banik_2024_WBT}.

\subsection{A million binaries from \emph{Gaia}~eDR3: sample selection and validation of \emph{Gaia} parallax uncertainties \citep{Badry_2021}}

The authors construct a \emph{Gaia}~eDR3-based catalogue of spatially resolved binary stars within 1~kpc, estimating chance alignment probabilities empirically. The catalogue includes 1.3 million likely bound pairs, including WD + MS and WD + WD binaries. They use this data to assess \emph{Gaia}~DR3 parallax uncertainties, providing a correction function and enabling diverse astrophysical applications. The study provides a useful base from which to download \emph{Gaia} data for the WBT.

\subsection{From Galactic Bars to the Hubble Tension: Weighing Up the Astrophysical Evidence for Milgromian Gravity \citep{Banik_Zhao_2022}}

This review of MOND discusses the WBT in some detail as an important future test. With regards to the available evidence, the authors conclude that ``there are compelling theoretical and empirical reasons for supposing that the EFE is an important aspect of MOND''. This is based partly on the WB results of \citet{Pittordis_2019} which we discussed earlier in Section~\ref{sec:Pittordis_2019}, but also based on several other considerations unrelated to the WBT, including theoretical considerations. The conclusion that only versions of MOND with the Galactic EFE remain empirically viable agrees with the results of other workers.

\subsection{Wide Binaries as a Modified Gravity test: prospects for detecting triple-system contamination \citep*{Manchanda_2023}}

The main challenge facing the WBT at the moment is the role played by undetected CBs (Section~\ref{Close_binaries}). While various strategies to reduce their prevalence and/or model the undetected CB population can help, this study focused on prospects for directly detecting this troublesome population. It found that direct, speckle, and coronagraphic imaging are particularly effective for larger CB separations, while the astrometric and RV acceleration induced by the CB orbit are more promising at smaller separations given the larger orbital acceleration. Only a small gap in parameter space remains difficult to detect with any of the considered methods. As a result, there are good prospects for characterizing the CB population in much more detail over the next decade. This will help to better understand how the WBT should be conducted and if trends in the CB population with WB separation might interfere with the predicted MOND signal.

\subsection{Statistical analysis of the gravitational anomaly in \emph{Gaia} wide binaries \citep{Hernandez_2024_statistical}}

This study reuses the WB sample from \citet{Hernandez_2023} and concludes that in the low-acceleration regime, gravity is enhanced over the Newtonian expectation by a factor close to 1.4. Considering the uncertainties, the deviation from Newtonian expectations was quoted as being significant at $2.6\sigma$.

\subsection{A critical review of recent \emph{Gaia} wide binary gravity tests \citep*{Hernandez_2024_critical}}
\label{sec:appendix_crit_review}


This study is largely a critical review of \citet{Banik_2024_WBT}. Several methodological concerns are apparent with \citet{Hernandez_2024_critical}. One notable aspect is in their figure~3, which suggests that the WBT cannot be reliably implemented because changes to the eccentricity distribution have a similar impact to changing the gravity law. However, the figure shows the opposite, in line with earlier results \citep{Banik_2018_Centauri, Pittordis_2018, Pittordis_2019}. There are several other aspects in \citet{Hernandez_2024_critical} that warrant further scrutiny, including the claim that the inferred fraction of CBs is too high because they expected values of about $0.25-0.5$. \citet{Banik_2024_WBT} obtain $1\sigma$ lower limits as low as 0.55 in some analysis variants where this was not fixed (see their table~3), arguing in their section~5.1.3 that further fine-tuning of the CB separation distribution could perhaps reduce it to 0.5, thus quite plausibly matching the upper end of the range considered reasonable by \citet{Hernandez_2024_critical}. Another aspect in their study that merits discussion is the claim regarding the step-like behaviour in the width of the $\widetilde{v}$ distribution with a sample that only goes down to $r_{\mathrm{sky}}/r_{_{\mathrm{M}}} = 0.2$, which the \citet{Banik_2024_WBT} sample reaches down to (see their figures~7 and 11). Our Figure~\ref{fig:v-tilde-r-rMOND} demonstrates that this is quite sufficient because at such a high acceleration of about $25 \, a_{_0}$, MOND predicts only negligible deviations from Newtonian gravity. Moreover, there is no assumption in \citet{Banik_2024_WBT} that WBs at any separation are perfectly Newtonian.

\citet{Hernandez_2024_critical} claim that the $\widetilde{v}$ distribution is appreciably broadened by measurement errors, even though these are much smaller than the intrinsic dispersion. The authors correctly note that if the intrinsic dispersion is 0.5 and the typical measurement error is 0.1, the observed distribution would have a width of $\sqrt{0.5^2 + 0.1^2} = 0.51$. This represents a 2\% broadening -- far smaller than the 20\% broadening expected in MOND. The authors then discuss how a system might be observed in a bin other than its actual bin using its latent $\widetilde{v}$. This is certainly true, but it ignores the fact that if the underlying distribution is approximately linear over the extent of each bin, then about as many systems would be scattered out of a bin into neighbouring bins, as would be scattered from neighbouring bins into the bin of interest. Consequently, there would hardly be any discernible \emph{net} impact on the number of WBs in each bin, and thus in the final $\widetilde{v}$ distribution. It should be obvious that if the underlying distribution is only broadened 2\% by measurement errors, then this will remain true after binning the distribution.

\begin{figure}
    \includegraphics[width=\columnwidth]{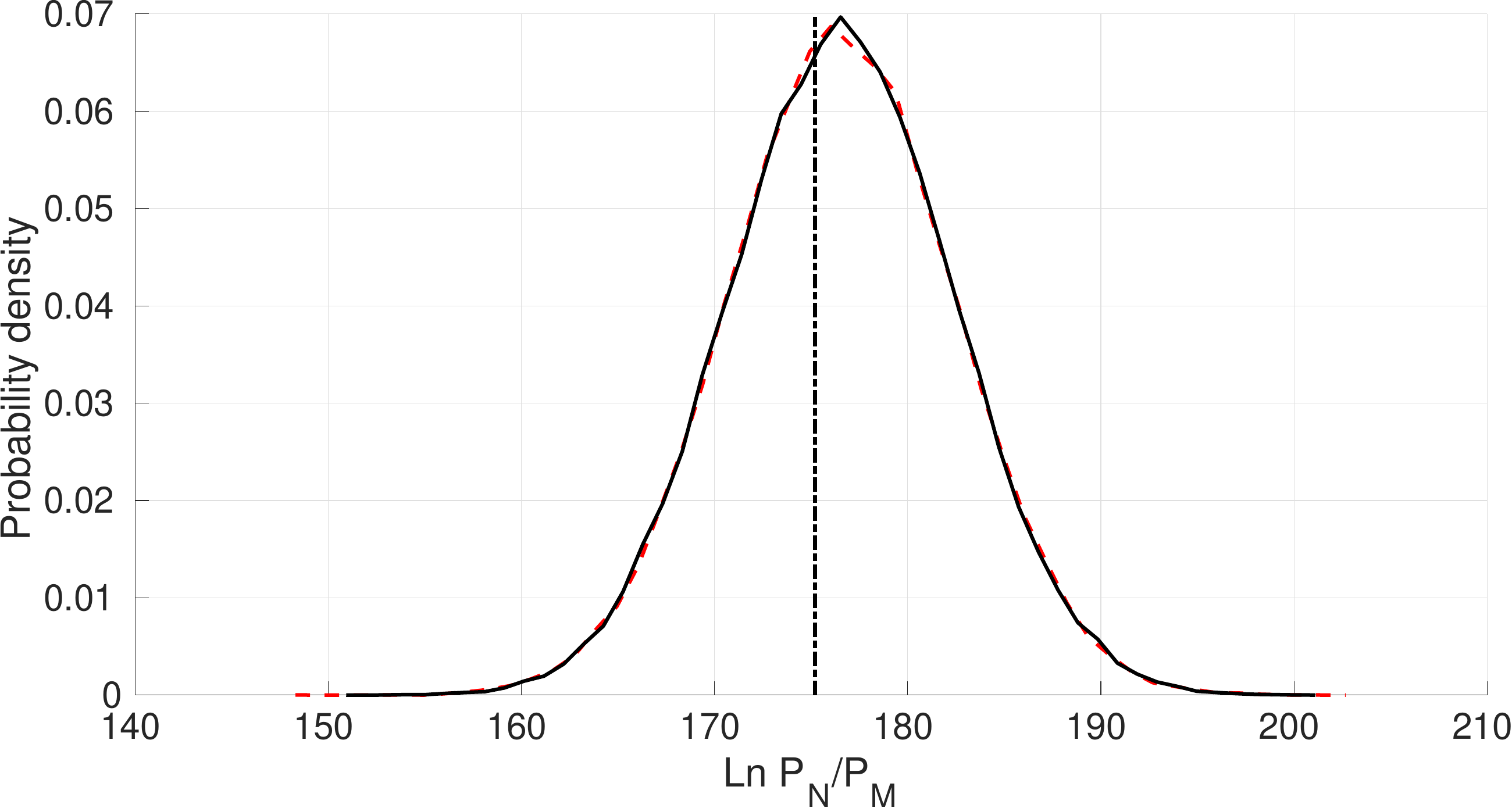}
    \caption{The ratio of binomial log-likelihoods between the best-fitting Newtonian and MOND models found in section~4.1 of \citet{Banik_2024_WBT}, both using the actual WB data as published in their table~2 (vertical black dot-dashed line) and in $10^5$ MC trials where we scatter the $\widetilde{v}$ of each WB by its measurement uncertainty, assuming a Gaussian distribution with standard deviation given by the MC trials discussed in their study (curves). Since the comparison uses binomial statistics with a binned $\widetilde{v}$ distribution covering the range $0-5$, some WBs can end up with $\widetilde{v}$ outside this range in the mock datasets. The solid black curve shows results with reflective boundary conditions on both ends, while the dashed curve has a reflective boundary at $\widetilde{v} = 0$ but allows systems to leak through the boundary at $\widetilde{v} = 5$, which would reduce the sample size. Both assumptions yield very similar results, which moreover are in good agreement with the result neglecting $\widetilde{v}$ measurement uncertainties.}
    \label{Delta_Ln_P_values_MC}
\end{figure}

To explore this issue further, we repeat the analysis in \citet{Banik_2024_WBT} using mock datasets where the $\widetilde{v}$ of each WB is scattered by its uncertainty, which we approximate as a Gaussian distribution with standard deviation given by the MC trials described in their section~2.4.6. We follow their approach of using binomial statistics with the same bin edges to compare each mock dataset to the best-fitting Newtonian and MOND models published in their table~2, thereby obtaining the log-likelihood $\ln P$ for both Newton and MOND, which we indicate using $N$ and $M$ subscripts, respectively. The use of binomial statistics that can handle even rather low number counts contradicts the claim of \citet{Hernandez_2024_critical} that ``the small average cell occupancy numbers of only 16'' were ``ignored'' by \citet{Banik_2024_WBT}. We handle any negative $\widetilde{v}$ values in the mock datasets by taking the modulus. Since their analysis uses bins which only go up to $\widetilde{v} = 5$, we also have to consider how we deal with systems that end up with $\widetilde{v} > 5$. Our nominal approach is to create a reflective boundary here, thereby mapping $5.1 \to 4.9$, etc. However, we also consider a leaky boundary, allowing WBs to be scattered up through the boundary and thereby be lost from the sample, reducing the overall sample size. Since there are hardly any WBs with $\widetilde{v}$ close to 5 and the measurement errors are so small, we find that both choices lead to almost identical results, as shown in Figure~\ref{Delta_Ln_P_values_MC}. Importantly, both sets of $10^5$ MC trials yield almost the same overwhelming preference for Newtonian gravity over MOND as the actual WB dataset, which gives $\Delta \ln P \equiv \ln P_N - \ln P_M = 175.2$ in favour of Newton \citep[see table~2 of][]{Banik_2024_WBT}. This demonstrates the expected result that neglecting the very small $\widetilde{v}$ uncertainties in their WB sample cannot have appreciably influenced their overall result or its high statistical significance. The underlying reason is that the width of the observed $\widetilde{v}$ distribution is almost exclusively intrinsic and not ``noise-variance-dominated'', as claimed by \citet{Hernandez_2024_critical}.

\subsection{A recent confirmation of the wide binary gravitational anomaly \citep{Hernandez_2025}}
\label{sec:appendix_recent_confirmation}

This study focuses on the findings of \citet{Cookson_2024}, highlighting its lack of mass scaling. This is a valid point, albeit the same issue arises in other studies as they also use the \textsc{bynary} software \citep{Hernandez_2022, Hernandez_2023}. This limitation in \citet{Cookson_2024} was spotted by Prof. Milgrom, leading to an amendment in the software. Part of our motivation for the present analysis is to consider the $\widetilde{v}$ parameter in order to account for possible variations in the typical WB mass with $r_{\mathrm{sky}}$, and more generally to ensure relative velocities are appropriately normalized to the predicted Newtonian $v_c$. The study of \citet{Cookson_2024} also has several other flaws which \citet{Hernandez_2025} does not mention (see Table~\ref{Review_table}). The main ones besides the lack of mass scaling are the use of the RMS statistic (which is particularly prone to outliers) and the lack of restriction to systems with $\widetilde{v} < 2.5$, which follows from not calculating $\widetilde{v}$ in the first place.

\end{appendix}

\bsp
\label{lastpage}
\end{document}